\documentstyle[eqsecnum,psfig,aps,prb,multicol]{revtex}
\begin{document} 
\draft

\title{Mott Insulators, Spin Liquids and Quantum Disordered
Superconductivity }
%\title{or, Putting ther Charge Back into Quantum magnetism
%}
\author{Matthew P. A. Fisher} 
\address{Institute for Theoretical Physics, University of 
California,
             Santa Barbara, CA 93106-4030}
\date{\today} 
\maketitle

\begin{abstract}

These introductory lecture notes describe recent results
on novel Mott insulating phases which are ``descendents"
of superconductors - obtained by ``quantum disordering".
After a brief overview of quantum magnetism, attention is focussed
on the spin-liquid phase of the two-leg Hubbard ladder
and the {\it nodal liquid} - a descendent of the $d_{x^2-y^2}$
superconductor.  These notes, which will
appear in a future Les Houches publication, are self-contained
and an effort has been made to keep them accessible.

\end{abstract}
\vskip 0.5 in
\begin{multicols}{2}
\section{Introduction}

At the foundation of the quantum theory of metals is the
theory of the non-interacting electron gas,  
in which the electrons move through the material
interacting only with the periodic potential
of the ions, and not with one another.  
Surprisingly, the properties of most metals
are quite well
described by simply ignoring the strong Coulomb repulsion
between electrons, essentially because Pauli exlusion
severely limits the phase space for
electron collisions.\cite{Mermin}\
But in some cases electron interactions can have
dramatic effects leading to a complete breakdown of the metallic 
state, even when the
conduction band is only partially occupied.
In the simplest such Mott insulator\cite{Mott}
there is only one electron per crystalline unit cell,
and so a half-filled metallic conduction
band would be expected.

With the discovery of the cuprate superconductors in 1986,\cite{Bednorz}\
there has been a resurgence of interest in Mott insulators.
There are two broad
classes of Mott insulators, distinguished by the presence or absence
of magnetic order.\cite{Fradkin,Auerbach}\  More commonly spin rotational invariance is
spontaneously broken, and long-range magnetic order, typically
antiferromagnetic, is realized.\cite{Affleck}\ There are then low
energy spin excitations, the spin one magnons.  Alternatively, in a
spin-liquid\cite{Fradkin} Mott insulator there are no broken symmetries.
Typically, 
the magnetic order is short-ranged and there is a gap to all spin
excitations : a spin-gap.

In the cuprates the Mott insulator is antiferromagnetically ordered,\cite{Ginsberg,Maple} 
but upon doping with holes the antiferromagnetism is rapidly 
destroyed, and above a certain level superconductivity occurs
with $d_{x^2-y^2}$ pairing symmetry.  
But at intermediate doping levels
between the magnetic and d-wave superconducting phases,
there are experimental signs of a spin gap 
opening below a crossover temperature
$T^*(x)$ (see Figure 1).
The ultimate nature of the underlying quantum ground
state in this portion of the phase diagram - 
commonly called the {\it pseudo-gap} regime - is an intriguing puzzle.
More generally,
the apparent connection between a spin-gap and 
superconductivity has been a source of motivation to search for Mott 
insulators of the spin-liquid variety.

Generally, spin liquids are more common 
in low dimensions where quantum fluctuations
can suppress magnetism.
Quasi-one-dimensional ladder materials\cite{Kojima95,Uehara96}\ are promising in this regard
and have received extensive attention, particularly the 
two-leg ladder.\cite{Dagotto96}\  The Mott insulating spin-liquid
phase of the two-leg
ladder can be understood by mapping to an appropriate spin-model - 
the Heisenberg antiferromagnet.  Spin-liquid
behavior results
from the formation
of singlet bond formation across the rungs of the ladder.\cite{Schulz86,Dagotto92}\

Almost without exception, theoretical studies of
spin-liquids {\it start} by mapping to an appropriate
spin-model, and the charge degrees of freedom
are thereafter ignored.  This represents an enormous
simplification, since spin models are so much easier
to analyze that the underlying interacting electron model. 
This approach to quantum magnetism has yielded tremendous
progress in the past decade.\cite{Auerbach}\  
But is the simplification to a spin-model always legitimate?
A central goal of these lectures is to analyze
a novel two-dimensional spin-liquid phase - called a nodal liquid\cite{Nodal,Nodal2} -  which {\it cannot} be described in terms of a spin model.  Although the nodal liquid
is a Mott insulator with a charge gap and has no broken symmetries,
it possesses gapless {\it Fermionic} degrees
of freedom which carry spin. 

\begin{figure}
\psfig{figure=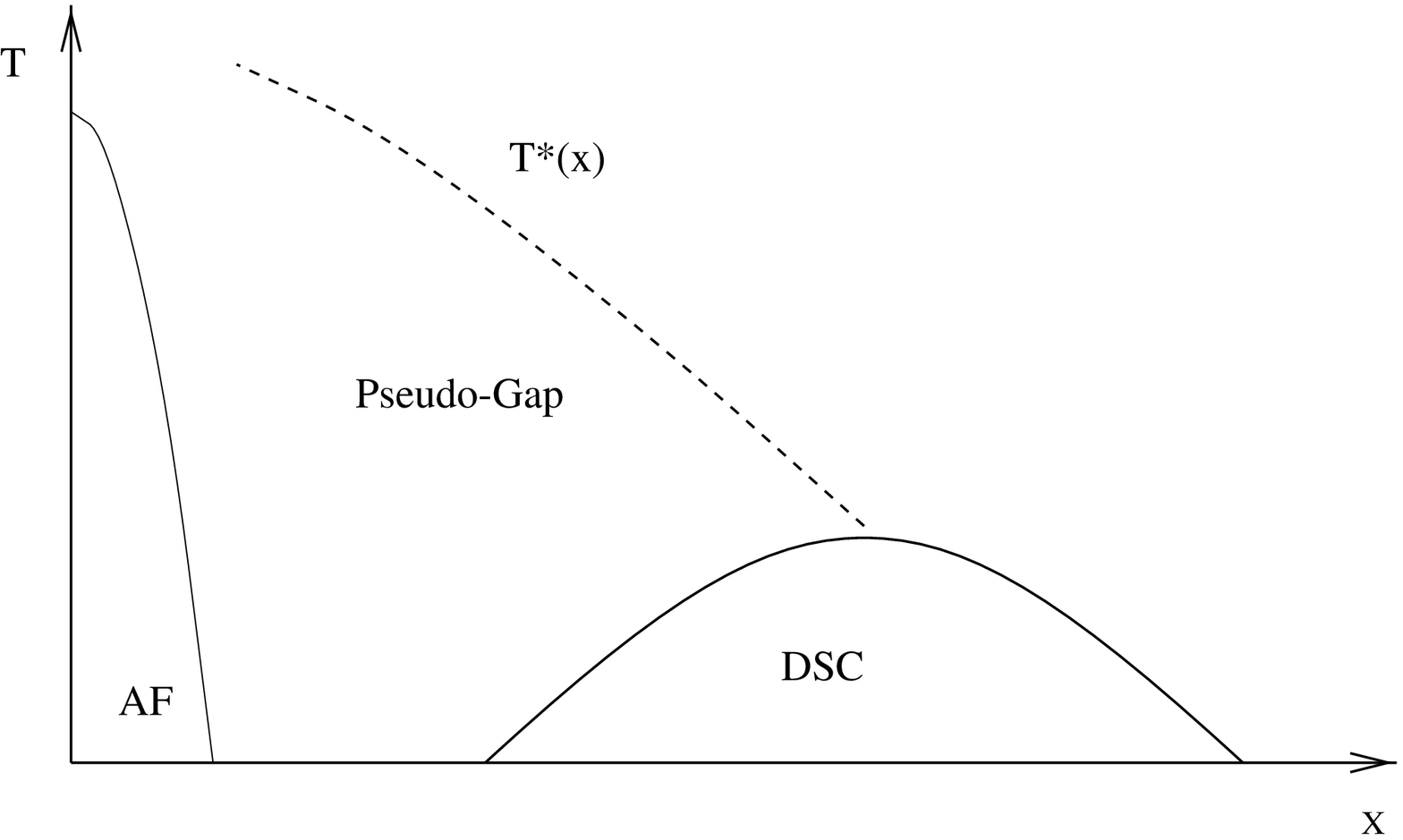,height=2.0in,angle=-90}
\vskip 0.2cm
{Fig.~1: Schematic phase diagram of a high-temperature superconductor
  as a function of doping $x$ and temperature $T$.}
\end{figure}

The standard route to the spin-liquid
invokes quantum fluctuations to
suppress the {\it magnetic} order of a quantum spin-model.\cite{Fradkin}\
The proximity of antiferromagnetism to d-wave superconductivity
in the cuprates suggests an alternate route.  
Indeed, as we shall see,
the nodal liquid
phase results when
a $d-wave$ {\it superconductor} is ``quantum disordered".
The gapless Fermionic excitations
in the nodal liquid are descendents
of the low energy quasiparticles
of the d-wave superconductor.

The spin-liquid phase
of the two-leg ladder gives us a simpler
example of a quantum disordered
superconductor.  To see this, we will
revisit the two-leg ladder, employing
a model of interacting
{\it electrons},\cite{SO8}\ rather than
truncating 
to a spin-model.  Retaining the
charge degrees of freedom will enable us to
show that the Mott-insulating  phase of the two-leg ladder
actually exhibits pairing, with
an approximate d-wave symmetry.
Moreover, upon doping, the two-leg ladder exhibits
quasi-long-range superconducting (d-wave) pairing correlations.
This behavior is reminiscent of that
seen in the underdoped cuprate superconductors.

These notes are organized as follows.
In Section~\ref{sec:Models} a simple tight binding model
of interacting electrons is introduced and it's symmetry properties
are discussed.
Section ~\ref{sec:Mott} specializes to the Mott insulating
state at half-filling, focussing
on the magnetic properties employing
the Heisenberg antiferromagnet spin-model.
In Section ~\ref{sec:Bosonize} the method of Bosonization
is briefly reviewed for the case of a one-dimensionless
spinless electron gas.  Section ~\ref{sec:Two-leg} is devoted
to an analysis of the Mott insulating state of the two-leg Hubbard ladder,
employing a weak coupling perturbative renormalization group
approach.  The remaining sections focus on the
two-dimensional d-wave superconductor, and the nodal liquid
phase which descends from it upon
quantum disordering.  Specifically, Section ~\ref{sec:d-wave} briefly
reviews BCS theory for a d-wave superconductor
focussing on the gapless quasiparticles.  An effective field
theory for this state, including quantum phase fluctuations,
is obtained in Section ~\ref{sec:Effective}.  Section ~\ref{sec:Duality}
implements a duality transformation of this effective
field theory, which enables a convenient description
of the nodal liquid phase in Section ~\ref{sec:Nodal}.

\section{Models and  Metals}
\label{sec:Models}

\subsection{Noninteracting electrons}

In metals the highest lying band
of Bloch states is only partially occupied,
and there are low energy electronic excitations
which consist of exciting electrons
from just below the Fermi energy
into unoccupied states.
These excitations can be thermally excited
and contribute to thermodynamic properties such as 
the specific heat, as well as to electrical conduction.\cite{Mermin}\
Tight binding models give a particularly simple
description of the conduction band.  In the simplest
case the states in the conduction band
are built up from a single atomic orbital
on each of the ions in the solid.
The conduction electrons are presumed to be moving
through the solid, tunnelling between ions.
We denote the creation and annihilation operators
for an electron with spin $\alpha = \uparrow,\downarrow$
on the ion at position $\bbox{x}$ by 
$c_{\alpha}^{\dagger}(\bbox{x})$ 
and $c_{\alpha}^{\vphantom\dagger}(\bbox{x})$.
These operators satisfy the canonical
Fermionic anti-commutation relations,
\begin{equation}
[c_\alpha(\bbox{x}) , c_\beta^\dagger (\bbox{x}') ]_- = \delta_{\alpha \beta}
\delta_{\bbox{x},\bbox{x}'}  .
\end{equation}
If the orbitals in question form
a simple Bravais lattice with, say, cubic symmetry,
then the appropriate tight binding Hamiltonian is,
\begin{equation}
\label{kineticHam}
H_0 = -  t \sum_{\langle \bbox{x}\bbox{x}' \rangle}
\left[c_\alpha^\dagger(\bbox{x})
    c_\alpha^{\vphantom\dagger}(\bbox{x}') + {\rm h.c.}\right] -
\mu \sum_{\bbox{x}} n(\bbox{x}) ,
\end{equation}
where the first summation is over
near neighbor sites.  Here $t$ is the tunnelling rate between
neighboring ions and for simplicity we have
ignored further neighbor tunnelling.  The electron density
$n(\bbox{x}) = c_\alpha^\dagger(\bbox{x})
c_\alpha^{\vphantom\dagger}(\bbox{x})$ can be adjusted by tuning the chemical potential, $\mu$.

In the Cuprate superconductors Copper and Oxygen
atoms form two dimensional sheets,\cite{Ginsberg}
with the Copper atoms 
sitting at the sites of a square lattice
and the Oxygen atoms sitting on the bonds,
as depicted schematically in Figure 2.
In the simplest one-band models the sites of the tight
binding model are taken as the Copper atoms,
and $c^\dagger(\bbox{x})$ {\it removes} an electron
(adds a hole) from a Copper $3d$ orbital.  In most of the materials
the $3d$ shell is almost filled
with roughly one hole per Copper atom,
so that the tight binding model is close
to half-filling with
$\langle n(\bbox{x}) \rangle \approx 1$.

The tight binding Hamiltonian is invariant
under translations by
an arbitrary real space lattice vector, $\bbox{R}$,
\begin{equation}
c_\alpha(\bbox{x}) \rightarrow c_\alpha(\bbox{x} + \bbox{R})  .
\end{equation}
This discrete symmetry implies the conservation of crystal momentum,
up to a reciprocal lattice vector, $\bbox{G}$, with
$exp(i\bbox{G} \cdot \bbox{R}) =1$.  
Being quadratic, the Hamiltonian can be diagonalized by transforming
to (crystal) momentum space by defining,
\begin{equation}
c_\alpha (\bbox{x}) = {1 \over {\sqrt{V}}} \sum_{\bbox{k}} c_{\bbox{k} \alpha}
e^{i\bbox{k} \cdot \bbox{x} }  .
\label{f-trans}
\end{equation}
Here $V$ denotes the ``volume" of the system,
equal to the total number of sites $N$ with the lattice spacing set to unity, and the sum is over crystal momentum within the first Brillouin zone
compatible with periodic boundary conditions. 
The momentum space creation and anihillation operators
also satisfy canonical Fermion anticommutation relations:
\begin{equation}
[c_{\bbox{k} \alpha} , c^\dagger_{\bbox{k}' \beta } ]_- = \delta_{\alpha \beta}
\delta_{\bbox{k} \bbox{k}'}   .
\end{equation} 
In momentum space the Hamiltonian takes the standard 
diagonal form,
\begin{equation}
H_0 = \sum_{\bbox{k} \alpha} \epsilon_{\bbox{k}} c_{\bbox{k}\alpha}^\dagger 
c_{\bbox{k} \alpha}  ,
\end{equation}
invariant under the discrete translation symmetry:
 $c_{\bbox{k} \alpha} \rightarrow e^{i \bbox{k} \cdot \bbox{R}} c_{\bbox{k} \alpha}$.  
For a 2d square lattice with near-neighbor hopping,
the energy is simply
\begin{equation}
\label{dispersion}
\epsilon_{\bbox{k}} = -2t[\cos k_x + \cos k_y] - \mu   .
\end{equation}
The ground state consists of filling those states
in momentum space with $\epsilon_{\bbox{k}}$ negative, leaving the
positive energy states unoccupied.  The Fermi surface,
separating the occupied from empty states,
is determined by the condition $\epsilon_{\bbox{k}} =0$.
For the 2d square lattice at half-filling with energy
dispersion Eqn.~\ref{dispersion} (at $\mu=0$),
the Fermi surface is a diamond, as shown in Figure 2.  
\begin{figure}
\hskip 0.4cm
\psfig{figure=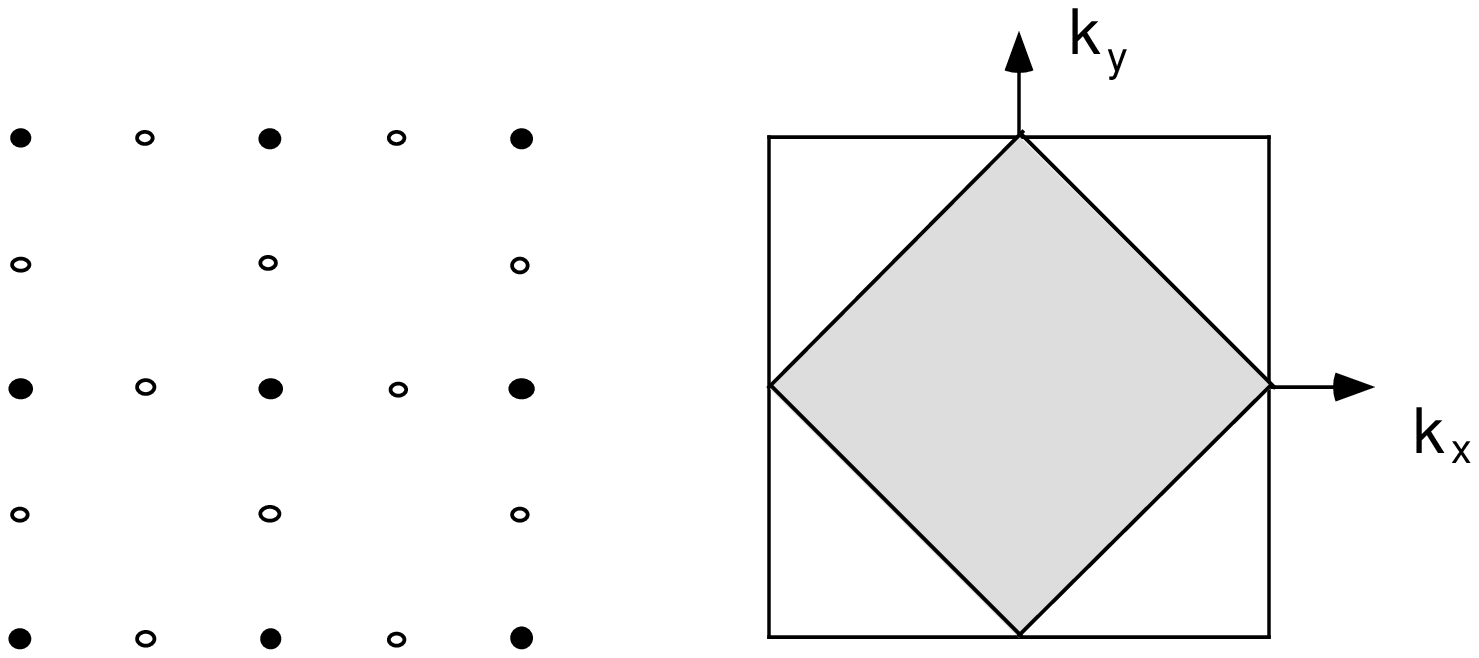,height=3.3cm,width=7.5cm}
\vskip 0.5cm
{Fig.~2: Schematic illustration
of a single Copper-Oxygen plane, consisting of a square
lattice of Copper atoms (solid points)
and Oxygen atoms (open circles).  Two-dimensional Brillouin zone
for the 2d square lattice tight binding
model with near neighbor hopping is shown at right.
At half-filling all states in the Fermi sea (shaded)
are occupied.}
\end{figure}

Particle/hole excitations
above the ground state consist
of removing an electron from within the full
Fermi sea, and placing it in an unoccupied positive
energy state.
In most metals the width of the conduction band (proportional
to $t$) is of order an electron volt (roughly $10^4 K$)
so that even at room temperature only ``low energy"
particle/hole states confined within close proximity to the Fermi surface
are thermally excited.        
In addition to being thermally active, these low
energy particle/hole excitations can be excited by an 
electric field, and lead to metallic
electrical conduction.  

In the band theory of solids, insulators occur whenever
the highest lying energy band is {\it fully} occupied.
Excited states then involve
promoting electrons into the next available band
which typically requires a very large energy (electron volts).
Not surprisingly, such band insulators are very poor
conductors of electricity.
By constrast, in Mott insulators the highest band
is only partially occupied, yet conduction is
blocked by strong electron interactions.

Before addressing the complications of electron interactions, it is
instructive to briefly consider the symmetries of the above Hamiltonian,
and the associated conserved quantities.
There are only two {\it continous} symmetries, associated
with conservation of charge and spin.    
The Hamiltonian is invariant under the global $U(1)$ charge symmetry,
\begin{equation}
\label{U1}
c_\alpha(\bbox{x}) \rightarrow e^{i \theta_0} c_\alpha(\bbox{x}),
\end{equation}
for arbitrary (constant) angle $\theta_0$.  Conservation of spin is
due to the global $SU(2)$ symmetry, 
$c_\alpha(\bbox{x}) \rightarrow U_{\alpha
\beta} c_\beta(\bbox{x})$,
with $U = \exp(i \bbox{\theta} \cdot \bbox{\sigma})$ and Pauli matrices $\bbox{\sigma}_{\alpha \beta}$.
The Hamiltonian is invariant under this transformation,
$H_0 \rightarrow H_0$, for {\it arbitrary}
spin rotations $\bbox{\theta}$.  Here and below we ignore
spin-orbit effects which (usually weakly) break
the continuous spin rotational symmetry.     

There are also a number of discrete symmetries.  The Hamiltonian
is real, $H_0^* = H_0$, a signature of time reversal invariance
(for models with spin-independent interactions).  For
a square lattice the Hamiltonian
is also invariant under reflection (or parity) symmetry,
$c_\alpha(\bbox{x}) \rightarrow c_\alpha(-\bbox{x})$.
This implies that $\epsilon_{\bbox{k}} = \epsilon_{-\bbox{k}}$.
On the square lattice,
a discrete particle/hole transformation is implemented by
\begin{equation}
c_\alpha (\bbox{x}) \stackrel{p/h}{\longrightarrow}
e^{i\bbox{\pi}\cdot\bbox{x}}c_{\alpha}^{\dagger}(\bbox{x}),
\end{equation}
with $\bbox{\pi} = (\pi,\pi)$.  At half-filling when $\mu =0$, $H_0$ is invariant 
under this symmetry,
but with further neighbor hopping terms the kinetic energy
will generally {\it not} 
be particle/hole symmetric.  
In momentum space the particle/hole transformation is implemented via
$c_{\bbox{k} \alpha} \rightarrow c^\dagger_{\bbox{\pi} - \bbox{k} \alpha}$
and invariance of the kinetic energy implies that
$\epsilon_{\bbox{k}} = - \epsilon_{\bbox{k}+ \bbox{\pi}}$.

\subsection{Interaction Effects}

Spin-independent density interactions can be included by
adding an additional term to the Hamiltonian:
\begin{equation}
H_1 = {1 \over 2} \sum_{\bbox{x},\bbox{x}'} v(\bbox{x} - \bbox{x}') n(\bbox{x})
n(\bbox{x}')   .
\end{equation}
For Coulomb interactions
$v(x) \sim e^2 /|x|$ is long-ranged.  For simplicity the long-ranged
interactions are often ignored.  In the Hubbard model\cite{Mott,Auerbach}
only the {\it on-site} repulsive interaction is retained,
\begin{equation}
H_u = u  \sum_{\bbox{x}} n_{\uparrow}(\bbox{x}) n_{\downarrow}(\bbox{x})  ,
\end{equation}
with $n_\alpha = c^\dagger_\alpha c_\alpha$.  
This can be re-cast into a manifestly spin-rotationally
invariant form:
\begin{equation}
H_u = {u \over 2} \sum_{\bbox{x}} n(\bbox{x})[n(\bbox{x}) - 1] .
\end{equation}
Despite the deceptive simplicity of these effective models,
they are exceedingly difficult to analyze.
Even the Hubbard Hamiltonian,
$H= H_0 + H_u$, which is parameterized by just two energy scales,
$t$ and $u$, is largely intractable,\cite{Fradkin}\ except in one-dimension.
Since the typical interaction scale $u$ is comparable to 
the kinetic energy $t$ there is no small parameter.
Moreover, one is typically interested in phenomena
occuring on temperature scales which are {\it much}
smaller than both $u$ and $t$.  

In most metals, the low energy properties are quite well
described by simply ignoring the (strong!) interactions.
This surprising fact can be understood
(to some degree) from Landau's Fermi-liquid theory,\cite{Mermin}\
and more recent renormalization group arguments.\cite{Shankar}\
The key point is that
the phase space available for collisions between
excited particles and holes {\it vanishes}
with their energy.  In metals the phase space
is evidently so restrictive that 
the surviving interactions do
not change the {\it qualitative} behavior of the 
low energy particle/hole excitations.  Indeed,
the quasiparticle excitations within Landau's
Fermi liquid theory have the same quantum numbers
as the electron (charge $e$ spin $1/2$ and momentum),
but move with a ``renormalized" velocity.
But some materials such as the Cuprates
are not metallic, even when band structure considerations
would suggest a partially occupied conduction band.
In these Mott insulators one must invoke electron interactions.

\section{Mott insulators and Quantum Magnetism}
\label{sec:Mott}

The Hubbard model at half-filling is perhaps the simplest
example of a Mott insulator.  
To see this, consider the behavior as the ratio $u/t$ is varied.
As discussed above,
for $u/t=0$ the model is diagonalized in momentum space, and exhibits
a Fermi surface.  But at half-filling
the model is also soluble when $u/t = \infty$.  Since the
onsite Hubbard energy takes the form, $u(n-1)^2/2$,
in this limit the ground state consists simply of one electron on each site.
The electrons are frozen and immobile, 
since doubly occupied and unoccupied sites 
cost an energy proportional to $u$.  
The state is clearly insulating  - a Mott insulator.

In this large $u$ limit
it is very costly in energy 
to add an electron, and the state exhibits a {\it charge gap}
of order $u$.  But there are many low energy {\it spin}
excitations, which consist of flipping the spin of an
electron on a given site.  For infinite $u$ this
spin-one excitation costs no energy at all,
and indeed the ground state is highly degenerate
since the spins of each of the $N$ localized electrons
can be either up or down.  
For large but finite $u/t$ one still expects
a charge gap, but the huge spin degeneracy
will be lifted.  

The fate of the spin degrees of freedom in the Mott insulator
is enormously interesting.  Broadly speaking,
Mott insulators come in two classes,
distinguished by the presence or absence of spontaneously
broken symmetries.  Often the spin rotational
invariance is spontaneously broken and the ground state
is magnetic, but $SU(2)$ invariant spin structures which
break translational symmetries are also possible.
In the second class, usually referred to as {\it spin liquid} states
there are {\it no} broken symmetries.

\subsection{Spin Models and Quantum Magnetism}

Traditionally, spin physics in the Mott insulating
states have been analyzed by studying simple
spin models.  These focus on the electron spin operators:
\begin{equation}
\bbox{S}({\bbox{x}}) = {1 \over 2} c^\dagger_\alpha (\bbox{x}) \bbox{\sigma}_{\alpha \beta}
c_\beta (\bbox{x} )  ,
\end{equation}
where $\bbox{\sigma}$ is a vector of Pauli matrices.
These spin operators satisfy
standard angular momentum commutation relations:
\begin{equation}
[ S_\mu (\bbox{x}) , S_\nu (\bbox{x}' ] = i \delta_{\bbox{x} \bbox{x}'}
\epsilon_{\mu \nu \lambda} S_\lambda   .
\end{equation}
They also satisfy,
\begin{equation}
\bbox{S}^2(\bbox{x})  = {3 \over 4} n(\bbox{x}) [2 - n(\bbox{x})]   .
\end{equation}
Within the restricted sector of
the full Hilbert space with exactly one electron per site,
these operators
are bone fide spin $1/2$ operators 
satisfying $\bbox{S}^2 = s(s+1)$ with $s=1/2$.
Their matrix elements in the restricted Hilbert space
are identical to the Pauli matrices:
$\bbox{\sigma} /2$.  

The simplest spin model consists of a (square) lattice of
spin $1/2$ operators
coupled via a near neighbor exchange interaction, $J$:
\begin{equation}
H = J \sum_{\langle \bbox{x} \bbox{x}' \rangle} \bbox{S}(\bbox{x}) \cdot
\bbox{S}(\bbox{x}')   .
\end{equation}
This spin model can be obtained from the half-filled
Hubbard model,\cite{Auerbach} by working perturbatively
in small $t/u$.  For $t/u=0$ the spins are decoupled,
but an antiferromagnetic exchange
interaction $J=4t^2/u$ is generated at second order in $t$.
Specifically, the matrix elements of the spin Hamiltonian
in the restricted Hilbert space
are obtained by using 
second order perturbation theory
in $t$.  The intermediate virtual states
are doubly occupied, giving an energy denominator $u$.

Mapping the Hubbard model to a spin model represents an {\it enormous}
simplification.  The complications due to the
Fermi statistics of the underlying
electrons have been subsumed into an exchange interaction.
The spin operators are essentially {\it bosonic},
commuting at different sites.
It should be emphasized that at higher order
in $t/u$ multi-spin exchange interactions
will be generated, also between further separated spins.
If $t/u$ is of order one, then it is by no means obvious
that it is legitimate to truncate to a spin model at all.
 
A central focus of quantum magnetism
during the past decade has been exploring the possible
ground states and low energy excitations of such spin models.\cite{Fradkin,Auerbach}
The above $s=1/2$ square lattice Heisenberg antiferromagnet is, of course,
only one member of a huge class of such
models.  These models can be generalized
to larger spin $s$, to different lattices and/or dimensionalities,
to include competing or frustrating interactions, to include
multi-spin interactions, to ``spins" in different groups
such as $SU(N)$, etc..
Not surprisingly, there is an almost equally rich set
of possible ground states.  

The main focus of these notes is the 2d ``nodal liquid", a spin-liquid
phase obtained by quantum disordering a d-wave
superconductor.  As we shall see in Sec.~\ref{sec:Nodal}, in the nodal
liquid the spin excitations are carried by {\it Fermionic}
degrees of freedom and cannot
be described by (Bosonic) spin operators.
In truncating to the restricted
Hilbert space with one electron per site,
one has effectively ``thrown out the baby with the bath water".
The nodal liquid phase {\it requires} retaining
the charge degrees of freedom.  

But spin models are {\it much} simpler than
interacting electron models, relevant to many if not most
Mott insulators (as well as other localized spin systems) and extremely
rich and interesting in their own right.
So I would like to
briefly summarize some of the possible ground states,
focussing on spin $1/2$ models on
bi-partite lattices.\cite{Auerbach,Affleck}\
Consider first those ground states
with spontaneously broken symmetries.  Most common
is the breaking of spin-rotational invariance.
If the spin operators are treated as {\it classical} fixed
length vectors, which is valid in the large spin limit
($s \rightarrow \infty$), the ground state
of the near neighbor square lattice antiferromagnet
is the Neel state (up on one sublattice, down
on the other) which breaks the $SU(2)$ symmetry.  
For finite $s$ the Neel state is {\it not}
the exact ground state, but the ground state
is still antiferromagnetically ordered, even
for $s=1/2$.  Quantum fluctuations play
a role in reducing the sub-lattice magnetization,
but (for the 2d square lattice) do not drive it to zero.
The low energy
excitations are {\it gapless} spin-waves (ie. magnons),
as expected when a continuous symmetry
is spontaneously broken.  

For some spin models the ground state
is spin rotationally invariant but
spontaneously breaks (discrete) translational symmetry.
The classic example is the Majumdar-Ghosh Hamiltonian,\cite{Auerbach}\
\begin{equation}
H_{MG} = J \sum_x [\bbox{S}(x) \cdot \bbox{S} (x+1) + {1 \over 2}
\bbox{S}(x) \cdot \bbox{S} (x+2)]  ,
\end{equation}
which describes a one dimensional $s=1/2$ Heisenberg
antiferromagnetic spin chain with a second neighbor
exchange interaction.  The exact ground state of this model
is a two-fold degenerate ``spin-Peierls" state:
\begin{equation}
|G\rangle_{\pm} = \prod_x [|\uparrow_{2x} \rangle |\downarrow_{2x \pm 1}\rangle
- |\downarrow_{2x} \rangle |\uparrow_{2x \pm 1}\rangle ]   .
\end{equation}
This state consists of a
product of ``singlet bonds"
formed from neighboring pairs of spins, and breaks 
invariance under translations by one lattice spacing.
Since the singlet bonds are rotationally invariant,
the $SU(2)$ symmetry remains unbroken.
The second neighbor interaction
has effectively suppressed the tendency towards
antiferromagnetic order.

\subsection{Spin Liquids}

{\it Spin liquid} ground states in which {\it no}
symmetries are broken generally occur more readily
in low dimensions where quantum fluctuations
are more effective at destroying magnetic order.     
The one-dimensional $s=1/2$ chain
with near neighbor antiferromagnetic exchange
exhibits power law magnetic correlations at the
antiferromagnetic wave vector $\pi$.\cite{Affleck}\  Although ``almost"
magnetically ordered the $SU(2)$ symmetry is {\it not}
broken in the ground state, which thus technically
qualifies as a spin liquid.  More dramatic
is the behavior of the $s=1/2$ antiferromagnetic
two-leg ladder, shown in Figure 3.  This model exhibits
a featureless spin-rotationally invariant ground state
with exponentially decaying spin correlation fuctions
and a non-zero energy {\it gap} for all spin excitations.\cite{Schulz86,Dagotto92}\
The physics can be best understood in the 
limit in which the exchange interaction across the rungs
of the ladder greatly exceeds the intra-leg
exchange:  $J_{\perp} >> J$.  When $J=0$ the ground state
consists of singlet bonds formed across the rungs of the ladder,
with triplet excitations separated by an energy gap of order $J_{\perp}$.
Perturbing in small $J$ will cause these singlet bonds
to ``resonate", but one expects the
spin gap to survive at least for $J << J_{\perp}$.
It turns out that the ground state
evolves adiabatically and smoothly with increasing $J$,
and in fact the spin-liquid survives for arbitrarily
large $J_{\perp}/J$.

\begin{figure}
\hskip0.5cm
\psfig{figure=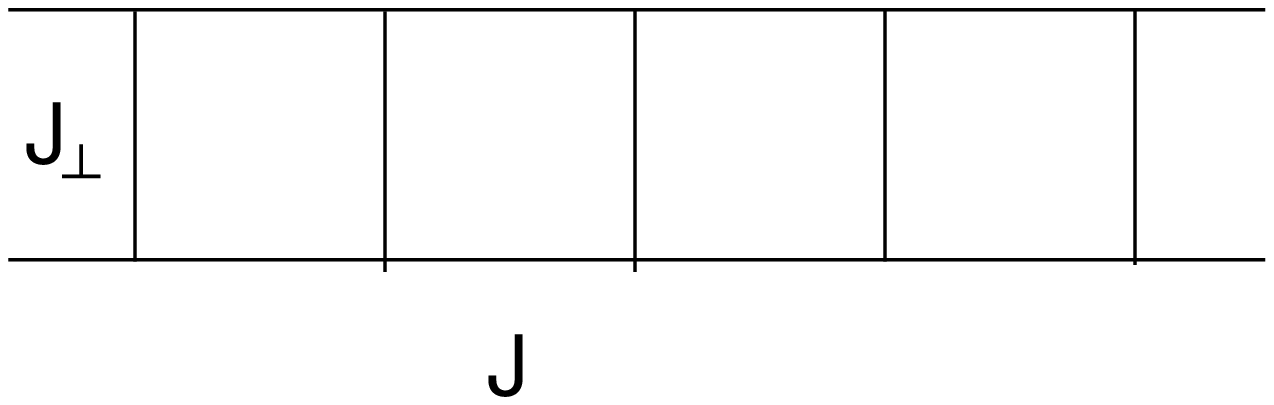,height=2.5cm,width=7cm}
\vskip 0.2cm
{Fig.~3:  Heisenberg spin model on a two-leg ladder.
Spin $1/2$ operators sit on the sites
of the ladder, interacting via
an antiferromagnetic exchange
$J$ along the ladder and $J_{\perp}$
across the rungs.
}
\end{figure}

There has been an enormous amount of theoretical effort
expended searching for {\it two-dimensional}
spin $1/2$ models which exhibit spin-liquid ground states
analogous to the two-leg ladder - but with little success.
The original motivation soon after
the discovery of superconductivity
in the Cuprates was based on Anderson's ideas\cite{Anderson}
that a Mott insulating spin-liquid exhibits ``pre-formed" Cooper
pairing.  Doping the Mott insulator would give
the Cooper pairs room to move and to condense into a
superconducting state,
presumed to have s-wave pairing symmetry.  But it soon became clear that
the undoped Mott insulator in the Cuprates is {\it not} a
spin-liquid, but actually antiferromagnetically ordered.
Moreover, recent experiments have established that
the pairing symmetry in the superconducting
phase is d-wave rather than s-wave.\cite{Wollman,Kirtley}\

However, recent theoretical work\cite{SO8} 
(see Section ~\ref{sec:Two-leg} below)
has established that the pairing in the spin-liquid
phase of the two-leg ladder actually has (approximate)
d-wave symmetry.  Moreover, doping this Mott insulator
does indeed give the pairs room to move,\cite{Balents96,Lin97}\
and they form a one-dimensional d-wave ``superconductor"
(with quasi-long-ranged pairing correlations).
The nodal liquid phase\cite{Nodal,Nodal2} discussed extensively below
is a two-dimensional analog of this spin-liquid
phase.  Indeed, we shall explicitly construct the nodal
liquid by quantum disordering a two-dimensional d-wave superconductor.
As we shall see, the resulting 2d nodal liquid posesses gapless
Fermionic excitations, which are descendents of the d-wave
quasiparticles.  These Fermions carry spin but no charge.
The nodal liquid presumably cannot
be the ground state of any
(Bosonic) spin-model.  To describe the nodal liquid one must
employ the underlying
interacting electron model which retains 
the charge degrees of freedom.

Recent experiment has focussed attention on
the underdoped regime of the Cuprate materials,\cite{Maple}
occuring between the antiferromagnetic and
superconducting phases (see Fig. 1).  In this pseudo-gap
regime insulating behavior is seen at
low temperatures,
and there are indications for a spin gap - behavior
reminiscent of a Mott insulating spin-liquid.
We have suggested\cite{Nodal} that this strange phase
can perhaps be understood in terms
of a doped nodal liquid.

Before discussing further the 2d nodal liquid, it is instructive
to revisit the spin liquid phase of the two-leg ladder
and analyze it directly with a model of interacting electrons.
Specifically, we consider weak interactions
(small $u/t$), a limit in which truncation
to a spin model is {\it not} possible.  This analysis
is greatly aided by ``Bosonization" - a powerful
method which enables an interacting electron model
in one dimension to be re-formulated in
terms of collective Bosonic degrees of freedom.
See Ref.~\onlinecite{Emery79}-\onlinecite{Delft}
as well as Fradkin's book\cite{Fradkin} for useful reviews of Bosonization.
First, in Sec.~\ref{sec:Bosonize}  we briefly review Bosonization
for the simplest case of a spinless one-dimensional
electron gas, before turning to 
the two-leg ladder in Sec.~\ref{sec:Two-leg}.

\section{Bosonization Primer}
\label{sec:Bosonize}

Consider the Hamiltonian for non-interacting
spinless electrons hopping on a 1d lattice,
\begin{equation}
H = -t \sum_x c^\dagger(x) c(x+1) + h.c. 
\label{Hamlatt}
\end{equation}
with hopping strength $t$. 
One can diagonalize this Hamiltonian by Fourier transforming to momentum
space as in Eqn.~\ref{f-trans}, giving
\begin{equation}
H = \sum_k \epsilon_k c_k^\dagger c_k  ,
\end{equation}
with energy dispersion 
$\epsilon_k = -t \cos(k)$ for
momentum $|k| < \pi$,
as shown in Figure 4.  In the ground state
all of the negative energy states with momentum  $|k| \le k_F$
are occupied.  At half-filling the Fermi wavevector
$k_F = \pi/2$.  An effective low energy theory for these excitations
can be obtained by focussing on momenta close to $\pm k_F$
and defining {\it continuum} Fermi fields:
\begin{equation}
\psi_{\scriptscriptstyle R}(q) = c_{k_F+q} ; \hskip0.75cm   \psi_{\scriptscriptstyle L}(q) = c_{-k_F + q}  .
\end{equation}
Here the subscripts $R/L$ refer to the right/left Fermi points,
and $q$ is assumed to be smaller than a momentum
cutoff, $|q| < \Lambda$ with $\Lambda << k_F$.
One can then linearize
the dispersion about the Fermi points, writing 
$\epsilon_{\pm k_F + q} = \pm v_F q$ with $v_F$ the Fermi velocity.
It is convenient to transform back to real
space, defining fields
\begin{equation}
\psi_{\scriptscriptstyle P}(x) = {1 \over \sqrt{V}} \sum_{|q| <\Lambda} e^{iqx} \psi_{\scriptscriptstyle P}(q) ,
\end{equation}
(with $P=R,L$) which vary slowly on
the scale of the lattice spacing.
This is equivalent to 
expanding the lattice electron operators
in terms of continuum fields,
\begin{equation}
c(x) \sim \psi_{\scriptscriptstyle R}(x) e^{ik_F x} + \psi_{\scriptscriptstyle L}(x) e^{-ik_Fx}  .
\label{continuum}
\end{equation}
After linearization, the effective low energy Hamiltonian
takes the form, $H = \int dx {\cal H}$, with Hamiltonian density,
\begin{equation}
{\cal H} = - v_F [
\psi^{\dag}_{{\scriptscriptstyle R}} i\partial_{x} 
\psi^{}_{{\scriptscriptstyle R}}
- \psi^{\dag}_{{\scriptscriptstyle L}}
i\partial_{x}\psi^{}_{{\scriptscriptstyle L}} ] .
\end{equation}
describing a one-dimensional relativistic
Dirac particle.  The associated Lagrangian density is simply
\begin{equation}
{\cal L} = \psi^{\dag}_{{\scriptscriptstyle R}} i\partial_{t} 
\psi^{}_{{\scriptscriptstyle R}} + \psi^{\dag}_{{\scriptscriptstyle L}}
i\partial_{x}\psi^{}_{{\scriptscriptstyle L}} - {\cal H} .
\end{equation}

\begin{figure}
\hskip 1.2cm
\psfig{figure=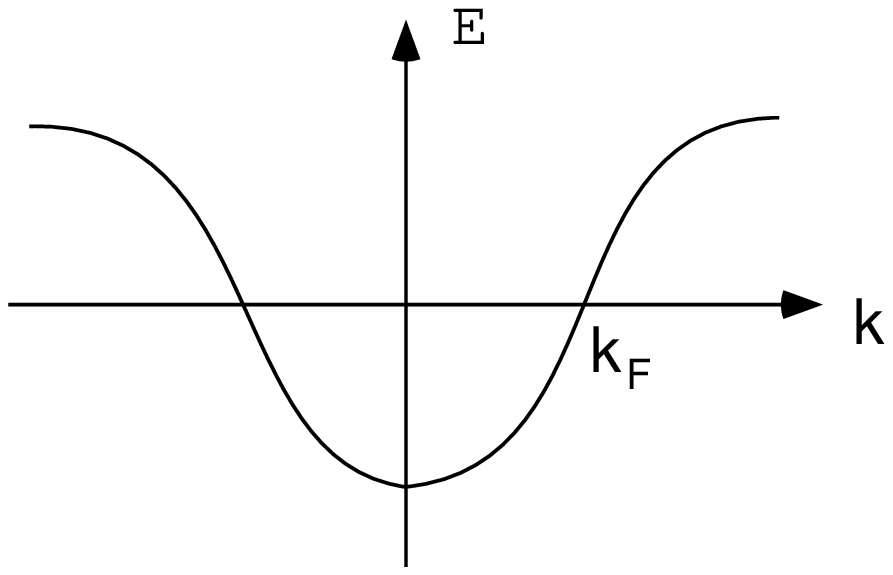,height=3cm,width=5cm}
\vskip 0.2cm
{Fig.~4:  Schematic energy dispersion for the
one-dimensional electron gas.  The negative energy
states are occupied, with momentum $|k| < k_F$.  The dispersion
can be linearized around $\pm k_F$, leading to a continuum Dirac
Fermion theory.}
\end{figure}

Consider a particle/hole 
excitation about the right
Fermi point, where an electron is removed from a state with $k<k_F$
and placed into an unoccupied
state with $k+q>k_F$. For small momentum change $q$, the energy of this
excitation is
$\omega_q = v_F q$. Together with the
negative momentum excitations
about the left Fermi point, this linear dispersion relation is identical to
that for phonons in
one-dimension. The method of Bosonization exploits this similarity by
introducing a phonon
displacement field, $\theta$, to decribe this linearly dispersing density
wave.\cite{Emery79,Shankar95}\  We follow the heuristic development of
Haldane\cite{Haldane}, which reveals the important physics, dispensing with
mathematical rigor. To this end, consider a Jordan-Wigner transformation\cite{Fradkin} which
replaces the electron operator, $c(x)$, by a (hard-core) boson operator,
\begin{equation}
c(x) = {\cal O}(x) b(x) \equiv exp[i\pi \sum_{x' < x} n(x')] b(x)    .
\end{equation}
where $n(x) = c^\dagger(x) c(x)$ is the
number operator. One can easily verify that the Bose operators commute at
different sites. Moreover, the lattice Hamiltonian 
Eqn.~\ref{Hamlatt} can be re-expressed
in terms of these Bosons, and takes the identical form with $c's$ replaced by $b's$. This transformation, exchanging
Fermions for Bosons, is a special feature of one-dimension. The Boson operators
can be (approximately) decomposed in terms of an amplitude and a phase, 
\begin{equation}
b(x) \rightarrow \sqrt{\rho} e^{i\varphi}  .
\end{equation}
We now imagine passing to the continuum
limit, focussing on scales long compared to the lattice spacing. In this limit
we decompose the total density as,
$\rho(x) = \rho_0 + \tilde{\rho}$, where the mean density, $\rho_0 = k_F/\pi$, and
$\tilde{\rho}$ is an operator measuring fluctuations in the density. As usual,
the density and phase are canonically conjugate quantum variables, taken to
satisfy
\begin{equation}
[ \varphi(x), \tilde{\rho}(x')] = i \delta(x-x') .
\label{commutation}
\end{equation}
Now we introduce a phonon-like displacement field, $\theta(x)$, via
$\tilde{\rho}(x) =\partial_x \theta(x) /\pi$.
The full density takes the form: $\pi \rho(x) = k_F +
\partial_x \theta$. The above commutation relations are satisfied if one
takes,
\begin{equation}
[\varphi(x),\theta(x^\prime)]=-i\pi \Theta(x^\prime-x)  .
\end{equation}
Here $\Theta(x)$ denotes the 
heavyside step function, not to be confused
with the displacement field $\theta$.
Notice that $\partial_x \varphi/\pi$ is the momentum conjugate to $\theta$. 

The effective (Bosonized) Hamiltonian density which describes the 1d density wave
takes the form:
\begin{equation}
{\cal H} = {v \over {2\pi}} [ g (\partial_x \varphi)^2 + g^{-1} (\partial_x \theta)^2] .
\label{BoseHam}
\end{equation}
This Hamiltonian describes a wave propagating at velocity $v$, as can be readily
verified upon using the commutation relations to obtain the equations of motion,
$\partial_t^2\theta = v^2
\partial_x^2 \theta$, and similarly for $\varphi$. 
Clearly one should equate
$v$ with the Fermi velocity, $v_F$.
The additional dimensionless parameter, $g$, 
can be determined as follows.
A small variation in density,
$\tilde{\rho}$, will lead to a change in energy, $E =\tilde{\rho}^2/2\kappa$,
where $\kappa = \partial\rho/\partial \mu$ is the compressibility. Since
$\partial_x\theta = \pi\tilde{\rho}$, one deduces from ${\cal H}$ that
$\kappa = g/\pi v$. But for a non-interacting electron gas, $\pi v \kappa = 1$,
so that $g=1$.  In the presence of (short-ranged) interactions
between the (spinless) electrons, one can argue that
the above Hamiltonian density remains valid, but with renormalized values
of both $g$ and $v$.  This Hamiltonian would
then describe a (spinless) Luttinger liquid,\cite{Haldane,Luttinger} rather than the free
electron gas.

The power of Bosonization relies on the ability
to re-express the electron operator $c(x)$
in terms of the Boson fields.
Clearly $c(x)$ must
remove a unit charge ($e$) at $x$, 
and satisfy Fermion anticommutation relations.
Consider first the {\it Bose} operator, $b \sim \exp (i\varphi)$,
which removes unit charge.
To see this,
note that one can write,
\begin{equation}
e^{i\varphi(x)} = e^{i \pi \int_{- \infty}^x dx' P(x')} ,
\label{shiftoperator}
\end{equation}
where $P =\partial_x \varphi/\pi$ is the momentum conjugate to $\theta$. Since
the momentum operator is
the generator of translations (in $\theta$), this creates a kink in
$\theta$ of height $\pi$ centered at
position $x$ - which corresponds to a localized unit of charge since the
density $\tilde{\rho} =\partial_x
\theta /\pi$.
To construct the (Fermionic) electron operator requires multiplying this Bose operator
by a Jordan-Wigner ``string": 
\begin{equation}
{\cal O}(x) = e^{i\pi \sum_{x'<x} n(x')}
\rightarrow e^{i \pi
\int^x \rho(x')} = e^{i(k_F x + \theta)}   .
\end{equation}
Since this string operator carries momentum $k_F$,
the resulting Fermionic operator ${\cal O} e^{i\varphi}$ 
should be identified with the right moving
continuum Fermi field, $\psi_{\scriptscriptstyle R}$.
We have thereby identified
the correct Bosonized form for the 
(continuum) electron operators:
\begin{equation}
\psi_{\scriptscriptstyle P}(x) = e^{i\phi_{\scriptscriptstyle P}(x)};
\hskip0.5cm 
\phi_{{\scriptscriptstyle P}} = 
 \varphi + P \theta  ,
\label{bosondef2}
\end{equation}
with $P=R/L =\pm$.
From Eqn.~\ref{commutation} the {\it chiral} Boson fields
$\phi_{\scriptscriptstyle P}$ can be shown to satisfy
the so-called Kac-Moody commutation realtions:
\begin{eqnarray}
[\phi_{{\scriptscriptstyle P}}(x),
\phi_{{\scriptscriptstyle P}}(x^\prime)]  =&  
i P \pi &{\rm sgn}(x-x^\prime),
\label{cm1}
\\
\left[\phi_{{\scriptscriptstyle R}}(x),
\phi_{{\scriptscriptstyle L}}(x^\prime)\right] =&
 i\pi  &.
\label{cm2}
\end{eqnarray}
These commutation relations can be used
to show that $\psi_{\scriptscriptstyle R}$ and 
$\psi_{\scriptscriptstyle L}$ anticommute.

It is instructive to re-express the Bosonized Hamiltonian
density in terms of the {\it chiral} boson fields,
\begin{equation}
{\cal H} = \pi v_F [n_{\scriptscriptstyle R}^2 + n_{\scriptscriptstyle L}^2 ] ,
\label{chiralHam}
\end{equation}
where we have defined right
and left moving densities
\begin{equation}
n_{\scriptscriptstyle P} = P {1\over {2\pi}} \partial_x\phi_{\scriptscriptstyle P}   ,
\label{leftright}
\end{equation}
which sum to give the total density, $n_R + n_L =
\tilde{\rho}$.  These chiral densities can be 
expressed in terms of the chiral
electron operators as,
\begin{equation}
n_{\scriptscriptstyle P} = :\psi_{\scriptscriptstyle P}^\dagger \psi_{\scriptscriptstyle P}: \equiv \psi_{\scriptscriptstyle P}^\dagger \psi_{\scriptscriptstyle P} - \langle \psi_{\scriptscriptstyle P}^\dagger \psi_{\scriptscriptstyle P}
\rangle   .
\end{equation}
Notice that the Bosonized Hamiltonian decouples 
into right and left moving sectors.  

An advantage of Bosonization is the ease with which
electron interactions can be incorporated.
Consider a (short-range) density-density interaction
added to the original lattice Hamiltonian.
Using Eqn.~\ref{continuum} this can be decomposed
into the continuum Dirac fields,
and will be quartic and spatially local.
Due to momentum conservation, only
three terms are possible: Two chiral terms
of the form $(\psi_{\scriptscriptstyle P}^\dagger \psi_{\scriptscriptstyle P})^2$ with $P=R/L$,
and a right/left mixing term of the form, $\psi_{\scriptscriptstyle R}^\dagger \psi_{\scriptscriptstyle R} \psi_{\scriptscriptstyle L}^\dagger \psi_{\scriptscriptstyle L}$.  Under Bosonization the chiral
terms are proportional to $(\partial_x \phi_ {\scriptscriptstyle P})^2$,
and can be seen to simply shift the Fermi velocity in
Eqn.~\ref{chiralHam}.  The right/left mixing term
also Bosonizes into a quadratic form proportional to
$(\partial_x \theta)^2 - (\partial_x \varphi)^2$.
When added to the Hamiltonian in Eqn.~\ref{BoseHam},
this term can be absorbed by shifting
{\it both} the Fermi velcocity {\it and} the dimensionless Luttinger
parameter, $g$, which is then no longer equal to one.
For repulsive interactions $g <1$,
whereas $g>1$ with attractive interactions.
This innocuous looking shift
in $g$ has profound effects
on the nature of the electron correlation functions.  In fact,
it leads to new chiral operators which
have fractional charge, $g e$.  The resulting
one-dimensional phase is usually called a ``Luttinger liquid".\cite{Haldane}
For electrons with spin or for 1d models with multiple
bands, the quartic Fermion operators can have even more
dramatic consequences, for example opening up
energy gaps as we shall see in Sec.~\ref{sec:Two-leg}.

The Lagrangian density in the Bosonized representation
takes the form of a free scalar field,
\begin{equation}
\label{1dLagrangian}
{\cal L} = {g \over 2} \kappa_\mu  (\partial_\mu \varphi)^2  ,
\end{equation}
with $g=1$ for the free Fermion gas,
and $g \ne 1$ in the interacting Luttinger liquid.
The Greek index $\mu$ runs over time and the spatial
coordinate, $\mu = 0,1 = t,x$.  Here $\kappa_0= 1/\pi v$
and $\kappa_1 = - v^2 \kappa_0$.  When re-expressed in terms of $\theta$   
the Lagrangian takes the {\it identical} form, except
with $g \rightarrow 1/g$ for the Luttinger liquid.
Changing from the $\varphi$ to the $\theta$ representation
can be viewed as a {\it duality} transformation.  In Sec.~\ref{sec:Duality}
we will consider
an analogous duality transformation in {\it two} spatial
dimensions.

\section{2 Leg Hubbard Ladder}
\label{sec:Two-leg}

\subsection{Bonding and antibonding bands}

We now consider electrons hopping on a two-leg ladder 
as shown in Fig. 5.
The kinetic energy takes the form,
\begin{equation}
H_0 = - t \sum_{\langle \bbox{x}\bbox{x}' \rangle}
\left[c_\alpha^\dagger(\bbox{x})
    c_\alpha^{\vphantom\dagger}(\bbox{x}') + {\rm h.c.}\right] -
\mu \sum_{\bbox{x}} n(\bbox{x})  ,
\end{equation}
where $n(\bbox{x}) = c_\alpha^\dagger(\bbox{x})
c_\alpha^{\vphantom\dagger}(\bbox{x})$, and
the summation is taken over near neighbors
on the two-leg ladder, with $y=1,2$.
Due to a parity symmetry under interchange
of the two legs of the ladder, it is convenient to consider
even and odd parity bonding and anti-bonding operators: 
\begin{equation}
b_{\alpha}(x) = {1 \over \sqrt{2}} [c_{\alpha}(x,y=1) + c_{\alpha}(x,y=2)]     , \end{equation}
\begin{equation}
a_{\alpha}(x) = {1 \over \sqrt{2}} [c_{\alpha}(x,y=1) - c_{\alpha}(x,y=2)]     , \end{equation}
which depend only on the coordinate $x$ along the ladder.
The Hamiltonian splits into even and odd
contributions,
$H_0 = H_a(a) + H_b(b)$.
Each is a one-dimensional tight binding model
which can be readily diagonalized by transforming to momentum space,
\begin{equation}
b(x) = {1 \over \sqrt{N} } \sum_k b_k e^{ikx}  ,
\end{equation}
and similarly for the anti-bonding operator.  Here
$N$ denotes the number of sites {\it along} the ladder.
The diagonal form is
\begin{equation}
H_0 = \sum_k [\epsilon^a_k a_{k\alpha}^\dagger a_{k\alpha} +
\epsilon^b_k b_{k\alpha}^\dagger b_{k\alpha} ] ,
\end{equation}
which describes {\it two} one-dimensional bands with dispersion
$\epsilon^{a/b}_k = -2t \cos k \pm t - \mu$.  These are sketched
in Figure 5.

\begin{figure}
\psfig{figure=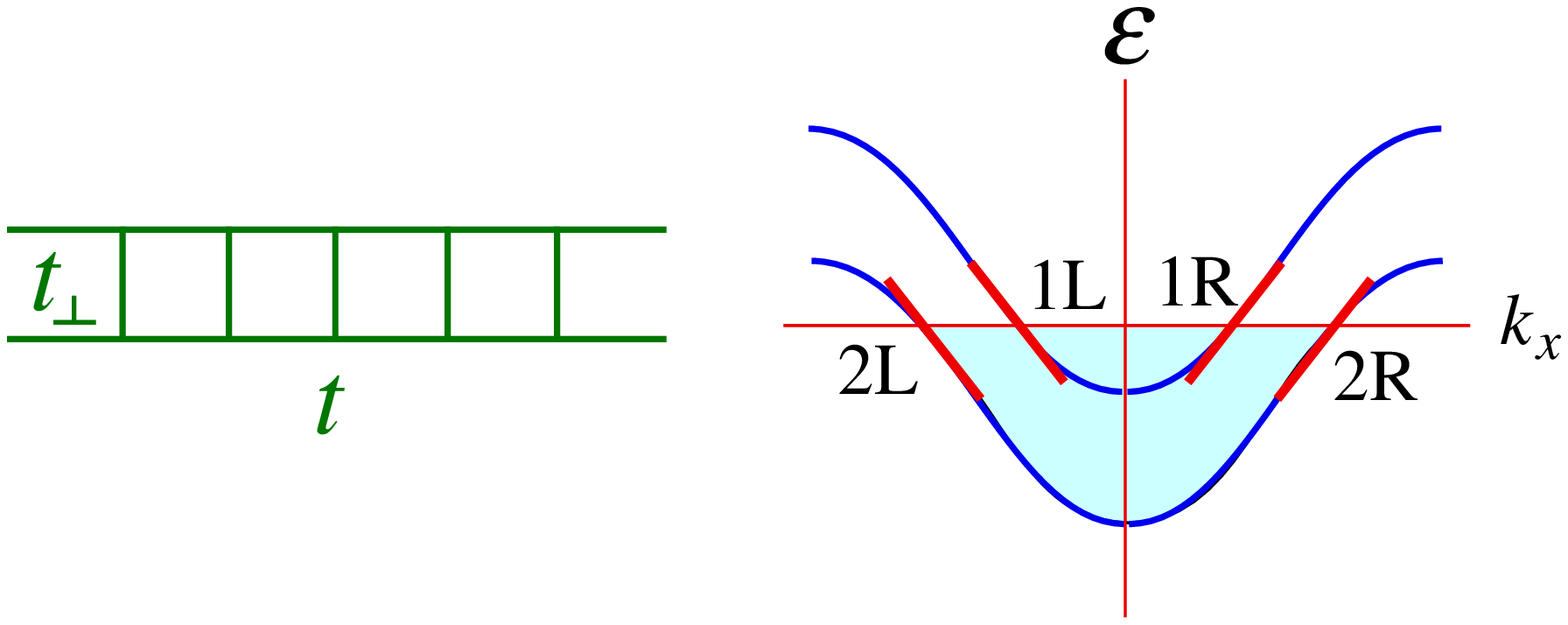,height=3.5cm,width=8.5cm,angle=0}
\vskip 0.2cm
{Fig.~5: A two-leg ladder and its band
structure.  In the low-energy limit, the energy dispersion is
linearized near the Fermi points. The two resulting 
relativistic Dirac Fermions are distinguished by pseudospin indices
$i=1,2$ for the anti-bonding and bonding bands, respectively.}
\end{figure}

Focussing on the case at half-filling with one electron per site
($\mu =0$), both 
bands intersect the Fermi energy, $\epsilon_F =0$.
There are four Fermi {\it points}
at $\pm k_{F1}$ and $\pm k_{F2}$, for the antibonding
and bonding bands, respectively.  
Gapless particle/hole excitations
exist at each of the four Fermi points.  
Due to particle/hole symmetry present with near neighbor hopping,
$\epsilon^a_k + \epsilon^b_{k+\pi}=0$, which
implies that $k_{F1} + k_{F2} = \pi$.  Moreover,
the Fermi velocity in each band is the same, hereafter
denoted as $v$.  It is instructive to plot these Fermi points
in {\it two-dimensional} momentum space, taking
transverse momentum $k_y = 0,\pi$ for the two
bands, as shown in Figure 6.  The four Fermi points can be viewed
as constant $k_y$ slices through a two-dimensional
Fermi surface. 

As we shall see, with even weak
electron interactions present the gapless Fermi points
are unstable, and a gap opens in the spectrum.  Of interest
are the properties of the resulting Mott insulator.   
As discussed in Section ~\ref{sec:Mott}, for {\it strong} interactions
mapping to a spin model is possible, and the
electron spins across the rungs of the ladder
are effectively locked into singlets:
\begin{equation}
|RS\rangle = {1 \over \sqrt{2}} [ c^\dagger_\uparrow (1) c^\dagger_\downarrow(2) - c^\dagger_\downarrow (1) c^\dagger_\uparrow(2) ] 
|0 \rangle ,
\end{equation}
where $y=1,2$ refers to the two legs of the ladder,
and we have suppressed the rung position $x$.
The state $|0 \rangle$ denotes a rung with no electrons.

\begin{figure}
\hskip 1.5cm
%\centerline{\epsfbox{figures/fold_BZ.eps}} \vspace{5pt} 
\psfig{figure=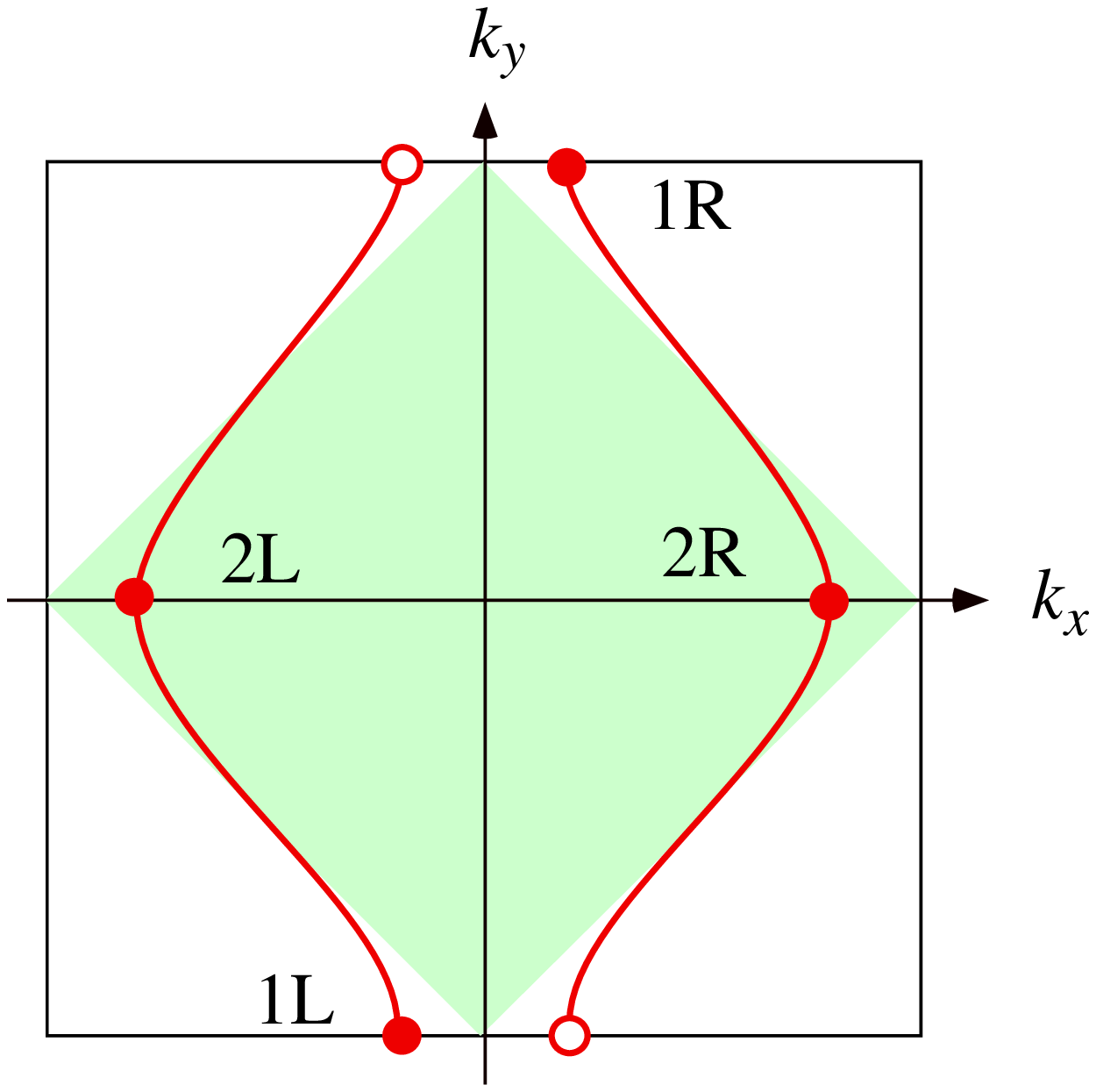,height=2.0in,angle=0}
\vskip 0.2cm
{Fig.~6:  Fermi points for the two-leg ladder
plotted in the two-dimensional Brillouin zone,
with the antibonding band (denoted $1$) at $k_y = \pi$
and the bonding band ($2$) at $k_y =0$.  The shaded region
represents the Fermi sea for a {\it two-dimensional}
square lattice model at half-filling.}
\end{figure}

It is extremely instructive to re-express this rung-singlet state
in terms of the bonding and anti-bonding operators.  One finds,
\begin{equation}
|RS \rangle = {1 \over \sqrt{2}} [ b^\dagger_\uparrow b^\dagger_\downarrow
- a^\dagger_\uparrow a^\dagger_\downarrow ] |0 \rangle  ,
\end{equation}
a linear combination of adding
a singlet (Cooper) pair into the bonding and antibonding orbitals.
This paired form is suggestive of superconductivity.
Indeed, when viewed in momentum space,
the ground state of a superconductor is a product
of singlet pairs with zero center of mass momentum at
different points around the Fermi surface.  
In an s-wave superconductor, the pairs are all
added with the {\it same} sign,
but if the pairs are formed with a relative angular momentum
(eg. d-wave) sign changes are expected.  
But notice the
most important relative {\it minus} sign in the rung singlet state!
The spin-liquid phase of the two-leg ladder is evidently
related to a paired superconductor with
{\it non-zero} angular momentum.  Since
pairing in the bonding band at $k_y=0$ has a positive
sign and pairing in the anti-bonding band at $k_y=\pi$
is negative, in the two-dimensional Brillouin zone
(see Figure 6) the sign is proportional to $k_x^2 - k_y^2$,
consistent with a so-called $d_{x^2 - y^2}$ pairing symmetry.

If the interactions are weak, it is legitimate to focus
on electronic states near the Fermi points.
As in Section \ref{sec:Bosonize}, the electron operators
can be conveniently decomposed into
continuum fields near the Fermi points which
vary slowly on the scale of the lattice.
Denoting $c_1 = a$ and $c_2 = b$, the bonding
and antibonding operators are expanded as,
\begin{equation}
  c^{}_{i\alpha} \sim \psi^{}_{{\scriptscriptstyle R} i \alpha}
 e^{ik_{{\scriptscriptstyle F}i}x} + 
  \psi^{}_{{\scriptscriptstyle L} i \alpha} 
e^{-ik_{{\scriptscriptstyle F}i}x},
  \label{decompose_electrons}
\end{equation}
with $i=1,2$.
Upon linearizing the spectrum around the four Fermi points
the kinetic energy takes the form, $H_0 = \int dx {\cal H}_0$, with 
Hamiltonian density,
\begin{equation}
{\cal H}_{0}= -v \sum_{i,\alpha} [
\psi^{\dag}_{{\scriptscriptstyle R}i\alpha} i\partial_{x} 
\psi^{}_{{\scriptscriptstyle R}i\alpha}
- \psi^{\dag}_{{\scriptscriptstyle L}i\alpha}
i\partial_{x}\psi^{}_{{\scriptscriptstyle L}i\alpha} ] .
\label{kinetics_f}
\end{equation}

This Hamiltonian describes massless Dirac Fermions, with four flavors labelled 
by band and spin indices.  Implicit in this theory
is a momentum cutoff, $\Lambda$,
whose inverse exceeds the lattice spacing.  
Only modes with momentum $|k| < \Lambda$
are included in these continuum fields.  Since the spectrum is massless,
this simple theory is ``critical" and scale invariant behavior
is expected.  This can be seen
by considering the (Euclidian) action,
written as a space-time integral of the Lagrangian density, 
\begin{equation}
S = \int d\tau dx {\cal L}_0  ,
\end{equation}
\begin{equation}
{\cal L}_0 = \sum_{{\scriptscriptstyle P}  \alpha} \psi^\dagger_{{\scriptscriptstyle P} i \alpha} i \partial_\tau \psi_{{\scriptscriptstyle P} i \alpha} + {\cal H}_0   ,
\end{equation}
with $P=R/L$, and $\tau$ denoting imaginary time.  
The partition function, $Z=Tr exp(-\beta H_0)$,
can be expressed as a (coherent state Grassman)
path integral,\cite{Negele}\
\begin{equation}
Z = \int [D\psi][ D\bar{\psi}] e^{-S(\bar{\psi},\psi)}   .
\end{equation}
A simple renormalization group can be implemented\cite{Goldenfeld,Shankar}
by first integrating out fields $\psi(k,\omega)$
with momentum $k$ lying in the interval
$\Lambda/b < |k| < \Lambda$, with rescaling parameter $b>1$.
Since modes
with different momentum and frequency are
not coupled, the action takes the same form
after this integration, except with a smaller
momentum cutoff, $\Lambda/b$.
The renormalization group transformation is completed
by a rescaling procedure which returns the cutoff
to it's original value:
\begin{equation}
\label{rescale}
x \rightarrow bx ; \hskip0.25cm \tau \rightarrow b \tau ;\hskip0.25cm 
\psi \rightarrow b^{-1/2} \psi  .
\end{equation}
The field rescaling has been chosen
to leave the action invariant.
This simple theory is at a renormalization group
{\it fixed point}.

\subsection{Interactions}

Electron-electron interactions scatter right-moving electrons into 
left-moving electrons and vice-versa.
We consider general finite-ranged spin-independent 
interactions, but 
assume that the typical interaction strength, $u$, is weak -- much smaller 
than the bandwidth.  We focus on the effects of the interactions to 
{\it leading} non-vanishing order in $u$.  In this limit it is 
legitimate to keep only those pieces of the interactions which 
scatter 
the low energy Dirac Fermions.  
A general four Fermion interaction on the
two-leg ladder (such as the Hubbard $u$)
can be readily decomposed in terms of
the continuum Dirac fields.  
It is instructive to see how these quartic
terms in $\psi(x)$ transform under the rescaling
transformation Eqn.~\ref{rescale}.  A simple
quartic term with no spatial gradients is seen
to be invariant, so that these operators
are ``marginal" under the renormalization group.
The corresponding interaction strengths will 
``flow" under the renormalization group transformation
due to non-linear interaction effects. 
On the other hand, a quartic term involving
gradients such as 
$u_2 (\psi^\dagger \partial_x \psi)^2$,
would rapidly scale to zero
under rescaling:  $u_2 \rightarrow u_2/b^2$,
and can thus be ignored.  Moreover, 
four-Fermion interactions which are 
chiral, say only scattering right movers, do {\it not}
renormalize to lowest order in $u$
and can thus also be neglected\cite{Balents96,Lin97}\.   
As discussed in Sec.~\ref{sec:Bosonize}, these terms
simply lead to small shifts in the Fermi velocity.
All of 
the remaining four-Fermion interactions can be conveniently expressed 
in terms of currents, defined as
\begin{eqnarray}
 J_{ij} =& \psi^{\dag}_{i\alpha} \psi^{}_{j\alpha}, 
 \qquad
 \bbox{J}_{ij}=&\frac12 \; \psi^{\dag}_{i\alpha} 
 \bbox{\sigma}_{\alpha\beta} \psi^{}_{j\beta};
 \label{J_def}
 \\
  I_{ij} =& \psi_{i\alpha} \epsilon_{\alpha\beta} \psi_{j\beta},
 \qquad
 \bbox{I}_{ij}=&\frac12 \; \psi_{i\alpha} 
(\epsilon \bbox{\sigma})_{\alpha\beta} \psi_{j\beta},
\label{I_def}
\end{eqnarray}
where the $R,L$ subscript has been suppressed. 
Both $J$ and $I$ are invariant under global
$SU(2)$ spin rotations, whereas 
$\bbox{J}$ and $\bbox{I}$ rotate as $SU(2)$ vectors.
Due to Fermi statistics, some
of the currents are (anti-)symmetric
\begin{equation}
I_{ij}=I_{ji} \qquad \bbox{I}_{ij} = -\bbox{I}_{ji},
\label{I_symm}
\end{equation}
so that 
$\bbox{I}_{ii}=0$ (no sum on $i$).

The full set of marginal momentum-conserving four-Fermion interactions
can be written
\begin{eqnarray} 
{\cal H}^{(1)}_{I} = && b^{\rho}_{ij} J_{{\scriptscriptstyle R}ij}
 J_{{\scriptscriptstyle L}ij} -
b^{\sigma}_{ij} \bbox{J}_{{\scriptscriptstyle R}ij} 
\cdot \bbox{J}_{{\scriptscriptstyle L}ij},
\nonumber\\
+&&f^{\rho}_{ij} J_{{\scriptscriptstyle R}ii} 
J_{{\scriptscriptstyle L}jj} -
f^{\sigma}_{ij} \bbox{J}_{{\scriptscriptstyle R}ii} 
\cdot \bbox{J}_{{\scriptscriptstyle L}jj}.
\label{int1}
\end{eqnarray} 
Here $f_{ij}$ and $b_{ij}$ denote the forward and
backward (Cooper) scattering amplitudes, respectively, between bands
$i$ and $j$.  Summation on $i, j=1,2$ is implied. To avoid double
counting, we set $f_{ii}=0$ (no sum on $i$). 
Hermiticity implies $b_{12}=b_{21}$ and parity symmetry
($R \leftrightarrow L$) gives $f_{12}=f_{21}$,
so that there
are generally eight independent couplings $b^{\rho,\sigma}_{11}$, 
$b^{\rho,\sigma}_{22}$, $b^{\rho,\sigma}_{12}$, 
and $f^{\rho,\sigma}_{12}$.
At half-filling with particle/hole symmetry
$b_{11} = b_{22}$.  Additional momentum non-conserving Umklapp
interactions of the form
\begin{equation} 
{\cal H}^{(2)}_{I} = 
u^{\rho}_{ij} I^{\dag}_{{\scriptscriptstyle R}ij} 
I_{{\scriptscriptstyle L}{\hat i}{\hat j}} -
u^{\sigma}_{ij} \bbox{I}^{\dag}_{{\scriptscriptstyle R}ij} \cdot 
\bbox{I}_{{\scriptscriptstyle L}{\hat i}{\hat j}}
+ {\rm h.c.}
\label{int2}
\end{equation} 
are also allowed, (here $\hat{1}=2, \hat{2}=1$). Because the currents
$(\bbox{I}_{ij}), I_{ij}$ are (anti-)symmetric, one can always choose
$u_{12} = u_{21}$ for convenience. We also take
$u^{\sigma}_{ii}=0$ since $\bbox{I}_{ii}=0$.
With particle/hole symmetry there are thus just three
independent Umklapp vertices, $u^{\rho}_{11}$, $u^{\rho}_{12}$, 
and $u^{\sigma}_{12}$.  Together with the six forward and backward
vertices, nine independent couplings are required to describe
the most general set of marginal non-chiral four-Fermion interactions
for a two-leg ladder with particle/hole symmetry at half-filling.

The renormalization group transformation described above can be implemented
by working perturbatively for small
interaction parameters\cite{Balents96,SO8}.
Upon systematically integrating out high-energy modes away from the Fermi 
points and then rescaling the spatial coordinate and Fermi fields, a 
set of renormalization group (RG) transformations can be derived for 
the interaction strengths.  Denoting the nine interaction strengths 
as $g_i$, and setting the
rescaling parameter $b=1 + d\ell$ with
$d\ell$ infinitesimal,
the leading order differential RG flow equations take the general form, 
\begin{equation}
\partial_\ell g_i = A_{ijk} g_j g_k ,
\end{equation}
valid up to order $g^3$.  The matrix of coefficients
$A_{ijk}$ is given explicitly in Ref.~\onlinecite{SO8}. 
 
These nine coupled non-linear differential equations
are quite complicated, but can be integrated numerically
starting with initial values appropriate to a lattice
interaction (such as the Hubbard interaction).
This integration reveals
that some of the couplings remain small, while others tend to 
increase, 
sometimes after a sign change, and then eventually diverge.  Quite 
surprisingly, though, the ratios of the growing couplings tend to 
approach fixed constants, which are {\it independent} of the 
initial 
coupling strengths, at least over a wide range in the nine 
dimensional 
parameter space.  These constants can be determined by inserting the 
Ansatz,
\begin{equation}
g_i(\ell) = {g_{i0} \over {(\ell_d - \ell)}}  ,
\label{power_law}
\end{equation}
into the RG flow equations, to obtain nine {\it algebraic} equations 
quadratic in the constants $g_{i0}$.  There are various distinct 
solutions of these algebraic equations, or rays in the nine-dimensional 
space, which correspond to different possible phases.  But for 
generic {\it repulsive} interactions between the electrons on the two-leg 
ladder, a numerical integration reveals that the flows are 
essentially 
always attracted to one particular ray.\cite{SO8}\  
This is the spin-liquid phase
of interest, which we refer to as a {\sl d-Mott} phase.
In the d-Mott phase,
two of the nine coupling
constants, $b^{\rho}_{11}$ and 
$f^{\sigma}_{12}$, remain small,
while the other seven grow large with fixed ratios:
\begin{eqnarray}
b^{\rho}_{12} = \frac14 b^{\sigma}_{12} = f^{\rho}_{12} 
= -\frac14 b^{\sigma}_{11} = 
\\
2u^{\rho}_{11} = 2u^{\rho}_{12} = \frac12 u^{\sigma}_{12} = g >0  .
\label{ratio}
\end{eqnarray}

Once the ratio's are fixed, there is a single remaining coupling 
contant, denoted $g$, which measures the distance from the origin 
along a very special direction (or ``ray") in the nine dimensional 
space of couplings.  The RG equations reveal that as the flows scale 
towards strong coupling, they are {\it attracted} to this special 
direction.  If the initial bare interaction parameters are 
sufficiently weak, the RG flows have sufficient ``time" to 
renormalize 
onto this special ``ray", before scaling out of the regime of 
perturbative validity.  In this case, the low energy physics, on the 
scale of energy gaps which open in the spectrum, is {\it universal}, 
depending only on the properties of the physics along this special 
ray, and independent of the precise values of the bare interaction 
strengths.

\subsection{Bosonization}

To determine the properties of the resulting d-Mott phase, it is
extremely helpful to Bosonize the theory.
As discussed in Sec.~\ref{sec:Bosonize} the (continuum) electron fields can expressed in terms
of Boson fields:
\begin{equation}
\psi^{}_{{\scriptscriptstyle P}i\alpha}= 
\kappa^{}_{i\alpha} 
e^{i\phi^{}_{{\scriptscriptstyle P}i\alpha} } ; \hskip0.5cm
\phi_{{\scriptscriptstyle P}i\alpha} = \varphi_{i\alpha} + P\theta_{i\alpha} ,
\label{bosonization}
\end{equation}
with $P=R/L = \pm$.  The displacement field 
$\theta_{i\alpha}$ and phase field 
$\varphi_{i\alpha}$ satisfy the commutation relations
\begin{equation}
[\varphi_{i\alpha} (x),\theta_{j \beta}(x^\prime)]=-i\pi \delta_{ij}
\delta_{\alpha \beta} \Theta(x^\prime-x)  .
\label{thetaphicom}
\end{equation}
Klein factors, satisfying
\begin{equation}
\{ \kappa^{}_{i\alpha}, \kappa^{}_{j\beta} 
\}=2\delta_{ij}\delta_{\alpha\beta},
\end{equation}
have been introduced so that the Fermionic operators 
in different bands or 
with different spins anticommute with one another.
When the Hamiltonian is Bosonized, the Klein factors
only enter in the combination,
$\Gamma = \kappa_{1 \uparrow} \kappa_{1 \downarrow} \kappa_{2 
\uparrow} \kappa_{2 \downarrow}$.
Since 
$\Gamma^2 = 1$, one can take $\Gamma = \pm 1$.  Hereafter, we will put 
$\Gamma = 1$.

The Bosonized form for the kinetic energy 
Eq.~\ref{kinetics_f} is
\begin{equation}
{\cal H}_{0} = \frac{v}{2\pi}\sum_{i,\alpha}  [
 (\partial_{x}\theta_{i\alpha})^{2}
+ (\partial_{x}\varphi_{i\alpha})^{2} ],
\end{equation}
which describes density waves propagating 
in band $i$ and with spin $\alpha$.

This expression can be conveniently separated into
charge and spin modes, by defining
\begin{eqnarray}
\theta_{i \rho} &=& (\theta_{i \uparrow} 
+ \theta_{i \downarrow} )/\sqrt{2}  
\label{bosondef1}\\
\theta_{i \sigma } &=& (\theta_{i \uparrow} 
- \theta_{i \downarrow} )/\sqrt{2} ,
\end{eqnarray}
and similarly for $\varphi$.  The $\sqrt{2}$ ensures that these new 
fields
satisfy the same commutators, Eq. (\ref{thetaphicom}).
It is also convenient to combine the fields in the two bands
into a $\pm$ combination, by defining
\begin{equation}
\theta_{\mu \pm} = ( \theta_{1 \mu} \pm \theta_{2 \mu} ) / \sqrt{2}  ,
\end{equation}
where $\mu = \rho,\sigma$, and similarly for $\varphi$.

The Hamiltonian density ${\cal H}_0$ can now be re-expressed in a 
charge/spin and flavor decoupled form,
\begin{equation}
{\cal H}_{0} = \frac{v}{2\pi}\sum_{\mu, \pm} [
 (\partial_{x}\theta_{\mu \pm})^{2}
+ (\partial_{x}\varphi_{\mu \pm})^{2} ].
\end{equation}
The fields $\theta_{\rho+}$ and $\varphi_{\rho +}$ describe the total 
charge and current fluctuations, since under Bosonization, 
$\psi^\dagger_{{\scriptscriptstyle P}i\alpha} 
\psi^{}_{{\scriptscriptstyle P}i\alpha} = 2\partial_x 
\theta_{\rho+}/\pi$ 
and $vP \psi^\dagger_{{\scriptscriptstyle P}i\alpha} 
\psi^{}_{{\scriptscriptstyle P}i\alpha} = 
2\partial_x 
\varphi_{\rho+}/\pi$.

While it is possible to Bosonize
the interaction Hamiltonians in full generality,\cite{SO8}\
we do not reproduce it here.
In addition to terms quadratic
in gradients of the Boson fields (as in ${\cal H}_0$),
the Bosonized interaction consists
of terms bi-linear
in $\cos 2\theta$ and $\cos 2\varphi$.
More specifically, of the eight non-chiral Boson
fields ($\theta_{\mu\pm}$ and $\varphi_{\mu \pm}$) only five enter as arguments of cosine terms.  In the momentum conserving terms these
are $\theta_{\sigma \pm}$, $\varphi_{\rho -}$ and $\varphi_{\sigma -}$.
The Umklapp terms also involve the overall
charge displacement field, via $cos 2\theta_{\rho +}$.
This can be understood by 
considering how the Boson fields
transform under a spatial translation, $x \rightarrow x + x_0$.
The chiral electron operators transform
as $\psi_{{\scriptscriptstyle P}i} \rightarrow \psi_{{\scriptscriptstyle P}i}
e^{ipk_{Fi} x_0} $,
which is equivalent to $\theta_{i\alpha} \rightarrow \theta_{i\alpha}
+ k_{Fi} x_0$.  Three of the charge/spin and flavor
fields are thus invariant under spatial translations, whereas 
$\theta_{\rho +} \rightarrow \theta_{\rho +} + \pi x_0$.  
The 
momentum conserving terms are invariant under
spatial translations, so {\it cannot}
depend on $\cos 2\theta_{\rho+}$.

The full interacting theory is invariant under spatially constant shifts of the
remaining three Boson fields - 
$\varphi_{\rho+},
\varphi_{\sigma+}$ and $\theta_{\rho-}$.
For the first two of these, the conservation
law responsible for this symmetry is readily apparent.  Specifically,
the operators $\exp(iaQ)$ and $\exp(iaS_z)$, with $Q$ the total
electric charge and $S_z$ the total z-component of spin, generate
``translations" proportional to $a$ in the two fields
$\varphi_{\rho+}$ and $\varphi_{\sigma+}$.  To see this, we note that
$Q = \int dx \rho(x)$ with $\rho(x) = 2\partial_x \theta_{\rho+}/\pi$
the momentum conjugate to $\varphi_{\rho+}$, whereas $S_z$ can be
expressed as an integral of the momentum conjugate to
$\varphi_{\sigma+}$.  Since the total charge is conserved, $[Q, H]=0$,
the full Hamiltonian must therefore be invariant under
$\varphi_{\rho+} \rightarrow \varphi_{\rho+} + a$ for arbitrary
constant $a$, precluding a cosine term for this field.  Similarly,
conservation of $S_z$ implies invariance under $\varphi_{\sigma+}
\rightarrow \varphi_{\sigma+} + a$.  

The five Boson fields entering as arguments of various 
cosine terms will tend to be pinned at the minima of these 
potentials.  Two of these 5 fields, $\theta_{\sigma-}$ and 
$\varphi_{\sigma -}$, are dual to one another so that the uncertainty 
principle precludes pinning both fields.  Since there are various 
competing terms in the potential seen by these 5 fields, minimization 
for a given set of bare interaction strengths is generally 
complicated.  
However, along the special ray in the nine dimensional
space of interaction parameters the nine independent coupling constants 
can be replaced by a {\it single} parameter $g$.
The resulting Bosonized
theory is found to reduce to a very simple and highly
symmetrical form when expressed in terms
of a new set of Boson fields, defined by
\begin{eqnarray}
(\theta, \varphi)_{1} =&(\theta, \varphi)_{\rho+},
\qquad 
(\theta, \varphi)_{2}=&(\theta, \varphi)_{\sigma+},
\nonumber\\
(\theta, \varphi)_{3}=&(\theta, \varphi)_{\sigma-}, 
\qquad
(\theta, \varphi)_{4}=&(\varphi, \theta)_{\rho-}.
\label{mott_field}
\end{eqnarray}
The first three are simply the charge/spin and flavor fields
defined earlier.  However, in the fourth pair of fields,
$\theta$ and $\varphi$ have been interchanged.

In terms of these new fields, the full interacting Hamiltonian
density along the special ray takes an exceedingly simple form:
${\cal H} = {\cal H}_0 + {\cal H}_I$, with
\begin{equation}
{\cal H}_0 = { v \over {2\pi}} \sum_a  [(\partial_x \theta_a )^2
+ ( \partial_x \varphi_a)^2 ]  ,
\label{free_boson}
\end{equation}
\begin{eqnarray}
{\cal H}_{I} &=& \frac{g}{2\pi^{2}} \sum_{a}
[(\partial_{x}\theta_a)^2 - (\partial_{x} \varphi_a)^2 ]
\nonumber\\
&&-4g \sum_{a \neq b} \cos 2\theta_{a} \cos 2\theta_{b}.
\label{int_boson}
\end{eqnarray}

\subsection{d-Mott Phase}

We now briefly discuss some of the general physical properties
of the d-Mott phase which 
follow from this Hamiltonian.  Ground state properties  can be inferred by 
employing semi-classical considerations.  Since the fields 
$\varphi_a$ 
enter quadratically, they can be integrated out
when the partition function is expressed as a path integral
over Boson fields.  This leaves an effective 
action in terms of the four fields $\theta_a$.  Since the single 
coupling constant $g$ is marginally relevant and flowing off to 
strong 
coupling, these fields will be pinned in the minima of the cosine 
potentials.  Specifically, there are two sets of semiclassical 
ground states with all $\theta_{a} =n_{a}\pi$ or all 
$\theta_{a} =(n_{a}+1/2)\pi$, where $n_{a}$ are integers.
It can be shown\cite{SO8} that these different solutions actually correspond
to the {\it same} physical state, so that the ground state
is unique.
Excitations will be 
separated from the ground state by a finite energy gap, since the 
fields are harmonically confined, and instanton excitations 
connecting 
different minima are also costly in energy.

Consider first those fields which are pinned
by momentum conserving interaction terms.
Since both $\theta_{\sigma \pm}$ fields are pinned, so are the 
spin-fields in each band, $\theta_{i \sigma}$ ($i=1,2$).  Since 
$\partial_x \theta_{i\sigma}$ is proportional to the z-component of 
spin in band $i$, a pinning of these fields implies that the spin in 
each band vanishes, and excitations with non-zero spin are expected 
to 
cost finite energy: the spin gap. This can equivalently be 
interpreted as singlet pairing of electron pairs in each band. It is 
instructive to consider the pair field operator in band $i$:
\begin{equation}
\Delta_i = \psi^{}_{{\scriptscriptstyle R}i \uparrow} 
\psi^{}_{{\scriptscriptstyle L}i \downarrow} = 
\kappa_{i \uparrow} \kappa_{i \downarrow} 
e^{ i \sqrt{2}  (\varphi_{i\rho} + \theta_{i\sigma})}  .
\end{equation}
With $\theta_{i \sigma} \approx 0$, $\varphi_{i\rho}$ can be 
interpreted as the phase of the pair field in band $i$.  The relative 
phase of the pair field in the two bands follows by considering the 
product
\begin{equation}
\Delta^{\vphantom\dagger}_1 \Delta^\dagger_2 = -\Gamma 
e^{i 2\theta_{\sigma -}} e^{i2\varphi_{\rho -}}  ,
\label{pairsign}
\end{equation}
with $\Gamma = \kappa_{1 \uparrow} \kappa_{1 \downarrow}\kappa_{2 
\uparrow} \kappa_{2 \downarrow} =1$.  Since $\theta_4 = \varphi_{\rho 
-}$ the relative phase is also pinned by the cosine potential, with a 
sign change in the relative pair field, $\Delta^{\vphantom\dagger}_1 
\Delta^\dagger_2 < 0$, corresponding to an approximate d-wave symmetry.  

To discuss the physics of the remaining overall charge
mode ($\theta_{\rho +}$), it is convenient to 
first imagine ``turning off" the Umklapp
interactions.  After pinning the other
three fields to the minima of the cosine
potentials, the pair field operator in band $i$ becomes
\begin{equation}
\Delta_i \sim (-1)^i e^{i \varphi_{\rho+}}   ,
\end{equation}
so that $\varphi_{\rho+}$ is the phase of the pair
field.
In the absence of Umklapp scattering,
the Lagrangian for this phase field
is simply,
\begin{equation}
{\cal L} = {1 \over 2} \kappa_\mu  (\partial_\mu \varphi_{\rho +})^2  .
\label{Lvarphi}
\end{equation}
Being in one-spatial dimension, these gapless
{\it phase} fluctuations lead to power law decay
of the pair field spatial correlation function,
$\Delta^*(x) \Delta(0) \sim 1/x^\eta$.  
A true superconductor (for $d>1$) exhibits (off-diagonal)
long-ranged order, and this correlation
function would not decay to zero even as $x \rightarrow \infty$.  
But in one-dimension a ``superconductor" can at best
exhibit power law decay, since
true off-diagonal long-ranged order is not possible.\cite{Auerbach}\
Thus, in the absence of Umklapp scattering the
2-leg ladder would be a one-dimensional d-wave
``superconductor".  

But what is the effect of the momentum
non-conserving Umklapp interactions?
Once the other three fields are pinned
in the minima of the cosine potentials in the above
Hamiltonian Eqn.~\ref{int_boson}, the 
Umklapp scattering terms take the simple form,
\begin{equation}
{\cal H}_u = -12g \cos 2\theta_{\rho +}  .
\end{equation}
This term tends to pin the field $\theta_{\rho +}$.
The pair field phase, $\varphi_{\rho +}$,
being the {\it conjugate} field
will fluctuate wildly.  These
quantum flucutations will destroy the power-law
1d ``superconducting" phase, leading to
an exponentially decaying pair-field correlation function.
What is the fate of this one-dimensional ``quantum disordered
d-wave superconductor"?

To see this, one simply has to consider the ``dual"
representation in terms of the $\theta_{\rho +}$ field,
rather than $\varphi_{\rho +}$.  
A {\it lattice} version of this duality transformation
is carried out in detail in the Appendix.
Alternatively, one can obtain the dual theory directly from the
Bosonized Hamiltonian Eqn.~\ref{free_boson}.
The appropriate Lagrangian dual
to Eqn.~\ref{Lvarphi} above, is simply
\begin{equation}
\label{Ltheta}
{\cal L} = {1 \over 2} \kappa_\mu  (\partial_\mu \theta_{\rho +})^2  ,
\end{equation}
which describes gapless {\it density} waves.  These
density flucutations will be pinned
by the Umklapp terms in $H_u$, leading to a Mott insulator
with a gap
to charge excitations.  Since there is also a spin-gap
this phase is equivalent to the spin-liquid,
discussed at strong coupling in terms 
of the Heisenberg model in Section ~\ref{sec:Mott}.  But we now see
that this spin-liquid phase exhibits superconducitng d-wave pairing
correlations, despite being an insulator.  
The spin-liquid phase can thus be described
as a quantum disordered one-dimensional d-wave
``superconductor".

The Euclidian action associated with the phase Lagrangian in Eqn.~\ref{Lvarphi}
is equivalent to the effective Hamiltonian
in the low temperature phase of the classical 2d xy model,
(with imaginary time playing the role of a second spatial coordinate).
The 2d xy model can be disordered
by introducing vortices into the phase of
the order parameter.\cite{Jose}\  For this it is convenient to
go to a dual representation.\cite{Amit}\
As shown explicitly in the Appendix, the
dual represention is equivalent to the $\theta_{\rho +}$ representation,
with the strength of the Umklapp term
playing the role of a vortex fugacity.
In Section ~\ref{sec:Duality}, we will quantum disorder
a {\it two-dimensional} d-wave superconductor,
and it will be extremely convenient to consider
a duality transformation - a three dimensional
version of the 2d $\theta \leftrightarrow \varphi$ duality discussed here.  
The resulting nodal liquid phase will be particularly
simple to analyze in the dual representation.

\subsection{Symmetry and Doping}

Due to the highly symmetric form of the Hamiltonian
in Eqn.~\ref{free_boson} and \ref{int_boson}, it is possible to
make considerable further progress in analyzing it's properties.
Indeed, as shown in Ref.~\onlinecite{SO8}, under a re-Fermionization
procedure this Hamiltonian
is equivalent to the $SO(8)$ Gross-Neveu model,\cite{Gross74}
which has been studied extensively by particle field theorists.
The $SO(8)$
Gross-Neveu model posesses a remarkable symmetry known as triality,\cite{Shankar80}
which can be used to equate the energies of various excited states.
In particular, the energy
of the lowest excited state with the quantum numbers of an electron
(charge $e$ and $s=1/2$) is {\it equal} to the 
energy of the lowest lying spinless charge $2e$ exited state (a Cooper pair).
This beautifully demonstrates  {\it pairing}
in the insulating d-Mott phase:  The energy
to add two electrons of opposite spin far apart
is twice as large as the energy to add them into
a Cooper pair bound state.  It turns out, moreover,
that the Gross-Neveu model is {\it integrable}\cite{Zamolodchikov79}
so it is possible to fully enumerate the energies
and quantum numbers of {\it all} the
low energy excited states\cite{SO8} (grouped into $SO(8)$ multiplets)
and compute exactly various correlation functions.\cite{Ludwig98}\

We finally briefly mention the effects of doping
the d-Mott phase away from half-filling.
This can be achieved by adding a chemical potential
term to the Hamiltonian in Eqn.~\ref{free_boson} and \ref{int_boson},
with $H_\mu = H -\mu Q$, where $Q$ is the {\it total}
electric charge:
\begin{equation}
Q = {2 \over \pi} \int \partial_x \theta_{\rho +} .
\end{equation}
Since the field $\theta_{\rho +}$ is pinned
in the cosine potential by the Umklapp interaction terms, $H_u$,
for small $\mu$ the density will stay fixed at half-filling.
Eventually, $\mu$ will pass through the Mott charge gap and
the density will change.  This occurs via
$\pi$ instantons in $\theta_{\rho +}$,  connecting
adjacent minima of the cosine potential.  Each instanton carries
charge $2e$, but no spin, so can be intepreted
as a Cooper pair.  In this doped phase, the Umklapp
scattering terms will no longer we able
to freeze the charge fluctuations, and one expects
gapless excitations in the density
and pair field phase, $\varphi_{\rho +}$.  This doped
phase will exhibit power-law d-wave superconducting
correlations.\cite{Balents96}

\section{d-Wave Superconductivity}
\label{sec:d-wave}

We now turn to the case of a two-dimensional superconductor
which exhibits a particular type of d-wave pairing
(denoted $d_{x^2-y^2}$) appropriate to the Cuprates.
Our ultimate goal is to quantum disorder this state
to obtain a description of the ``nodal liquid".
There are two main distinctions
between the 2d d-wave superconductor
and it's one-dimensional
counterpart considered above.  Firstly,
a 2d superconductor exhibits {\it true} (off-diagonal)
long-ranged order at $T=0$.  But more importantly, due to sign changes in the pair wave function, the $d_{x^2-y^2}$
superconductor exhibits {\it gapless} quasiparticle excitations.
We first briefly review BCS theory which gives one a powerful framework
to describe d-wave pairing and the gapless quasiparticles.
In Section ~\ref{sec:Effective} below we incorporate
{\it quantum flucutations}
of the order parameter phase to obtain a complete
{\it effective}
low-energy theory of the $d_{x^-y^2}$ phase.
In Section ~\ref{sec:Duality} a dual represention
is derived, and used to quantum disorder the superconductor
in Section ~\ref{sec:Nodal}.

\subsection{BCS Theory Re-visited}

It is instructive to briefly review BCS theory,\cite{Schrieffer}\
focussing on the symmetries of the pair wave function
and the superconducting order parameter.
In particular, it is important to emphasize the important
distinction between the wave function for the center of mass
of the Cooper pair (often ignored) and the wavefunction
for the relative coordinate.

Consider a Hamiltonian expressed as a sum of
kinetic energy and interaction terms,
$H=H_0 + H_{int}$, with $H_0$ given in Eqn.~\ref{kineticHam}.
We consider a rather general form for the electron interactions:
\begin{equation}
H_{int} = {1 \over {2V}} \sum_{\bbox{k},\bbox{k}^\prime \bbox{q}} v_{\bbox{q}}(\bbox{k},\bbox{k}^\prime) c^\dagger_{\bbox{k}+\bbox{q}\alpha}
c^\dagger_{-\bbox{k}+\bbox{q}\beta} 
c_{-\bbox{k}^\prime + \bbox{q}\beta} c_{\bbox{k}^\prime +\bbox{q}\alpha}   ,
\end{equation}
which is invariant under global charge $U(1)$ and spin $SU(2)$ symmetries.  For simplicity Umklapp interaction terms have been
ignored, so that the crystal momentum is conserved.
The interaction term describes a two electron scattering process 
with $2\bbox{q}$ the total conserved momentum 
of the pair.  For a density-density interaction
in real space, such as the Coulomb interaction,
$v_{\bbox{q}} (\bbox{k},\bbox{k}') = v(|\bbox{k} - \bbox{k}'|)$,
so is independent of $\bbox{q}$.  

Superconductivity within BCS theory requires an attractive
interaction (in the appropriate
angular momentum channel) between electrons.
But the bare Coulomb interaction is of course strongly repulsive.  
In traditional low temperature superconductors, phonons are believed
to drive the pairing, inducing a retarded attractive
interaction at low energies below the deBye energy.  Superconductivity in the high temperature Cuprates is probably of electronic origin.
In this case, retardation leading to an attractive interaction
at low energies would be due to virtual interactions
via high energy electron states well away from $E_F$.
These processes
can be studied via a renormalization group
procedure,\cite{Shankar} which consists of ``integrating out"
high energy electron states, and seeing how the
remaining interactions between those electrons near the
Fermi energy are modified. 
This is precisely what we implemented in detail
for the two-leg ladder in Sec.~\ref{sec:Two-leg}.
One thereby arrives at an effective low energy theory
involving electron states within a small energy range
of width $2\Lambda$
around $E_F$, scattering off one another
with an {\it effective} (or renormalized) interaction potential.
In the following, we view $v_{\bbox{q}} (\bbox{k},\bbox{k}')$
as an effective low energy interaction.  
For the two-leg ladder the renormalized
potential is given by putting the nine coupling
contants equal to their values along the special ray.
Upon Bosonization, the effective
potential is given explicitly in Eqn.~\ref{int_boson}.
More generally, the form
of the renormalized potential
will be constrained by the original symmetries of the
Hamiltonian.  Specifically, time reveral and parity symmetries
imply that $v_{\bbox{q}} (\bbox{k},\bbox{k}')$ is real, and
odd in it's arguments: $v_{\bbox{q}} (\bbox{k},\bbox{k}') =    
v_{-\bbox{q}} (-\bbox{k},-\bbox{k}')$.  Hermiticity implies
$v_{\bbox{q}} (\bbox{k},\bbox{k}') =    
v_{\bbox{q}} (\bbox{k}',\bbox{k})$.
The summation over momentum is now understood to be
constrained, involving only
electron operators with energy in a shell
of width $2 \Lambda$ about $E_F$.

BCS theory can be implemented by considering the operator,
\begin{equation}
P^{\alpha \beta}_{\bbox{k}} (\bbox{q}) = c_{-\bbox{k} + \bbox{q} \alpha} c_{ \bbox{k} + \bbox{q} \beta}  ,
\end{equation}
which destroys a pair of electrons, with total momentum $2 \bbox{q}$.
For $\bbox{k}$ near the Fermi surface,
and $|\bbox{q}| << k_F$,
$[P_{\bbox{k}}(\bbox{q}), P^\dagger_{\bbox{k}}(\bbox{q}')]=0$
for $\bbox{q} \ne \bbox{q}'$, so that the pair operator resembles
a boson operator, $b(\bbox{q})$.  By analogy with Bose condensation,
in the superconducting phase
one expects a non-zero expectation value for the
pair operator:  $\langle P \rangle \ne 0$.
The pair operators entering into $H_{int}$
are expressed as $P = \langle P \rangle + \delta P$,
and the fluctuations $\delta P = P - \langle P \rangle$ are
presumed to be small.  
Upon ignoring
terms quadratic in $\delta P$, $H_{int}$ can be written
(dropping additive constants),
\begin{equation}
H_1 = {1 \over {2V}} \sum_{\bbox{k},\bbox{q}} [ c^\dagger_{\bbox{k} + \bbox{q} \alpha} 
c^\dagger_{-\bbox{k} + \bbox{q} \beta } \Delta^{\beta \alpha}_{\bbox{k}}(\bbox{q}) + h.c.  ]  ,
\end{equation}
where we have introduced the (complex) superconducting
order parameter (or ``gap"), $\Delta$, defined as,
\begin{equation}
\label{fullself}
\Delta^{\alpha \beta}_{\bbox{k}}(\bbox{q}) = \sum_{\bbox{k}'} v_{\bbox{q}}(\bbox{k},\bbox{k}') \langle
c_{-\bbox{k}' + \bbox{q} \alpha} c_{ \bbox{k}' + \bbox{q} \beta} \rangle .
\end{equation}
BCS is a self-consistent mean field theory:  The full
mean field (or quasiparticle) Hamiltonian, $H_{qp} = H_0 + H_1$,
which depends on $\Delta$, is employed to compute the
expectation value $\langle
c_{-\bbox{k}' + \bbox{q} \alpha} c_{ \bbox{k}' + \bbox{q} \beta} \rangle$.  Upon insertion in
Eqn.~\ref{fullself} one obtains a self-consistent equation which determines
$\Delta$ - the celebrated BCS gap-equation.  
Notice that $H_{qp}$ is bi-linear in electron operators
and hence tractable, although it does involve ``anomalous"
terms involving pairs of creation or annihilation operators.

Before carrying through this procedure, it is instructive
to consider the form for the pair wavefunction which follows
from a non-zero expectation value of the pair operator, $\langle P
\rangle \ne 0$.
Consider removing a pair of electrons, at positions $\bbox{R} \pm \bbox{r} /2$,
with $\bbox{R}$ the center of mass position and $\bbox{r}$ the relative coordinate.
The pair wave function can be defined as,
\begin{equation}
\Phi^{\alpha \beta} ( \bbox{R} , \bbox{r} ) = \langle c_\alpha (\bbox{R} - \bbox{r} /2) c_\beta (\bbox{R} + \bbox{r} /2) \rangle ,
\end{equation}
which depends on the {\it spin} of the electrons as well
as the (center of mass and relative) positions.
Upon transforming the electron operators into momentum space,
one finds that
\begin{equation}
\Phi^{\alpha \beta} ( \bbox{R} , \bbox{r} ) =
\sum_{\bbox{Q}} e^{i \bbox{Q} \cdot \bbox{R}} \Phi^{\alpha \beta}( \bbox{Q} , \bbox{r} )   ,
\end{equation}
with $\bbox{Q}$ the center of mass momentum and
\begin{equation}
\Phi^{\alpha \beta}( \bbox{Q} , \bbox{r} ) = {1 \over N} \sum_{\bbox{k}} \langle
P^{\alpha \beta}_{\bbox{k}} ( \bbox{Q} /2) \rangle 
e^{i \bbox{k} \cdot \bbox{r} }   .
\end{equation}
Notice that the wavefunction in the {\it relative} coordinate,
involves a Fourier transform with respect
to the relative pair momentum, $\bbox{k}$.

It is also instructive to define a {\it spatially varying}
superconducting order parameter by Fourier transforming
the gap function, $\Delta_{\bbox{k}}(\bbox{q})$:
\begin{equation}
\Delta_{\bbox{k}}^{\alpha \beta} ( \bbox{x} ) = \sum_{\bbox{Q}} e^{i \bbox{Q}
\cdot \bbox{x} } \Delta_{\bbox{k}}(\bbox{Q}/2)  .
\end{equation}
In the superconducting phase
one can often ignore the
spatial dependence of the complex order
parameter $\Delta_{\bbox{k}} (\bbox{x} )$,
and indeed in BCS theory this $\bbox{x}$ dependence is dropped.
However, if one wishes to include the effects of
quantum fluctuations (to quantum disorder the superconductor)
it is necessary to consider a spatially varying order parameter
as discussed in Sec.~\ref{sec:Effective} below.

By analogy with Bose condensation, one
expects the Cooper pairs to be condensed into a
state of {\it zero} momentum, $\bbox{Q} =0$.  This requires 
\begin{equation}
\langle P^{\alpha \beta}_{\bbox{k}} (\bbox{q}) \rangle
= \delta_{\bbox{q} , \bbox{0} } \langle c_{-\bbox{k} \alpha} c_{ \bbox{k}
\beta} \rangle   ,
\end{equation}
which gives a {\it relative} pair wavefunction,
$\Phi (\bbox{r}) \equiv \Phi ( \bbox{Q} =0 , \bbox{r} )$ of
the form,
\begin{equation}
\Phi^{\alpha \beta} (\bbox{r}) = {1 \over N} \sum_{\bbox{k}} e^{i \bbox{k}
\cdot \bbox{r} } \Phi^{\alpha \beta}_{\bbox{k}};  \hskip0.5cm 
\Phi^{\alpha \beta}_{\bbox{k}} = \langle c_{- \bbox{k} \alpha }
c_{ \bbox{k} \beta } \rangle   .
\end{equation}
Due to the electron anticommutation relations one has
$\Phi^{\alpha \beta}_{\bbox{k}} = - \Phi^{\beta \alpha}_{- \bbox{k}}$,
which implies 
that the pair wavefunction is {\it antisymmetric} under exchange of the two electrons:
$\Phi^{\alpha \beta} (\bbox{r}) = - \Phi^{\beta \alpha} (- \bbox{r} )$.

When the Cooper pairs are condensed into a 
state with zero momentum, the superconducting order parameter
becomes spatially
uniform: $\Delta^{\alpha \beta}_{\bbox{k}} (\bbox{x}) \equiv \Delta^{\alpha \beta}_{\bbox{k}}$, as seen from Eqn.~\ref{fullself}.  The mean field
Hamiltonian then takes a rather simpler form:
\begin{equation}
H_1 = {1 \over {2}} \sum_{\bbox{k}} [ c^\dagger_{\bbox{k} \alpha} 
c^\dagger_{-\bbox{k} \beta } \Delta^{\beta \alpha}_{\bbox{k}} + h.c.  ]  ,
\end{equation}
whereas the self-consistentcy condition becomes,
\begin{equation}
\Delta^{\alpha \beta}_{\bbox{k}} = {1 \over V} \sum_{\bbox{k}'} v_{0}(\bbox{k},\bbox{k}') \langle
c_{-\bbox{k}' \alpha} c_{ \bbox{k}' \beta} \rangle .
\end{equation}
    
Since the full model has a conserved $SU(2)$ spin symmetry,
the relative pair wavefunction can be expressed
as the product of an orbital and a spin wavefunction:
$\Phi^{\alpha \beta}_{\bbox{k}} = \phi_{\alpha \beta} \Phi_{\bbox{k}}$.
The spin piece can be chosen as an eigenfunction of
the total spin of the pair, that is a singlet with $S=0$
or a triplet with $S=1$.  In conventional low temperature
superconductors and in the Cuprates the Cooper pairs are singlets
with, 
\begin{equation}
\phi_{\alpha \beta} = \delta_{\alpha \downarrow} \delta_{\beta
\uparrow} - \delta_{\alpha \uparrow} \delta_{\beta
\downarrow}  ,
\end{equation}
in which case the orbital wavefunction is symmetric:
$\Phi_{\bbox{k}} = \Phi_{-\bbox{k}} = \langle c_{- \bbox{k} \downarrow}
c_{\bbox{k} \uparrow } \rangle$.  (In the superfluid
phases of $3-He$ on the other hand,
the Cooper pairs have $S=1$.)  The superconducting order parameter
is then also a singlet;  $\Delta^{\alpha \beta}_{\bbox{k}} \equiv
\phi_{\alpha \beta} \Delta_{\bbox{k}}$, with $\Delta_{\bbox{k}} = \Delta_{-\bbox{k}}$ satisfying
\begin{equation}
\label{self}
\Delta_{\bbox{k}} = {1 \over V} \sum_{\bbox{k}'} v_{0}(\bbox{k},\bbox{k}') \langle
c_{-\bbox{k}' \downarrow} c_{ \bbox{k}' \uparrow} \rangle .
\end{equation}
For singlet pairing,
the final mean field (quasiparticle) Hamiltonian becomes,
$H_{qp} = H_0 + H_1$ with,
\begin{equation}
H_1 = \sum_{\bbox{k}} [ \Delta_{\bbox{k}} c^\dagger_{\bbox{k} \uparrow} 
c^\dagger_{-\bbox{k} \downarrow }  +  \Delta^*_{\bbox{k}} c_{-\bbox{k} \downarrow} 
c_{\bbox{k} \uparrow } ]  .
\end{equation}

To complete the self-consistency requires diagonalizing the quasiparticle
Hamiltonian.  This is usually done in a way which
masks the spin rotational invariance.\cite{Schrieffer}\  
We prefer to keep the spin rotational
invariance explicit, by defining a new set of Fermion operators, for $k_y >0$:
\begin{equation} 
\chi_{1 \alpha} (\bbox{k}) = c_{\bbox{k} \alpha} ; \hskip0.5cm
\chi_{2 \alpha} (\bbox{k}) = i\sigma^y_{\alpha \beta} c^\dagger_{-\bbox{k}
\beta}   ,  
\end{equation}
which satisfy canonical Fermion anti-commutation relations:
\begin{equation}
[\chi_{a \alpha}(\bbox{k}) , \chi^\dagger_{b \beta} (\bbox{k}') ]_-
= \delta_{ab} \delta_{\alpha \beta} \delta_{\bbox{k} \bbox{k}' }  .
\end{equation}
The first index $a,b=1,2$ acts in the particle/hole subspace.
The $\sigma^y$ in the definition of $\chi_{2 \alpha}$
has been introduced so that these new operators transform like
$SU(2)$ spinors under spin rotations:  $\chi_{a\alpha} \rightarrow U_{\alpha \beta} \chi_{a\beta}$, 
with $U = \exp(i \bbox{\theta} \cdot \bbox{\sigma})$  a global
spin rotation.

In these variables, the quasiparticle Hamiltonian becomes 
\begin{equation}
\label{qpHam}
H_{qp} = \left. \sum_{\bbox{k}} \right.^\prime \chi^\dagger(\bbox{k}) 
[ \tau^z \epsilon_{\bbox{k}} + \tau^+ \Delta_{\bbox{k}}
+ \tau^- \Delta^*_{\bbox{k}} ] \chi(\bbox{k})  ,
\end{equation}
where the prime on the summation denotes over $k_y$ positive, only,
and we have introduced a vector of Pauli matrices, $\vec{\tau}_{ab}$
acting in the particle/hole subspace.  Also, we
are employing the notation $\tau^{\pm} = (\tau^x \pm i \tau^y)/2$.
To evaluate the self-consistency condition Eqn.~\ref{self}
we need the anomalous average of two electron fields (the orbital
piece of the relative pair wavefunction),
which is re-expressed as,
\begin{equation}
\label{pairwf}
\Phi_{\bbox{k}} \equiv {1 \over 2} \sum_{\pm} \langle
c_{\mp \bbox{k} \downarrow} c_{\pm \bbox{k} \uparrow} \rangle =
{1 \over 2} \langle
\chi^\dagger (\bbox{k}) \tau^+ \chi (\bbox{k} ) \rangle .
\end{equation}

Diagonalization is now achieved by performing an $SU(2)$ rotation
in the particle/hole subspace,
by defining rotated Fermion fields:  $\chi (\bbox{k}) \equiv
U(\bbox{k}) \tilde{\chi} (\bbox{k})$,
with $U(\bbox{k}) = e^{-i \bbox{\theta}_{\bbox{k}} \cdot \bbox{\tau} }$.
Assuming for simplicity that 
$\Delta_{\bbox{k}}$ is real, the appropriate rotation is
around the y-axis by an angle $\theta_{\bbox{k}}$,
$U(\bbox{k}) = e^{- i \theta_{\bbox{k}} \tau^y /2 }$, with
\begin{equation}
sin(\theta_{\bbox{k}}) = { \Delta_{\bbox{k}} \over E_{\bbox{k}} } ;
\hskip0.75cm  E_{\bbox{k}}  = \sqrt{  \epsilon^2_{\bbox{k}} + \Delta^2_{\bbox{k}} }.
\end{equation}
In terms of the rotated Fermion fields,
$\tilde{\chi}$, the quasiparticle Hamiltonian is diagonal,
\begin{equation}
H_{qp} = \left. \sum_{\bbox{k}} \right.^\prime E_{\bbox{k}} \tilde{\chi}^\dagger(\bbox{k}) 
\tau^z  \tilde{\chi}(\bbox{k})  ,
\end{equation}
with $E_{\bbox{k}}$ the quasiparticle energy.
Finally, we define a set of rotated electron operators via
\begin{equation} 
\tilde{\chi}_{1 \alpha} (\bbox{k}) = a_{\bbox{k} \alpha} ; \hskip0.5cm
\tilde{\chi}_{2 \alpha} (\bbox{k}) = i\sigma^y_{\alpha \beta} a^\dagger_{-\bbox{k}
\beta}   ,  
\end{equation}
and the quasiparticle Hamiltonian can be re-expressed
in standard form,
\begin{equation}
H_{qp} = \sum_{\bbox{k}} E_{\bbox{k}} a^\dagger_{\bbox{k} \alpha}
a_{\bbox{k} \alpha}   ,
\end{equation}
where we have dropped an additive constant. 
Notice that the quasiparticle energy
$E_{\bbox{k}} \ge 0$ for all momentum.
On the Fermi surface, $\epsilon_{\bbox{k}} =0$ and
the quasiparticle energy is given by $|\Delta_{\bbox{k}}|$ - the
energy gap.

To complete the self-consistentcy,
the anomalous electron average (or relative orbital pair wavefunction
from Eqn.~\ref{pairwf}) is expressed
in terms of the quasiparticle operators.
Upon using the fact that $[U^\dagger \tau^+ U]_{diag} = sin(\theta)
\tau^z /2$ one obtains,
\begin{equation}
\label{wfqp}
\Phi_{\bbox{k}} = {\Delta_{\bbox{k}} \over {2 E_{\bbox{k}}} } [\langle a^\dagger_{\bbox{k} \alpha}
a_{\bbox{k} \alpha} \rangle - 1 ]  ,
\end{equation}
which reduces to
$\Phi_{\bbox{k}} = - \Delta_{\bbox{k}}/2 E_{\bbox{k}}$ at zero temperature.
At finite temperature the number of quasiparticles
is simply a Fermi function: $\langle a^\dagger_{\bbox{k} \alpha} a_{\bbox{k} \alpha} \rangle = 2f(E_{\bbox{k}})$, 
with $f(E) = [exp(\beta E) +1 ]^{-1}$.  One thereby obtains the celebrated
BCS gap equation:
\begin{equation}
\label{gapeq}
\Delta_{\bbox{k}}  = - {1 \over V} \sum_{\bbox{k}'} v_{0}(\bbox{k},\bbox{k}') 
{ \Delta_{\bbox{k}'} \over { 2E_{\bbox{k}'} } } [ 1 - 2f(E_{\bbox{k}'}) ]  .
\end{equation}

\subsection{d-wave Symmetry}

In a system with rotational invariance the orbital
piece of the pair wavefunction, proportional to $\Delta_{\bbox{k}}$
from Eqn.~\ref{wfqp}, can be
chosen as an eigenstate of angular momentum,
a spherical harmonic $Y_{l m}$ in three dimensions.
The simplest case is s-wave, with $\Delta_{\bbox{k}}$
a constant over the (spherical) Fermi surface.  
Real materials
of course do not share the full continuous rotational symmetry
of free space.  Nevertheless, a superconductor in which
$\Delta_{\bbox{k}}$ is everywhere positive over the Fermi surface
is (loosely) referred to as having s-wave pairing - a property
of all conventional low temperature superconductors.
Since $|\Delta_{\bbox{k}}|$ is the quasiparticle energy
on the Fermi surface, there are no
low energy electronic excitations
in an s-wave superconductor - the Fermi surface
is {\it fully} gapped.  Within BCS theory
the magnitude of the (zero temperature)
energy gap is related to the superconducting transition
temperature: $2|\Delta| \approx 3.5 k_B T_c$.  The presence
of an energy gap leads to thermally activated behavior for various
low temperature properties, such as the electronic specific heat
and the magnetic penetration length.

It is clear from the self-consistent gap equation Eqn.~\ref{gapeq} that
a purely repulsive effective interaction,
$v_0(\bbox{k},\bbox{k}') > 0$, precludes s-wave pairing
within BCS theory (since $1-2f(E_{\bbox{k}}) \ge 0$).
In conventional superconductors, phonons are believed
to drive s-wave pairing,\cite{Schrieffer}\ generating
an effective attractive interaction at low energies.  

Recent experiment\cite{Wollman,Kirtley} has established that in the 
high temperature superconductors the orbital
pairing symmetry is a particular form of d-wave, usually
denoted as $d_{x^2 - y^2}$.  
Here $x$ and $y$ refer to the directions
along the crystalline axis of  
a single Cu-O sheet, within which the Cu atoms form a square
lattice.  In terms of the corresponding two dimensional
momentum, $\bbox{k} = (k_x,k_y)$, the angular dependence
of the gap function
in this state is $\Delta_{\bbox{k}} \sim k^2_x - k^2_y$,
and from Eqn.~\ref{wfqp}
the orbital piece of the relative pair wave function
has the same d-wave symmetry.

A novel feature of the $d_{x^2 - y^2}$ state is that the
gap function $\Delta_{\bbox{k}}$ {\it vanishes}
along lines in $k-space$ with $k_x = \pm k_y$,
corresponding to {\it nodes} in the relative pair
wave function.
These lines
intersect the (two-dimensional) Fermi surface at four
{\it points} in momentum space.  Near these four points
(or ``nodes") in momentum space there are electronic
excitations with arbitrary low energy, in striking constast
to the fully gapped s-wave case.
These low energy quasiparticle excitations dominate the physics
of the $d_{x^2 - y^2}$ superconductor at temperatures well
below $T_c$, leading to power law temperature corrections
in such quantities as the electronic specific heat
and the magnetic penetration length.  

\subsection{Continuum description of gapless quasiparticles}

It is convenient to obtain a {\it continuum}
description of the gapless d-wave quasiparticles,
analogous to
the Dirac theory desription of the low energy properties
of the 1d free Fermions employed in Section ~\ref{sec:Bosonize}.
A continuum form can be obtained directly from the general quasiparticle
Hamiltonian Eqn.~\ref{qpHam} by specializing to $d_{x^2-y^2}$ symmetry
and then focussing on those 
momenta close to the four nodes
where the quasiparticle energy
$E_{\bbox{k}}=0$ (see Figure 7).    
For a model with particle/hole symmetry $\epsilon_{\bbox{k}} = - \epsilon_{\bbox{k}+ \bbox{\pi}}$, which together with parity symmetry
implies that the four nodes occur at
the special wavevectors $\pm \bbox{K}_{j}$, with
$\bbox{K}_{1} = (\pi/2,\pi/2)$ and 
$\bbox{K}_{2} = (-\pi/2,\pi/2)$. 
It is convenient to introduce {\it two} continuum fields
$\Psi_j$, one for each pair of nodes, expanded around 
$\pm\bbox{K}_{1},\pm\bbox{K}_{2}$:
\begin{equation}
\Psi_{ja\alpha}(\bbox{q}) = \chi_{a\alpha}(\bbox{K}_{j}+\bbox{q}) .
\end{equation}
Here, the wavevectors $\bbox{q}$ are assumed to be small, within
a circle of radius $\Lambda$ around the origin.
With this definition, the particle/hole transformation is extremely simple, 
\begin{equation}
\label{ph-Psi}
\Psi  \rightarrow \Psi^{\dagger}.
\end{equation}
For this reason it is convenient to {\sl always} define the continuum
fields $\Psi$ around $\pm\bbox{K}_{j}$, and account for deviations of
the node momenta from these values by a particle/hole
symmetry-breaking parameter $\lambda$.

\begin{figure}
\hskip 1 cm
\psfig{figure=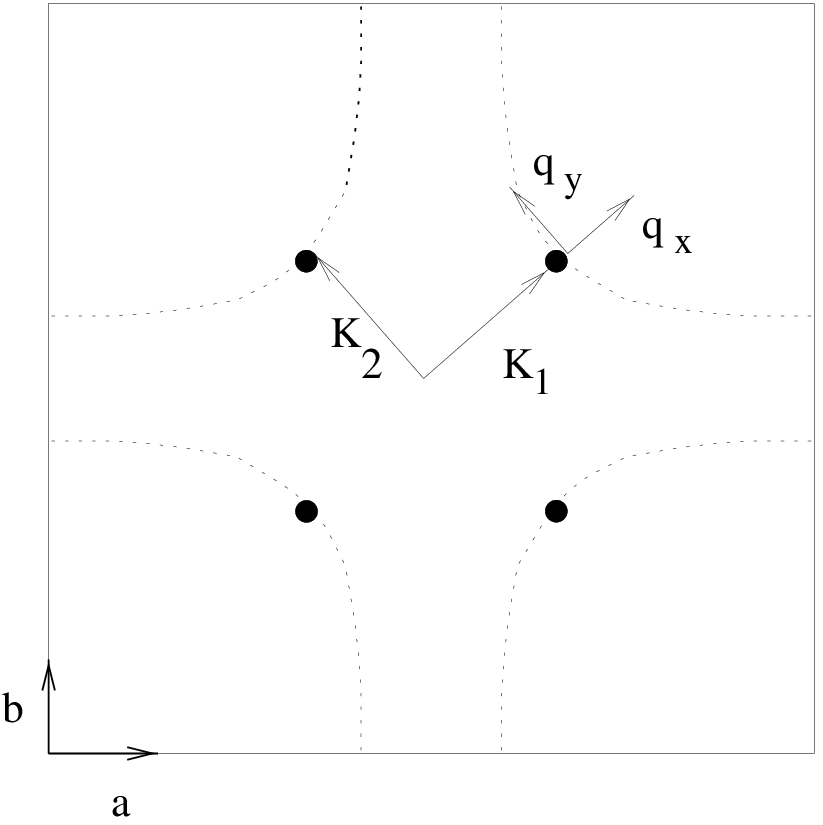,height=2.0in,angle=-90}
\vskip 0.5cm
{Fig.~7: In the $d_{x^2-y^2}$ superconductor
the quasiparticle energy vanishes at four points
($\pm \bbox{K}_1$ and $\pm \bbox{K}_2$) in the Brillouin zone.  The dotted line
represents the Fermi surface.  The wavevector $\bbox{q}$
is rotated with respect to the $a$ and $b$ axis of the
square lattice.
}
\end{figure}

Once we have restricted attention to the momenta near the nodes, it is
legitimate to linearize in the quasiparticle Hamiltonian.  The
resulting theory is more conveniently written in coordinates
perpendicular and parallel to the Fermi surface, so we perform the
rotation via $x \rightarrow (x-y)/\sqrt{2}$ and $y \rightarrow
(x+y)/\sqrt{2}$, correspondingly transforming the momenta $q_{x}$ and
$q_{y}$ (see Figure 7).  Linearizing near the nodes, we put
$\epsilon_{\bbox{K}_1 + \bbox{q}} = v_F q_x$ where $v_F$ is the Fermi velocity and
\begin{equation}
\Delta_{\bbox{K}_1 + \bbox{q}} = \tilde{\Delta} q_y + O(q^2), 
\end{equation}
where $\tilde{\Delta}$ has
dimensions of a velocity.  An identical linearization is possible
around the second pair of nodes, except with $q_x \leftrightarrow
q_y$.  It is finally convenient
to Fourier transform back into real space by defining,
\begin{equation}
\Psi_j(\bbox{x}) = {1 \over {\sqrt{V}} } \sum_{\bbox{q}} e^{i\bbox{q} \cdot
\bbox{x} } \Psi_j(\bbox{q})   ,
\end{equation}
where the momentum summation is for $q < \Lambda$.
The continuum fields $\Psi(\bbox{x})$ describe long lengthscale variations
of the quasiparticles, on scales greater than $\Lambda^{-1}$.  
We thereby arrive at a compact
form for the continuum quasiparticle Hamiltonian in
a $d_{x^2 - y^2}$ superconductor: ${\cal H}_{qp} = {\cal H}_\Psi + {\cal
H}_\lambda$ with
\begin{eqnarray}
\label{cqpHam}
  {\cal H}_{\Psi} & = & \Psi_1^\dagger [v_F\tau^z i \partial_x +
  (\tilde{\Delta} \tau^+ + \tilde{\Delta}^* \tau^-) i \partial_y ] \Psi_1 
  \nonumber \\
  & & + (1 \leftrightarrow 2; x \leftrightarrow y)  ,
\end{eqnarray}
and the particle/hole symmetry breaking term,
\begin{equation}
{\cal H}_\lambda = \lambda \Psi_j^\dagger \tau^z \Psi_j  .
\end{equation}

The quasiparticle Hamiltonian takes the form of (four) Dirac equations
in $2+1$ space-time dimensions,
and can be readily diagonalized.  For the first pair of nodes
one obtains the relativistic dispersion, 
\begin{equation}
\label{E(q)}
E_1(\bbox{q}) = \sqrt{ (v_F q_x + \lambda)^2 + |\tilde{\Delta}|^2 q_y^2} , 
\end{equation}
and a similar expression is obtained
for $E_2$ except with $q_x$ and $q_y$ interchanged. 
As usual in Dirac theory, the negative
energy single particle states with energy $-E_j(\bbox{q})$ 
are filled but positive energy holes states can be created.
As expected, the quasiparticle
energy vanishes at the nodes ($\bbox{q} = 0$ with
particle/hole symmetry $\lambda =0$), so the
``relativistic" particle is massless.
Notice that non-zero $\lambda$ indeed shifts the positions of
the nodes.

In this continuum description $\tilde{\Delta}$ serves as a complex
superconducting order parameter for the $d_{x^2 -y^2}$
state.  Indeed, when $\tilde{\Delta}=0$ one recovers the metallic Fermi surface
and the quasiparticle Hamiltonian describes gapless excitations
for all $q_y$.  Below we will include quantum fluctuations
by allowing  $\tilde{\Delta}$ to depend on space and time.
Before doing so, it is convenient
to see how $\tilde{\Delta}$ transforms under a particle/hole
transformation.  From the transformation properties
of the electron fields one deduces that
the gap transforms as, $\Delta_{\bbox{k}} \rightarrow - \Delta^*_{-\bbox{k}
+ \bbox{\pi}}$, which is equivalent to complex
conjugation for the (linearized) order parameter,
\begin{equation}
\tilde{\Delta} \rightarrow \tilde{\Delta}^* .
\end{equation}
Together with Eqn.~\ref{ph-Psi} this implies that
the quasiparticle Hamiltonian in Eqn.~\ref{cqpHam}
is indeed particle/hole symmetric: ${\cal H}_\Psi \rightarrow {\cal H}_\Psi$.

\section{Effective Field Theory}
\label{sec:Effective}

\subsection{Quasiparticles and Phase Flucutations}

Our goal in this section is to obtain a complete low-energy
effective theory for the $d_{x^2 - y^2}$ superconductor. 
This task is complicated by the existence of {\it additional}
gapless excitations, besides the quasiparticles.
Specifically, since the global $U(1)$ charge
conservation symmetry ($c_\alpha \rightarrow e^{i\theta_0} c_\alpha$) 
is spontaneously broken by the existence
of a non-zero order parameter, $\tilde{\Delta} \ne 0$,
gapless Goldstone modes are expected.
(In a three-dimensional superconductor these
modes are actually gapped, due to the presence
of long-ranged Coulomb interactions,
but would be gapless for a thin 2d film.)
These modes propogate in the {\it phase} of the complex order
parameter.
Thus a correct low energy theory for the $d_{x^2 -y^2}$ state
requires consideration of a {\it spatially}
varying order parameter, $\tilde{\Delta}(\bbox{x})$.    
Generally, both the magnitude and the phase of the complex
order parameter can vary, but we will focus exclusively
on the phase fluctuations,
writing 
\begin{equation}
\tilde{\Delta}(\bbox{x})  = |\Delta| e^{i\varphi(\bbox{x})} ,
\end{equation}
with $|\Delta|$ a (real) {\it constant}.
Since amplitude fluctuations are costly in energy
this should suffice in the superconducting phase,
and will also allow us to describe the nodal liquid
in which superconductivity is destroyed by {\it phase} fluctuations.
The desired low energy effective theory can be obtained
from symmetry considerations, and the
form of the continuum quasiparticle Hamiltonian.
A more microscopic approach, discussed
briefly below, would entail integrating out high
energy degrees of freedom in a functional integral representation.

\subsubsection{Symmetry considerations}

Since the BCS gap equation has a degenerate manifold of solutions
for arbitrary phase $\varphi$, the energy should only
depend on {\it gradients} of $\varphi(\bbox{x})$.
The appropriate Lagrangian which describes
the {\it fluctuations} of the phase of the superconducing
order parameter can thus be developed as
a gradient expansion, with lowest order terms of the form,
\begin{equation}
\label{applag}
  {\cal L}_\varphi = {1 \over 2} \kappa_\mu  (\partial_\mu \varphi)^2 ,
\end{equation}
where the Greek index $\mu$ runs over time and two spatial
coordinates: $\mu = 0,1,2 = t,x,y$. 
Here $\kappa_0$ is equal to the
compressibility of the condensate 
(ignoring for the moment long-ranged Coulomb forces)
and $\kappa_j = - v_c^2 \kappa_0$
(for $j=1,2=x,y$) with $v_c$ the superfluid sound velocity.
This form is largely dictated by symmetry.
Parity and four-fold rotational symmetry {\it determine}
the form of the spatial gradient terms.  
The stiffness coefficients, $\kappa_\mu$, can be estimated
as follows.  The pair compressibility $\kappa_0$ should be roughly
one half the
electron compressibility  -- at least for weak interactions.
If the pairing is electronic in origin, 
the Fermi velocity sets
the scale for $v_c$.

In general a Berry's phase term\cite{Fradkin} {\it linear} in $\partial_t \varphi$
is allowed,
\begin{equation}
\label{Berry}
{\cal L}_{Berry} = n_0 \partial_t \varphi  ,
\end{equation}
where $n_0$ is a two-dimensional number density.
For a model
with particle/hole symmetry which must
be invariant under
\begin{equation}
\label{ph-trans}
\varphi \rightarrow -\varphi ,
\end{equation}
(which follows from the particle/hole transformation properties
of the order parameter $\tilde{\Delta} \sim e^{i\varphi}
\rightarrow \tilde{\Delta}^*$) it naively
appears that the number density $n_0$ must vanish.
However, this is not the case\cite{Nodal2}.  To see this it is necessary
to return to the lattice where the term in the (Euclidian) action
which follows from ${\cal L}_{Berry}$ is,
\begin{equation}
S_{Berry} = i n_0 \int_0^{\beta \hbar} d\tau \sum_i \partial_\tau \varphi_i  ,
\end{equation}
where $i$ labels the sites of a square lattice with
lattice spacing set to one and $\beta = 1/k_BT$ .
The partition
function is expressed as a functional integral
of $exp(-S)$ over configurations
$\varphi_i(\tau)$, with 
$\beta$ periodic boundary conditions
on the fields $e^{i\varphi}$.  This implies the boundary conditions,
\begin{equation}
\varphi_i(\beta) = \varphi_i(0) + 2\pi N_i ,
\end{equation}
with {\it integer} winding numbers $N_i$.   
We thus see that the Berry's phase term contributes
a multiplicative piece to the partition function
(in each winding sector) of the form;
\begin{equation}
\label{expBerry}
exp(-S_{Berry}) = e^{i2\pi n_0 N_W}  ,
\end{equation}
with integer $N_W = \sum_i N_i$.  Under the particle/hole transformation
Eqn.~\ref{ph-trans}, 
the winding numbers
change sign, $N_W \rightarrow - N_W$.  The Berry's
phase term is thus invariant under the particle/hole transformation
{\it provided} $n_0$ is integer or half-integer.

The appropriate value for $n_0$ can be readily determined by obtaining
the lattice Hamiltonian associated with the Lagrangian density
${\cal L}_\varphi + {\cal L}_{Berry}$.  The first contribution
can be conveniently regularized on the lattice as,
\begin{equation}
L_\varphi = -t \sum_{<i,j>} cos(\varphi_i - \varphi_j) 
- {1 \over u} \sum_i (\partial_t {\varphi}_i)^2  .
\end{equation}
Upon inclusion of the Berry's phase
term this gives the lattice Hamiltonian,
\begin{equation}
\label{Cooplatt}
H_\varphi = -t \sum_{<i,j>} cos(\varphi_i - \varphi_j) +
u \sum_i (n_i - n_0)^2  .
\end{equation}
Here $n_i$ denotes a Cooper-pair number {\it operator},
canonically conjugate to the phase fields:
\begin{equation}
[\varphi_i ,n_j ] = i \delta_{ij}  .
\end{equation}
The first term in $H_\varphi$ describes the hopping 
of charge $2e$ (spinless) Cooper
pairs between neighboring sites of the lattice, and the second term
is an onsite repulsive interaction.  The parameter $n_0$ 
plays the role of an ``off-set" charge and determines
the average number of Cooper pairs per site.  For the Hubbard
model at half-filling with one electron per site, the number of Cooper pairs
clearly equals {\it one-half} the number of lattice sites.
This is especially apparent in the limit of very large
attractive Hubbard interaction when the electrons pair into on-site singlets,
but is expected to be more generally valid.  Thus,
it is clear that one should take $n_0 =1/2$.
Tuning away from particle/hole symmetry
with a chemical potantial $\mu$, corresponds to
changing $n_0$ away from one-half.

In the superconducting phase one expects that the winding
numbers will all vanish, since the phase field $\varphi$ is essentially
constant in space and time, and the Berry's term plays no role.
But when the superconductor is ``quantum disordered",
the phase field fluctuates wildly with signifigant
winding, and inclusion of the Berry's phase term is expected
to be important (but see Section \ref{sec:Duality} below).

It remains to couple these phase fluctuations to
the gapless quasiparticles.
Since the order parameter $\tilde{\Delta}$ directly enters
the quasiparticle Hamiltonian Eqn.~\ref{cqpHam}, 
one can readily guess the appropriate coupling.
We should simply replace
$\tilde{\Delta} \rightarrow v_{\scriptscriptstyle\Delta} e^{i\varphi}$
with $v_{\scriptscriptstyle\Delta}$ real.
Since $\varphi$ varies spatially, some care is required.
In the quasiparticle Hamiltonian we let,
\begin{equation}
  \tilde{\Delta} \tau^+ i\partial_y \rightarrow v_{\scriptscriptstyle\Delta} 
\tau^+ e^{i\varphi/2}
  (i \partial_y) e^{i\varphi/2}    ,
\label{dgauge}
\end{equation}
and similarly for the $\tau^-$ term.
This ``symmetric" form leads to an hermitian Hamiltonian, physical
currents, and respects the symmetries of the problem. A careful
derivation of Eq.~\ref{dgauge}\ is given below. With
this prescription, the quasiparticle Hamiltonian becomes
\begin{eqnarray}
\label{hqp}
  {\cal H}_{qp} &= & \sum_{s=\pm} \Psi_1^\dagger [v_F\tau^z i \partial_x +
  v_{\scriptscriptstyle\Delta} \tau^s  e^{is\varphi/2}
  (i \partial_y) e^{is\varphi/2} ] 
  \Psi_1 \nonumber \\
  & & + (1 \leftrightarrow 2; x \leftrightarrow y)  .
\end{eqnarray}
Since $\varphi$ can also fluctuate with time, it will convenient
to consider the time dependence via a Lagrangian formulation.  The
Lagrangian density is
\begin{equation}
  {\cal L}_{qp} = \Psi_j^\dagger i \partial_t \Psi_j - {\cal H}_{qp}   .
\end{equation}
The full low-energy effective Lagrangian in the d-wave superconductor
is obtained by adding the two contributions:
${\cal L}_\varphi + {\cal L}_{qp}$.

\subsubsection{Microscopic approach}

To illustrate how one might try to ``derive"
this effective theory from a more microscopic starting point,
we briefly consider a simple model Hamiltonian,
\begin{equation}
H = H_0 -  V\sum_{\langle
    \vec{x}\vec{x}'\rangle}
  c_\alpha^\dagger(\vec{x})c_\beta^\dagger(\vec{x}') 
  c_\beta^{\vphantom\dagger}(\vec{x}') 
  c_\alpha^{\vphantom\dagger}(\vec{x}) ,
\end{equation}
where $H_0$
is the usual kinetic energy describing hopping on a 2d square lattice and
we have added an {\it attractive}
near-neighbor interaction with strength $V$.
To derive the effective field theory, it is
convenient to express the partition function $Z= {\rm Tr}
e^{-\beta H}$, as an {\sl imaginary time} coherent state
path integral,\cite{Negele}
\begin{equation}
  Z = \int [{\cal D}c] [{\cal D}\overline{c}]  e^{-S},
\end{equation}
where $c$ and $\overline{c}$ are Grassman fields
and the Euclidean action is simply
\begin{equation}
  S = \int \! d\tau \left\{ \sum_{\vec{x}}\overline{c}_\alpha(\vec{x})
    \partial_\tau c_\alpha(\vec{x}) 
    + H[\overline{c},c] \right\}.
\end{equation}
We consider here only $T=0$, for which the $\tau$ integration domain
is infinite.  The possibility of a d-wave superconducting
phase can be entertained
by decoupling the above action using a Hubbard-Stratonovich
transformation:
\begin{equation}
  Z = \int  [{\cal D}c] [{\cal D}\overline{c}] [{\cal D}\Delta ]
[ {\cal D} \Delta^* ]
  e^{-S_1}, 
\end{equation}
with $S_1 = \int \! d\tau [\sum_{\vec{x}}\overline{c}_\alpha(\vec{x})
    \partial_\tau c_\alpha(\vec{x}) +H_{eff}]$.  The effective
    Hamiltonian can be decomposed into $H_{eff} = H_0+H_{int} + H_{\scriptscriptstyle\Delta}$, with 
\begin{equation}
 H_{int} =   \sum_{\langle \vec{x}\vec{x}'\rangle} \left[
   \Delta^{\alpha\beta}_{\vec{x}\vec{x}'}
    \overline{c}_\alpha(\vec{x})\overline{c}_\beta(\vec{x}') + h.c. \right],
  \label{Heffint} 
\end{equation}
\begin{equation}
 H_{\scriptscriptstyle\Delta} = {1 \over V} \sum_{\langle
    \vec{x}\vec{x}'\rangle}
 |  \Delta^{\alpha\beta}_{\vec{x}\vec{x}'} |^2 . 
\label{Heffdelta}
\end{equation}

Eqs.~\ref{Heffint}-\ref{Heffdelta}\ form a basis for studying the
original electron model.  
At this stage BCS mean field theory could be implemented by
integrating out the electron degrees of freedom
to obtain an effective action only depending on $\Delta$,
$S_{eff}(\Delta)$.
Minimizing this action with respect to $\Delta$
would give the gap equation.  One could imagine including
fluctuations by expanding about the saddle point solution.
But for a d-wave superconductor this procedure is problematic,
since integrating out 
{\it gapless} quasiparticles will generate {\it singular} long-ranged interactions
in $S_{eff}(\Delta)$.  It is preferable to {\it retain}
the gapless quasiparticles in the effective theory, and only
integrate out the {\it high frequency} electron modes which are well away
from the nodes.  In this way, the dynamics and interactions
generated for     
the order parameter $\Delta$ will be {\it local}.

Rather than trying to implement this procedure, we content ourselves
with arguing that the ``symmetric" prescription
adopted above indeed gives the correct form for the phase-quasiparticle 
coupling term.
To this end
we focus on singlet pairing, defining 
\begin{equation}
\Delta^{\alpha\beta}_{\vec{x}\vec{x}'} = \Delta(\vec{x},\vec{x}')
(\delta_{\alpha\uparrow}\delta_{\beta\downarrow} -
\delta_{\alpha\downarrow}\delta_{\beta\uparrow}) .
\end{equation}
The triplet pieces
of $\Delta$ are presumed to be massive,
so that they can be safely integrated out.
Since $\Delta$
lives on the bonds, it is convenient to associate two such fields with 
each site on the square lattice, i.e.
\begin{eqnarray}
  \Delta_1(\vec{x}) & \equiv & \Delta(\vec{x},\vec{x}+\hat{e}_1),
  \label{pv1} \\
  \Delta_2(\vec{x}) & \equiv & \Delta(\vec{x},\vec{x}+\hat{e}_2),\label{pv2}
\end{eqnarray}
where $\hat{e}_1, \hat{e}_2$ are unit vectors along the $a$ and $b$
axes of the square lattice, respectively. 
The interaction Hamiltonian becomes,
\begin{eqnarray}
  H_{int} & = & \sum_{j,\vec{x}} \Bigg\{ \Delta_j(\vec{x})
  \left[c^\dagger_\uparrow(\vec{x})
    c^\dagger_\downarrow(\vec{x}+\hat{e}_j) -
    \uparrow\leftrightarrow\downarrow\right] + {\rm h.c.}\Bigg\} ,    
  \label{d12}
\end{eqnarray}
where the sum includes all lattice sites and $j=1,2$.  
The magnitudes of $\Delta_1$ and
$\Delta_2$, as well as their relative sign,
are determined by the effective action generated upon
integrating out the high-energy modes.  
For a d-wave superconductor
the effective action will be minimized
for $\Delta_1 = -\Delta_2 = \Delta_0 e^{i\varphi}$,
up to massive modes.  We can now take the continuum limit.
For agreement with Sec.~\ref{sec:d-wave}, we define
$v_{\scriptscriptstyle\Delta} = 2\sqrt{2}\Delta_0$, or $\Delta_1 =
-\Delta_2 = \tilde\Delta/2\sqrt{2}$.  In addition, we take the
continuum limit of the electron fields, using the decompositions
\begin{eqnarray}
  c^\dagger_\uparrow & \sim & \Psi_{111}^\dagger i^{x+y} \!-\!
  \Psi_{122}^{\vphantom\dagger}(-i)^{x+y}\!+\!\Psi_{211}^\dagger(-i)^{x-y} 
  \!-\!\Psi_{222}^{\vphantom\dagger}i^{x-y}, \nonumber \\
  c^\dagger_\downarrow & \sim & \Psi_{112}^\dagger i^{x+y} \!+\!
  \Psi_{121}^{\vphantom\dagger}(-i)^{x+y}\!+\!\Psi_{212}^\dagger(-i)^{x-y} 
  \!+\!\Psi_{221}^{\vphantom\dagger}i^{x-y},\nonumber
\end{eqnarray}
and the hermitian conjugates of these equations.  Inserting these into
Eq.~\ref{d12}, gradient-expanding the $\Psi$ fields, and rotating $45$ 
degrees to $x-y$ coordinates along the $(\pi,\pi)$ and $(-\pi,\pi)$
directions, one obtains
$H_{int} = \int\!d^2{x} {\cal H}_{int}$, with
\begin{eqnarray}
  {\cal H}_{int} & = & \left[ {\tilde\Delta \over 2} \left(\Psi_1^\dagger \tau^+
    i\partial_y \Psi_1^{\vphantom\dagger} -
    (i\partial_y\Psi_1^\dagger) \tau^+
    \Psi_1^{\vphantom\dagger}\right) + {\rm h.c.}\right] \nonumber \\
& & + (1 \leftrightarrow 2, x\leftrightarrow y).
\end{eqnarray}
This form is identical to the $\tilde\Delta$ term in Eq.~\ref{cqpHam}\ when the
order parameter $\tilde\Delta$ is constant, but the symmetric
placement of derivatives is important in the presence of phase
gradients.  In particular, now let $\tilde\Delta =
v_{\scriptscriptstyle\Delta}e^{i\varphi}$ and integrate by parts to
transfer the derivative in the second term from the $\Psi^\dagger$ to
the $\tilde\Delta\Psi$ combination.  Upon using the operator identity
\begin{equation}
  {1 \over 2} \left(e^{i\varphi}i\partial_y + i\partial_y
    e^{i\varphi}\right) = e^{i\varphi/2} i\partial_y e^{i\varphi/2},
\end{equation}
this becomes identical to the symmetrized form of the
phase-quasiparticle interaction hypothesized in Eq.~\ref{hqp}.

\subsection{Nodons}

Treatment of quantum phase
fluctuations is complicated by the 
coupling between the quasiparticle Fermion
operators, $\Psi$, and exponentials
of the phase $\varphi$,
as seen explicitly in ${\cal H}_{qp}$ in Eqn.~\ref{hqp}.
The form of the coupling is determined by the electric charge
carried by $\Psi$, which is uncertain - being built
from electron {\it and} hole operators.
To isolate the uncertain charge of $\Psi$
it is extremely convenient to
perform a change of variables,\cite{Nodal}
defining a new set of fermion fields $\psi_j$ via
\begin{equation}
  \psi_j = \exp(-i\varphi \tau^z/2) \Psi_j  .
  \label{transform}
\end{equation}
In the superconducting phase, and in the absence of quantum flucutations
of the order-paramater phase, one can set $\varphi = 0$,
and these new fermions are simply the d-wave quasiparticles.
However, when the field $\varphi$ is dynamical and fluctuates
strongly this change of variables is non-trivial.  In particular,
the new fermion fields $\psi$ are electrically {\sl neutral},
invariant under a global $U(1)$ charge transformation
(since $\varphi \rightarrow \varphi + 2\theta_0$ under
the $U(1)$ charge transformation in Eqn.~\ref{U1}).  As we shall see,
when the d-wave superconductivity is quantum disordered,
these new fields will play a fundamental role, describing
low energy gapless excitations, centered at the former
nodes.  For this reason, we refer to these fermions as {\sl nodons}.
For completeness, we quote the symmetry properties of the nodon 
field under a particle/hole transformation.
Since $\varphi \rightarrow -\varphi$, one has simply
\begin{equation}
\psi \rightarrow \psi^{\dagger}.
\end{equation}

%while under time-reversal,
%\begin{equation}
%{\rm T}:\;\; \psi \rightarrow i\tau^{y}\psi, \qquad \overline\psi 
%\rightarrow \overline\psi i\tau^{y}.
%\end{equation}
%

The full Lagrangian in the d-wave superconductor, ${\cal L} = {\cal
  L}_{\varphi} + {\cal L}_{qp}$, can be conveniently re-expressed in
terms of these nodon fields since ${\cal L}_{qp} = {\cal L}_{\psi} +
{\cal L}_{int} + {\cal L}_\lambda$ with a free nodon piece,
\begin{eqnarray}
  {\cal L}_{\psi} & = & \psi_1^\dagger [ i \partial_t - v_F \tau^z
  i\partial_x - v_{\scriptscriptstyle\Delta} 
  \tau^x i\partial_y ] \psi_1 \nonumber \\
  & & + (1 \leftrightarrow 2,x \leftrightarrow y)  ,
  \label{Lpsi}
\end{eqnarray}
interacting with the phase of the order-parameter:
\begin{equation}
  {\cal L}_{int} = \partial_\mu \varphi J_\mu .
\end{equation}
Here the electrical 3-current $J_\mu$ is given by
\begin{equation}
  J_0 = {1 \over 2} \psi_j^\dagger \tau^z \psi_j^{\vphantom\dagger}
  ,
  \label{eq:J0}
\end{equation}
\begin{equation}
  J_j = {v_F \over 2} \psi_j^\dagger \psi_j^{\vphantom\dagger}  .
  \label{eq:Jj}
\end{equation}
Because the transformation in Eq.~\ref{transform}\ is local, identical
expressions hold for these currents in terms of the quasiparticle
fields, $\Psi$.  The form of the particle/hole asymmetry term
remains the same in terms of the nodon fields:
\begin{equation}
{\cal L}_\lambda = \lambda \psi^\dagger_j \tau^z \psi_j .
\label{Llambda}
\end{equation}

It is instructive to re-express the components of
the currents $J_\mu$ back in terms of the original electron operators.  One finds
\begin{equation}
  J_0 = {1 \over 2} (c_{\bbox{K}_j}^\dagger c_{\bbox{K}_j}^{\vphantom\dagger} +
  c_{-\bbox{K}_j}^\dagger c_{-\bbox{K}_j}^{\vphantom\dagger}) , 
\end{equation}
(with an implicit spin summation) which corresponds physically
to the total electron density living at the nodes,
in units of the Cooper pair charge.  Similarly,
\begin{equation}
  J_j = {v_F \over 2} (c_{\bbox{K}_j}^\dagger c_{\bbox{K}_j}^{\vphantom\dagger} - c_{-\bbox{K}_j}^\dagger c_{-\bbox{K}_j}^{\vphantom\dagger}) 
\end{equation}
corresponds to the {\it current} carried by the electrons at the nodes.
Thus, $J_\mu$ can be correctly interpreted as
the quasiparticles three-current. 

To complete the description of a quantum mechanically fluctuating order
parameter phase interacting with the gapless fermionic excitations at the
nodes, we minimally couple to an external electromagnetic field,
$A_\mu$.  Since the nodon fermions are neutral, the only coupling is
to the order-parameter phase, via the substitution $\partial_\mu \varphi
\rightarrow \partial_\mu \varphi - 2 A_\mu$.  Here we have set the
electron charge $e=1$, with a factor of $2$
appropriate for Cooper pairs.
The final Lagrangian then takes the form ${\cal L} = {\cal
  L}_{\varphi} + {\cal L}_{\psi} + {\cal L}_{int} + {\cal L}_\lambda$, with
\begin{equation}
\label{Lphi}
  {\cal L}_{\varphi} = {1\over 2} \kappa_\mu (\partial_\mu \varphi - 2 A_\mu)^2 ,
\end{equation}
\begin{equation}
\label{Lint}
  {\cal L}_{int} = (\partial_\mu \varphi - 2 A_\mu ) J_\mu  ,
\end{equation}
and ${\cal L}_{\psi}$ still given by Eq.~\ref{Lpsi}.  Here we have
dropped the Berry's phase term, which is not expected
to play an important role in the superconducting phase.  
Long-ranged Coulomb interactions could be readily incorporated
at this stage by treating $A_0$ as a dynamical field and adding a term
to the Lagrangian of the form, ${\cal L}_{coul} = (1/2) (\partial_j
A_0)^2$.  The spatial components of the electromagnetic field, $A_j$,
have been included to keep track of the current operator.  

\subsubsection{Symmetries and Conservation Laws}

If the full effective Lagrangian ${\cal L}$ is to correctly describe the low energy physics it must exhibit the same symmetries as the original
electron Hamiltonian - the most important being
charge and spin conservation.  Since the $\psi$ operators
are electrically neutral the full $U(1)$
charge transformation is implemented by $\varphi \rightarrow 
\varphi + 2\theta_0$ for constant $\theta_0$,
and ${\cal L}$ is indeed invariant.  Moreover,
the Lagrangian is invariant under $\psi_\alpha \rightarrow U_{\alpha \beta} \psi_{\beta}$ for arbitrary (global) $SU(2)$ spin 
rotations $U=exp(i\bbox{\theta} \cdot \bbox{\sigma})$.
Since the Cooper pairs are in spin singlets,
{\it all} of the spin is carried by the
nodons.

As usual, associated with each continuous symmetry is
a conserved ``charge" which satisfies a continuity equation
(Noether's theorem).  Since the Lagrangian only depends
on gradients of $\varphi$, the Euler-Lagrange equation of motion
reduces to the continuity equation,
\begin{equation}
\partial_\mu J_\mu^{tot} = 0  ,
\end{equation}
where the total electric 3-current
is given by $J_\mu^{tot} = \partial {\cal L}/\partial(\partial_\mu \varphi)
= - \partial {\cal L} / \partial A_\mu$.
This gives,
\begin{equation}
  J^{tot}_\mu = \kappa_\mu
  (\partial_\mu \varphi - A_\mu) + J_\mu ,
  \label{Jtot}
\end{equation}
where the first term is the Cooper pair 3-current and the second
the quasiparticles current.  

The analogous conserved {\it spin} currents can be
obtained by considering infinitesimal spin rotations,
\begin{equation}
U=1 + i\bbox{\theta}(\bbox{x},t) \cdot \bbox{\sigma} ,
\end{equation}
for {\it slowly} varying $\bbox{\theta}(\bbox{x},t)$.
Under this spin rotation the Lagrangian transforms as,
\begin{equation}
{\cal L} \rightarrow {\cal L} + \partial_\mu \bbox{\theta} \cdot \bbox{j}_\mu  ,
\end{equation}
with $\bbox{j}_\mu$ given below.
After an integration by parts, invariance of the action $S$
under global spin rotations implies continuity
equations $\partial_\mu \bbox{j}_\mu = 0$
for {\it each} of the three {\it spin} polarizations, $\bbox{j}$.
The space-time components of the conserved spin currents are given explicitly by,
\begin{equation}
\label{spindens}
\bbox{j}_0 = {1 \over 2} \psi_1^\dagger \bbox{\sigma} \psi_1 + (1 \rightarrow 2) ,
\end{equation}
\begin{equation}
\bbox{j}_x = {1 \over 2} v_F \psi_1^\dagger \bbox{\sigma} \tau^z \psi_1 + 
{1 \over 2} v_{\scriptscriptstyle\Delta} \psi_2^\dagger \bbox{\sigma} \tau^x \psi_2  ,
\end{equation}
and $\bbox{j}_y$ the same as $\bbox{j}_x$ except with $\psi_1 \leftrightarrow \psi_2$.  Notice that in contrast
to the electrical current, the spin current operator
has a contribution which is proportional
to the velocity tangential to the Fermi surface,
$v_{\scriptscriptstyle\Delta}$, which is anomalous
when re-expressed in terms of the original electron operators.

Surprisingly, the effective Lagrangian exhibits
{\it additional} continuous symmetries, {\it not}
present in the original Hamiltonian.
Firstly, ${\cal L}$ is invariant under {\it separate}
$SU(2)$ spin rotations on the
two pairs of nodes, $\psi_j$ for $j=1,2$.
Moreover, the Lagrangian is also invariant 
under two additional $U(1)$ transformations 
$\psi_{j} \rightarrow e^{i\theta_{j}}
\psi_{j}$ for arbitrary constant phases, $\theta_{j}$.
These latter symmetries imply two new conserved
``charges", $\psi^\dagger_j \psi_j$ (no sum on $j$).  
We refer to these conserved quantities as ``nodon charges". 
The associated conserved nodon 3-currents
take the same form as the spin currents above,
except replacing $\bbox{\sigma}/2$ by the identity.     
As seen from Eq.~\ref{eq:Jj}, 
the conserved nodon charges are proportional to the quasiparticle electrical
{\it current}, since $J_j = (v_F/2)
\psi_j^\dagger \psi_j$.  

It is possible
to add to ${\cal L}$ additional
interaction terms which are consistent
with the original $U(1)$ and $SU(2)$ symmetries,
but do {\it not} conserve the ``nodon charge".
Specifically, anomalous quartic interaction terms
of the form $\psi^4$ arise from Umklapp
scattering processes in the original
electron Hamiltonian and clearly change the nodon charge.
However, such interactions
are unimportant at low energies
due to severe phase space restrictions.
To see this, consider how
the action, $S = \int d^2  x dt {\cal L}$ transforms
under a renormalization group (RG) rescaling transformation,
\begin{equation}
\label{rescale2+1}
x_\mu \rightarrow b x_\mu ; \hskip 0.5cm \psi \rightarrow b^{-1} \psi ;
\hskip 0.5cm \varphi \rightarrow b^{-1/2} \varphi  ,
\end{equation}
with rescaling parameter $b>1$.
By construction, this leaves the quadratic pieces
$S_\psi$ and $S_\varphi$ invariant,
but interaction terms such as $u\psi^4$
scale to zero under the RG ($b \rightarrow \infty$)
since $u \rightarrow u/b$.  It is the $T=0$ 
``fixed point" theory described by the quadratic terms
which exhibits the additional symmetries.
Incidentally,
the coupling term ${\cal L}_{int}$ above
also scales to zero (as $b^{-1/2}$) under the renormalization group.
In the resulting quadratic theory the quasiparticles
and phase fluctuations actually decouple.

\subsubsection{Superfluid stiffness}

The above effective theory
is particularly convenient for examining
very low temperatures properties of the $d_{x^2-y^2}$
state.  Of interest are charge response functions such as
the electrical conductivity and the superfluid
stiffness (measureable via the penetration length).
The spin excitations (carried by the quasiparticles)
can also be probed via resonance techniques, such as NMR and ESR.
Impurity scattering can be readily incorporated
by coupling a random potential to the electron density
(which can be re-expressed as a nodon bi-linear).

For illustrative purposes we briefly consider
the quasiparticle contribution to the low temperature
superfluid stiffness and 
extract the famous $T$-linear dependence.  
For a Galilean
invariant system of mass $m$
bosons the superfluid stiffness $K_s$ equals the 
superfluid density divided by $m$.   
But more generally $K_s$ can be extracted
rather directly by considering the response of the system to a
{\it transverse} vector potential.\cite{Pines}  We set $A_0=0$
and decompose the {\it static} vector potential
$A_j$ into longitudinal and transverse pieces:
\begin{equation}
A_j = A_{\ell,j} + A_{t,j}  ,
\end{equation}
with $\partial_j A_{t,j} =0$ and $\epsilon_{ij} A_{\ell,j} =0$.
The superfluid stiffness is then given by,
\begin{equation}
K_s = {1 \over V} { {\partial^2 F}  \over {\partial A_{t,x}^2 } }  ,
\end{equation}
where $F=-k_B T ln Z$ is the Free energy and $V \rightarrow \infty$ is the
area of the 2d system.  Here $A_{t,x}$ can be taken
spatially constant.  

To extract $F$ the partition function can be written
as an imaginary time coherent state path integral,\cite{Negele}
\begin{equation}
Z = \int [{\cal D} \varphi ][ {\cal D} \psi ][ {\cal D} \overline{\psi}]
exp(-S_{\scriptscriptstyle E})   ,
\end{equation}
with Euclidian action $S_{\scriptscriptstyle E} = \int d^2x d\tau {\cal L}_{\scriptscriptstyle E}$.  The longitudinal vector potential,
which can be expressed as a gradient of a scalar field
$A_{\ell,j} = \partial_j \Lambda$, can be eliminated
entirely by shifting $\varphi \rightarrow \varphi + \Lambda$.
Moreover, the crossterm between $\partial_j \varphi$
and $A_{t,j}$ vanishes since $A_t$ is
divergenceless.  The Gaussian integral over 
$\varphi$ can then be readily perfomed
and simply generates an irrelevant interaction term
($J \sim (\psi^\dagger \psi)^2$) which can be
ignored.  One thereby arrives
at an effective action depending only
on $\psi$ and $A_j$ with associated Hamiltonian
density of the form:  ${\cal H}_{eff} = {\cal H}_\psi + {\cal H}_{\scriptscriptstyle A}$,
with ${\cal H}_\psi$ the free nodon Hamiltonian and
\begin{equation}
{\cal H}_{\scriptscriptstyle A} = {1 \over 2} K_s^0 A_{t,j}^2 + A_{t,j} J_j  .
\end{equation}
Here $K_s^0 = \kappa_0 v_c^2$ is the superfluid stiffness
from the Cooper pairs, and $J_j=(v_{\scriptscriptstyle F} /2) \psi_j^\dagger \psi_j$.  Notice that the
(transverse) vector potential acts as an
{\it effective} chemical potential for the ``nodon charge" density, $\rho_n = \psi_j^\dagger \psi_j$.  Thus, the superfluid stiffness can be expressed
in terms of the nodon ``compressibility" as
\begin{equation}
K_s = K_s^0 - (v_{\scriptscriptstyle F} /2)^2 \kappa_n  ,
\end{equation}
where $\kappa_n = \partial \rho_n/\partial \mu_n$ and
$\mu_n = (v_{\scriptscriptstyle F}/2) A_{t,x}$ is the nodon ``chemical potential".

The nodon compressibility can be extracted by diagonalizing
the Hamiltonian, $H_\psi$.
From the first pair of nodes one obtains the free Fermion form, 
\begin{equation}
H_\psi = \sum_{\bbox{q}} E_1(\bbox{q}) [a^\dagger_{\bbox{q}} a_{\bbox{q}}
+ b^\dagger_{\bbox{q}} b_{\bbox{q}}] ,
\end{equation} 
where $E_1(\bbox{q})$ is given in Eqn.~\ref{E(q)} and we have
suppressed the spin index.
Here $a$ and $b$ are particle and hole operators, respectively.
The nodon charge is simply,
\begin{equation}
\rho_n = {1 \over V} \sum_{\bbox{q}} [\langle a^\dagger_{\bbox{q}} a_{\bbox{q}}
\rangle - \langle b^\dagger_{\bbox{q}} b_{\bbox{q}} \rangle ],
\end{equation}
where the averages are taken with ${\cal H}_\psi - \mu_n \rho_n$.
At finite temperatures one obtains
\begin{equation}
\rho_n = 2 \int { {d \bbox{q}} \over {(2\pi)^2 }}  [f(E_1(\bbox{q})-\mu_n)
- f(E_1(\bbox{q}) + \mu_n) ] ,
\end{equation}
where $f(E)$ are Fermi functions, and the factor of $2$ is from
the spin sum.  Finally, upon differentiating with respect
to $\mu_n$ and performing the momentum integral
one extracts the desired result for
the low temperature superfluid stiffness:
\begin{equation}
K_s(T) = K_s^0 - c {v_{\scriptscriptstyle F} \over v_{\scriptscriptstyle \Delta}} k_B T   ,
\end{equation}
with the dimensionless constant $c=(ln 2/2\pi)$.

\section{VORTICES}
\label{sec:Duality}

\subsection{hc/2e versus hc/e vortices}

Having successively incorporated
phase fluctuations into the effective low
energy description of the $d_{x^2-y^2}$ state,
we now turn to a more interesting task - quantum
disordering the superconductivity to obtain
the nodal liquid phase, a novel Mott insulator.
The superconductivity is presumed
to be destroyed by strong {\it quantum}
fluctuations of the order parameter
phase $\varphi$ driven by vortex
excitations.   In two-dimensions vortices are 
simply whorls of current
swirling around a core region.
But in a superconductor the circulation
of such vortices is {\it quantized},
since upon encircling the core
the phase $\varphi$ can only change
by integer multiples of $2\pi$.  
Inside the core of a vortex the {\it magnitude}
of the complex order parameter $|\tilde{\Delta}|$ vanishes,
but is essentially constant outside.
In the superconducting phase, the size of the core is set
by the coherence length - 
roughly $10 \AA$ in
the Cuprate materials.  Such vortices are
thus tiny ``point-like" objects, with a truly microscopic size
in the Cuprate materials.

The ``elementary"
vortex has a phase winding of $\pm 2\pi$.
When a superconductor is placed in an external magnetic field,
the currents circulating around the core of a vortex
tend to screen out the magnetic field, except within
a region of the penetration length, $\lambda$, from the vortex core.
(In the cuprate materials $\lambda$ is in the range of a thousand
angstroms.)  In addition to the circulation,
the total magnetic flux
near a vortex is {\it quantized} - in units of the flux quantum $hc/2e$.
An ``elementary" vortex quantizes precisely $hc/2e$
of magnetic flux, and will thus henceforth be
referred to as an $hc/2e$ vortex.  As we shall argue,\cite{Nodal2}
to obtain the nodal liquid phase it will be necessary
to ``liberate" double-strength $hc/e$ vortices,
keeping the $hc/2e$ vortices ``confined".

Generally, the position of these ``point-like" vortices
can change with time, and their dynamics
requires a quantum mechanical description.
Thus a collection of many vortices can be viewed as a 
many body system of ``point-like" particles.
Since positive ($+1$) and negative ($-1$) circulation vortices can
annihilate - and disappear (just as for real elementary particles 
like electrons and positrons), they behave
as ``relativistic" particles.  There is a conserved vortex ``charge" in this process, namely the total circulation, and an associated current.  
Since the Cooper pairs are Bosons, one anticipates
that the ``dual particles" - the vortices - are also
Bosonic forming a relativistic Boson system, and this is indeed the case.\cite{Duality}

However, in the superconducting phase at zero temperature
there are {\it no} vortices present - this phase
constitutes a
``vacuum" of vortices.  More precisely, due to quantum fluctuations
vortices are present as 
short-lived ``virtual" fluctuations, popping out of the ``vacuum"
in the form of small tightly bound (neutral) pairs.  For the low energy
properties of the superconductor these fluctuations can
be largely ignored.
But what happens if these
virtual pairs unbind into a proliferation of free
mobile vortices?  Vortex motion is very effective
at scrambling the 
phase $\varphi$ of the superconducting order,
so that mobile vortices will in fact
destroy the superconductivity.  
Since the vortices are Bosonic, once they are
free and mobile they will ``Bose condense",
at least at zero temperature.  
One thereby obtains a non-superconducting
insulating state, with the ``vortex-condensate"
serving as an appropriate order parameter.
As we shall see, it will be extremely
convenient to pass to a ``dual" representation\cite{Duality,fDuality} 
in which the vortices
are the basic ``particles" - rather than the Cooper pairs.

Consider first unbinding and condensing the ``elementary"
$hc/2e$ vortices.\cite{Nodal2}  When a Cooper pair is taken around such
a vortex it's wave function acquires a $\pm 2\pi$ phase change.
Likewise, when an $hc/2e$ vortex is taken around a Cooper pair,
the {\it vortex} wavefunction acquires the $2\pi$ phase change.
Thus, $hc/2e$ vortices ``see" Cooper pairs as a source
of ``dual flux", each carrying one unit.  (This notion
can be made precise by performing a duality
transformation - see below and the Appendix.)  
For a Hubbard model of electrons at half-filling, on average
there is one-half of a Cooper pair per site,
as seen explicitly in the effective lattice Cooper pair
Hamiltonian, Eqn.~\ref{Cooplatt}, which has offset charge
$n_0 =1/2$.  Thus, these elementary vortices
``see" a dual ``magnetic field", with one-half
of a dual flux-quantum per plaquette.   When the $hc/2e$ vortices
unbind and condense, they will quantize this dual
flux, in precisely the same way that the condensation
of Cooper pairs in a real superconductor will quantize an applied
magnetic field - forming an Abrikosov flux-lattice (if Type II).    
The analog of the Abrikosov flux-lattice
for the  $hc/2e$ {\it vortex condensate} is
an ordered lattice of {\it Cooper pairs}.  In this ``crystal"
state at half-filling,
the Cooper pairs will preferentially sit on one of the two
equivalent sub-lattices of the square lattice.  
This state can be described as a
commensurate charge-density-wave with ordering wavevector 
(${\bf Q} = \pi,\pi)$, which spontaneously
breaks the discrete symmetry under translation
by one lattice spacing.  Such ordering implies a considerable
degree of double occupancy for the electrons,
and thus seems most reasonable for a Hubbard type model
with an attractive on-site interaction (negative $u$).  In the Cuprate
materials there is a strong on-site {\it repulsion}, and moreover
there is no evidence for ``charge-ordering"
near ${\bf Q}$.  Thus, for a description of the pseudo-gap
regime in the Cuprate materials, we can rule out
the $hc/2e$ vortex-condensate on phenomenological grounds.

Instead, we consider the possibility of unbinding
and condensing double-strength $hc/e$ vortices,
keeping the elementary $hc/2e$ vortices confined.\cite{Nodal2}
When an $hc/e$ vortex is taken around a Cooper pair
it acquires a $4\pi$ phase change.  A $2\pi$ phase
change corresponds to taking such an $hc/e$ vortex around ``half"
of a Cooper pair - which has charge $e$.
Thus, a condensation of $hc/e$ vortices should correspond
to a ``crystal" of such charge $e$ objects.  But at half-filling
with charge $e$ per lattice site, this should correspond
to a state {\it without} charge ordering
or translational symmetry breaking.
As we shall see, for a $d_{x^2 - y^2}$ superconductor
the resulting $hc/e$ ``vortex-condensate" gives a description
of the nodal liquid phase. 

This procedure - keeping the elementary $hc/2e$ vortices confined
and only liberating the $hc/e$ vortices -
is responsible for the remarkable properties
of the nodal liquid.\cite{Nodal2}  To see why, consider first the Berry's phase
term in Eqn.~\ref{Berry}.  With only $hc/e$ vortices present,
the Cooper pair phase, $\varphi$, only winds
by integer multiples of $4\pi$ - {\it not} $2\pi$.  
At half-filling (with $n_0=1/2$) the Berry's phase term
will {\it not} contribute to the partition function
(see Eqn.~\ref{expBerry}) and can thus be dropped
entirely in the description of the nodal liquid. 
This can be implemented by defining a
new phase field:
\begin{equation}
\phi = \varphi /2 ,
\end{equation}
and only allowing vortices
in $\phi(x)$ with circulation $2\pi$ times an integer.
This restriction precludes $hc/2e$ vortices,
and guarantees that the field
\begin{equation}
b = e^{i \phi}  ,
\end{equation}
is {\it single-valued}.  As an operator,
$b$ creates a spinless excitation with charge $e$.
When re-written in terms of
$\phi$, the effective Lagrangian for a d-wave superconductor
with quantum phase fluctuations (from Eqns.~\ref{Lphi},\ref{Lint}) becomes 
${\cal L} = {\cal
  L}_{\phi} + {\cal L}_{int} + {\cal L}_{\psi} $
with 
\begin{equation}
\label{LphiA}
  {\cal L}_{\phi} +  {\cal L}_{int} = 
{1\over 2} \kappa_\mu (\partial_\mu \phi -  A_\mu + \kappa_\mu^{-1}
J_\mu)^2 ,
\end{equation}
and ${\cal L}_{\psi}$ given in Eq.~\ref{Lpsi}.  The Berry's
phase term has been dropped, since it plays no role
when $exp(i\phi)$ is a single valued field.
Here, we have absorbed a factor of two
into $\kappa_\mu$ and also completed the square
with the nodon current, $J_\mu$, dropping
order $J_\mu^2$ terms which are irrelevant as discussed
after Eq.~\ref{rescale2+1}.  
Notice that the coefficient of $A_\mu$ is one - as
expected for a charge $e$ operator $exp(i\phi)$.
By precluding $hc/2e$ vortices, we see the emergence
of a new bosonic field, $exp(i\phi)$, with exotic
quantum numbers - charge $e$ but spin zero - 
which will be referred to as a ``holon".
This is the first hint of spin-charge separation\cite{Anderson,krs,Affleck}
in the nodal liquid.

As we shall see, another remarkable
consequence of precluding
$hc/2e$ vortices,
is that the charge neutral spin one-half
nodons survive under $hc/e$ vortex
condensation
into the nodal liquid.
To see why this is {\it not}
the case if elementary $hc/2e$ vortices
are condensed\cite{Nodal2} (as in the charge-density-wave), it is very instructive
to consider the transformation
which relates the nodons to
the d-wave quasiparticles, Eq.~\ref{transform},
which can be written in terms of the new field $\phi$ ($=\varphi/2$)
as:
\begin{equation}
\psi = exp(-i\tau_z \phi) \Psi  .
\end{equation}
In the presence of vortices,
the nodon field $\psi$ only remains
single-valued if $hc/2e$ vortices are excluded (so
that  $exp(\pm i\phi)$ is single valued).
Indeed, when a nodon is taken around an
$hc/2e$ vortex, it's wavefunction
{\it changes sign}, since $\phi$ winds by $\pi$.  This
implies a very strong and long-ranged ``statistical" interaction between
nodons and $hc/2e$ vortices.  If $hc/2e$
vortices proliferate and condense,
it will clearly be very difficult
for the nodons to propogate coherently.
In fact, we have argued recently\cite{Nodal2} that in this case
the nodons are bound (actually ``confined")
to the holons, leaving only the electron in the spectrum
of the charge-density-wave.

\subsection{Duality}

We now consider implementing the procedure of unbinding and condensing
$hc/e$ vortices in the $d_{x^2-y^2}$ superconductor.  To this end, it is extremely
convenient to pass to the ``dual" representation\cite{Duality,fDuality}
in which the vortices are the basic ``particles",
rather than the Cooper pairs.  
The most straightforward way to incorporate
$hc/e$ vortices is by placing the 
(single-valued) field $exp(i\phi)$ 
on the {\it sites} of a lattice,\cite{Duality} 
so that
vortices can exist in the {\it plaquettes}.
A lattice duality transformation can be implemented
in which the phase $\phi$ is replaced
by a dual field, $\theta$, which is the phase
of a vortex complex field, $\Phi \sim e^{i\theta}$.  
In a Hamiltonian description,
$\Phi$ and $\Phi^\dagger$ can be viewed as vortex quantum field
operators - which destroy and create $hc/e$ vortices.
On a $2+1$-dimensional Euclidian space-time lattice,
the appropirate model corresponding
to the phase Lagrangian Eq.~\ref{LphiA} is essentially
a classical 3d-xy model with an effective 
gauge field:  
\begin{equation}
\label{eff-A}
A^{eff}_\mu = A_\mu - \kappa_\mu^{-1} J_\mu    .
\end{equation}
The lattice duality transformation for the 3d-xy model
with gauge field is implemented 
in some detail
in the Appendix.  An alternative
method which we sketch below, involves
implementing the duality transformation
directly in the continuum.\cite{fDuality}

To this end we introduce a vortex 3-current, $j^v_\mu$,
which satisfies,
\begin{equation}
  j^v_\mu = \epsilon_{\mu \nu \lambda} \partial_\nu \partial_\lambda
  \phi.
  \label{jv}
\end{equation}
In the presence of $hc/e$ vortices, $\phi$ is multi-valued,
$\partial_\mu\phi$ is not curl-free, and $j_\mu^v$ is non-vanishing.
Even in the dual vortex representation
the total electrical charge must be conserved.
This can be achieved by expressing the {\it total} electrical
3-current (in units of the electron charge $e$) as a curl,
\begin{equation}
  J^{tot}_\mu = \epsilon_{\mu \nu \lambda}  \partial_\nu a_\lambda  ,
  \label{gauge}
\end{equation}
where we have introduced a ``fictitious" dynamical gauge field,
$a_\mu$.  (In the Appendix 
the electrical 3-current is expressed
as a lattice curl of $a_\mu$.)  
Upon combining Eqn.~\ref{Jtot} with \ref{jv} and \ref{gauge}, one can
eliminate the phase field, $\phi$, and relate
$a_\mu$ to the vortices:
\begin{equation}
  j^v_\mu  = \epsilon_{\mu \nu \lambda} \partial_\nu [ \kappa_\lambda^{-1}
  \epsilon_{\lambda \alpha \beta} 
  \partial_\alpha a_\beta +  A_\lambda - 
  \kappa_\lambda^{-1} J_\lambda ]  ,
  \label{eq:jva}
\end{equation}
where $J_\mu$ is the quasiparticle 3-current defined earlier in
Eqs.~\ref{eq:J0}-\ref{eq:Jj}.

In this continuum approach to duality,
a dual description is obtained by constructing a 
Lagrangian, ${\cal L}_D$, 
depending on $a_\mu$, $J_\mu$ and $j^v_\mu$, whose
equation of motion, obtained by differentiating the action with respect
to $a_\mu$, leads
to the above equation.  It is convenient
to first express the vortex 3-current in terms of
a complex field, $\Phi$, which can be viewed as an
$hc/e$ vortex destruction operator.
The dual Lagrangian is constructed to have an an
associated $U(1)$ invariance under $\Phi \rightarrow e^{i\alpha}
\Phi$, which guarantees that $j^v_\mu$ is indeed conserved.
When an $hc/e$ vortex is taken around
a Cooper pair it aquires
a $4\pi$ phase change ($2\pi$ around a charge $e$ ``holon").
In the dual representation the vortex wavefunction
$\Phi$ should acquire a $4\pi$ phase
change (or $2\pi$ for a ``holon").  This can be achieved by minimally
coupling derivatives af $\Phi$
to the ``fictitious" vector potential  
$a_\mu$.
 
The appropriate dual Lagrangian can be conveniently decomposed as
${\cal L}_D = {\cal L}_\psi + {\cal L}_v + {\cal L}_a$,
where ${\cal L}_\psi$ is given in Eq.~\ref{Lpsi}.  
The vortex piece has the Ginzburg-Landau form,\cite{Tinkham}\
\begin{equation}
\label{L-vortex}
  {\cal L}_v = {\kappa_\mu \over 2} |(\partial_\mu - ia_\mu)\Phi|^2 - V_{\Phi}
  (|\Phi|)   ,
\end{equation}
as constructed explicitly with lattice duality in the Appendix.
The vortex 3-current, following from 
$j^v_\mu = - \partial {\cal L}_v /
\partial a_\mu$, is
\begin{equation}
  j^v_\mu = \kappa_\mu {\rm Im}[\Phi^*(\partial_\mu - ia_\mu)\Phi] .
\end{equation}
For small $|\Phi|$ (appropriate close to a second order transition)
one can expand the potential as, $V_\Phi(X) = r_\Phi X^2 + u_\Phi X^4$.
The remaining piece of the dual Lagrangian is
\begin{equation}
\label{La-dual}
{\cal L}_a = {1 \over {2\kappa_0}} (e_j^2 - b^2) + a_\mu \epsilon_{\mu \nu 
\lambda} 
\partial_\nu (A_\lambda - \kappa_\lambda^{-1} J_\lambda)  ,
\label{Lgauge}
\end{equation}
with dual ``magnetic" and ``electric" fields: $b=\epsilon_{ij}
\partial_i a_j$ and $e_j = v_c^{-1} (\partial_j a_0 - \partial_0
a_j)$.  It can be verified that the dual Lagrangian has
the desired property that Eq.~\ref{eq:jva}\  follows from the equation of
motion $\delta S_D / \delta a_\mu =0$.

\section{Nodal Liquid Phase}
\label{sec:Nodal}

In this section we employ the dual representation of
the $d_{x^2-y^2}$ superconductor to analyze the quantum disordered
phase - the {\it nodal liquid}.
The dual representation comprises a complex vortex field,
which is minimally coupled to a gauge field, as well as a set
of neutral nodon fermions.  Without the nodons and in imaginary
time, the
dual Lagrangian is formally equivalent to
a classical three-dimensional superconductor at finite temperature,
coupled to a fluctuating electromagnetic field.
To disorder the d-wave superconductor, we must order
the dual ``superconductor" -- that is, condense the $hc/e$ vortices.
The nature of the resulting phase
will depend sensitively on doping, since upon doping, the dual
``superconductor" starts seeing an applied ``magnetic field".
Below, we first consider the simpler case of half-filling.
We then turn to the doped case, where two scenarios are possible
depending on whether the dual ``superconductor" is Type I or Type
II.\cite{Tinkham}\

\subsection{Half-filling}

Specialize first to the case of electrons at half-filling, with particle-hole
symmetry.  In the dual representation,
the ``magnetic field", $b$, is equal to the
deviation of the total electron density 
from half-filling.
Thus at half-filling $\langle b \rangle = 0$ and the dual Ginzburg-Landau
theory is in zero applied field.
The quantum disordered phase corresponds to condensing the
$hc/e$ vortices, setting $\langle \Phi \rangle = \Phi_0 \ne 0$.
In this dual Meissner phase the vortex Lagrangian
becomes
\begin{equation}
{\cal L}_v = {1 \over 2} \kappa_\mu \Phi_o^2 (a^t_\mu)^2  ,
\end{equation}
where $a^t$ represents the transverse piece of $a_\mu$.
It is then possible to integrate out the field $a_\mu$
which now enters {\it quadratically} in the Lagrangian.
Equvalently, $a_\mu$ can be eliminated
using the equation of motion which
follows from   $\delta S_D/\delta a_\mu =0$.
The full Lagrangian in the nodal liquid phase
is then
\begin{equation}
{\cal L}_{nl} = {\cal L}_\psi + A_\mu I_\mu  +
{\epsilon_0 \over 2} E_j^2 - {B^2 \over {2 \mu_0}}+ O\left[(\partial 
J)^2\right]   ,
\label{Lnodliqu}
\end{equation}
where we have introduced the physical magnetic and electric fields:
$B = \epsilon_{ij} \partial_i A_j$ and $E_j = \partial_j A_0 - \partial_t A_j$.  
The last two terms describe a dielectric, with
magnetic permeability  $\mu_0 = \kappa_0 \Phi_0^2$
and dielectric constant $\epsilon_0 = (\mu_0 v_c^2)^{-1}$,
with the sound velocity entering, rather than the speed of light.
The external electromagnetic field is coupled to the 3-current
$I_\mu$, which can be expressed as a bi-linear of the nodon fermions as,
\begin{equation}
I_\mu = {\epsilon_0 \over {\kappa_0^2 v_c^2}} [ \kappa_\nu \partial_\nu^2
J_\mu - \kappa_\mu \partial_\mu (\partial_\nu J_\nu)] .
\end{equation}
Notice that this 3-current is automatically conserved:
$\partial_\mu I_\mu =0$.  

The order $(\partial J)^2$ terms which we have not written
out explicitly are quartic in the 
fermion fields, and also involve two derivatives.  Since ${\cal L}_\psi$
describes Dirac fermions in $2+1$ space-time dimensions,
these quartic fermion terms are highly irrelevant, and rapidly vanish
under the rescaling transformation in Eqn.~\ref{rescale2+1}.
Thus, the low energy description
of the nodal liquid phase is
exceedingly simple.  It consists of four neutral
Dirac fermion fields --  two spin polarizations ($\alpha = 1,2$) for each of 
the two pairs of nodes.
Despite the free fermion description, the nodal liquid phase is highly
{\it non-trivial} when re-expressed in terms of the underlying
electron operators.  Indeed, the $\psi$ fermion operators are built
from the quasiparticle operators $\Psi$ in the d-wave superconductor,
but are electrically neutral, due to the ``gauge transformation" in
Eq.~\ref{transform}. 
 
In addition to the gapless nodons, one expects exotic {\it charged}
excitations at finite energy in the nodal liquid.  To see this,
imagine applying an external dual ``magnetic field" to the
Ginzburg-Landau ``superconductor", which
corresponds to a non-zero chemical potential for
the electrons.  Being in the Meissner state,
this ``field" will be screened out, so that the
internal field, $b$, which corresponds
to deviations in the electron charge density from half-filling,
will vanish.  Clearly, this corresponds to a Mott insulator\cite{Fisher89}
with the Mott gap being proportional to the dual
critical field.  In a Type II superconductor,
an internal magnetic field will be ``quantized"
into flux-tubes carrying a quantum of flux\cite{Tinkham}.
For the dual Ginzburg-Landau theory,
this corresponds to a quantization of electric
charge, with a flux tube corresponding
to charge $e$.  Thus, in the nodal liquid one expects
the presence of gapped finite energy excitations
with charge $e$.  These ``holon" excitations
are exotic since they carry {\it no} spin.
The holon is the basic topological excitation that can be created
in the $hc/e$ vortex-condensate.  The existence of
a spin one-half neutral nodon excitation and
a spinless charge $e$ holon excitation in the nodal liquid,
is a dramatic demonstration of spin-charge separation.\cite{Anderson,krs,Affleck}  The excitations in the nodal liquid
have the same quantum numbers as in the spin-charge separated
gauge theories,\cite{PALee} but are weakly interacting,
rather than strongly coupled by a gauge field.

\subsubsection{Spin response}

Although the nodons are electrically neutral they do carry {\it spin},
so the low-energy spin response in the nodal liquid can be computed from the
Dirac Lagrangian ${\cal L}_\psi$.  Moreover,
since ${\cal L}_\psi$
was not altered under the duality transformation,
the spin properties of the nodal liquid are essentially
identical to those in the $d_{x^2 -y^2}$ superconducting
phase.  As a simple example, consider the uniform
magnetic spin susceptibility, $\chi$. 
The uniform part of the electron spin operator
is given as the conserved spin
density in Eqn.~\ref{spindens}:
\begin{equation}
{\bf S} (\bbox{x}) = {1 \over 2} \psi^\dagger_{ja}(\bbox{x}) \bbox{\sigma} 
\psi_{ja}^{\vphantom\dagger} (\bbox{x})  .
\end{equation}
Being bi-linear in nodon operators
spin correlation and response functions can be 
readily computed
from the free nodon theory.  For example, the uniform 
spin susceptibility is given by
\begin{equation}
\chi = \int_0^\infty \!\! dE (-\partial f/\partial E) \rho_n(E)  ,
\end{equation}
where the nodon density of states is $\rho_n(E) = (const) E/v_F 
v_{\scriptscriptstyle\Delta}$,
and $f(E)$ is a Fermi function.  One finds $\chi \sim T/v_F 
v_{\scriptscriptstyle\Delta}$.
There are also low energy spin excitations at wavevectors which
span between two different nodes.  The associated spin operators
can be obtained by re-expressing the electron spin operator,
\begin{equation}
{\bf S}_q = {1 \over 2} \sum_k c^\dagger_{k + q} \bbox{\sigma} 
c_{k}^{\vphantom\dagger} ,
\end{equation}
in terms of the nodons.
For example, the staggered magnetization operator, ${\bf 
S}_{\bbox{\pi}}$, is found to be
\begin{equation}
  {\bf S}_{\bbox{\pi}} = {1 \over 2} \left[\psi^\dagger (\tau^y
  \bbox{\sigma} \sigma^y)  
  \psi^\dagger + {\rm h.c.}\right] .
  \label{Spi}
\end{equation}
Notice that this operator is actually ``anomalous" in terms of the conserved
nodon charge.  

In addition to carrying spin, the nodons carry energy, and so
will contribute to the thermal transport.
In the absence of scattering processes (such as Umklapp)
the finite temperature
nodon thermal conductivity
is infinite.  In practive,
impurities will scatter the nodons
and lead to a finite
thermal conductivity.  In fact, impurity scattering should also
play an important role in modifying the
spin response of the nodal liquid.

\subsubsection{Charge response}

The electrical charge properties of the nodal liquid are 
of course very different than in the superconductor.
To see this, imagine changing the chemical potential
away from $\mu=0$ which corresponds to
applying an external ``magnetic" field
to the dual Ginzburg-Landa theory: ${\cal L}_\mu = -\mu b$.
Being in the ``Meissner" phase, the electron
density will stay ``pinned" at half-filling
for 
$\mu \le \mu_c$, with $\mu_c$ the
Ginzburg-Landau critical field.
Despite the presence of this charge gap, there are low energy
current fluctuations in the nodal liquid.  Indeed,
in this phase the electrical current operator
is $I_\mu$, which is bi-linear in the nodon fermions, $\psi$.
To compute the electrical
conductivity in the nodal liquid requires
computing a two-point correlator of $I_\mu$ at zero wavevector (say in the $x-$direction)
$I_x(q=0) = (\epsilon_0/\kappa_o v_s^2) \partial_t^2 J_x(q=0)$.
But notice that $J_x(q=0)$ is proportional to a globally conserved
nodon charge, since
$J_x(\bbox{x})  = (v_F/2) \psi_1^\dagger \psi_1^{\vphantom\dagger}$.
Thus, when the nodon number is conserved one has $I_x(q=0)=0$,
and the nodons {\it do not} contribute
to the electrical conductivity. 
When impurity (or Umklapp) scattering is present, however,
the nodon number is no longer conserved, and
the nodons will contribute to the real
part of the electrical conductivity,
but only at finite frequencies.

It is instructive to briefly consider the behavior of the
electron Green's function, which can be accessed
in photo-emission and tunneling experiments.  The electron operator
$c_\alpha(\bbox{x})$ can be decomposed as a product of nodon 
and holon operators.
For example, near the node at $\bbox{K}_j$
one can write,
\begin{equation}
c_\alpha(\bbox{x}) = e^{i \bbox{K}_j \cdot \bbox{x}} 
e^{i\phi(\bbox{x})}
\psi_{j1\alpha} (\bbox{x})  + ...
\end{equation}
where $\psi$ is a nodon operator and $exp(i\phi)$
can be interpreted as a holon destruction operator.
In the nodal liquid phase,
the electron Green's function, $G(\bbox{x}, t) = \langle c^\dagger(\bbox{x},
t) c(\bbox{0},0) \rangle$ factorizes as,
\begin{equation}
G(\bbox{x}, t) = e^{i \bbox{K}_j \cdot \bbox{x}} \langle 
e^{-i\phi(\bbox{x},t)} e^{i \phi(\bbox{0},0)} \rangle
G_n(\bbox{x},t)    ,
\end{equation}
where the nodon Green's function is,
\begin{equation}
G_n(\bbox{x},t)   = 
\langle \psi^\dagger_{j1\alpha}(\bbox{x},t) 
\psi_{j1\alpha}(\bbox{0},0) \rangle  .
\end{equation}
Although $G_n(\bbox{x},t)$ decays as a power law $|x|^{-2}$ and $t^{-2}$, since creating a holon costs a finite energy
the holon Green's function is expected
to be short-ranged, decaying exponentially
in space and time.
This indicates a gap in the electron spectral function at the Fermi
energy.  

\subsection{Doping the Nodal Liquid}

We briefly discuss the effects of doping charge into
the nodal liquid phase.  In a grand canonical ensemble
this is achieved
by changing the chemical potential, $\mu = A_0$.  
In the dual Ginzburg-Landau description of the vortices,
a chemical potential acts as an applied dual field, as seen
from Eq.~\ref{Lgauge}, since 
\begin{equation}
{\cal L}_\mu = -  \mu b   .
\end{equation}
The dual magnetic field, $b = \epsilon_{ij} \partial_i a_j$,
is the total electric charge in units of $e$.
Provided the applied dual field, $\mu$, is smaller
than the critical field ($\mu_c$) of the Ginzburg-Landau theory,
the dual superconductor stays in the Meissner phase -- which
is the nodal liquid phase at half-filling.  But for
$\mu \ge \mu_c$ dual flux will penetrate the Ginzburg-Landau superconductor,
which corresponds to doping the nodal liquid.  The form
of the dual flux penetration will depend critically on
whether the dual Ginzburg-Landau theory is Type I or Type II.
Within a mean-field treatment this is determined
by the ratio of the dual penetration length, $\lambda_v$,
to the dual coherence length, $\xi_v$ (where the subscript
$v$ denotes vortices).  In particular, Type II behavior
is expected if $\lambda_v/\xi_v \ge 1/\sqrt{2}$, and Type I
behavior otherwise.  In the Ginzburg-Landau description
$\lambda_v$ determines the size of a dual flux tube,
which is essentially the size of a Cooper pair.  We thus expect
that $\lambda_v$ will be roughly equal to
the superconducting coherence length, $\xi$, which is 
perhaps $10-15 \AA$ in the cuprates.  On the other
hand, $\xi_v$ is the size of the ``vortex-core" in the dual
vortex field, and  presumably can be no smaller than the microscopic
crystal lattice spacing, $\xi_v \ge 3-5\AA $.
This reasoning suggest that $\lambda_v/\xi_v$ is probably
close to unity in the cuprates, so that either Type I or Type II behavior
might be possible - and could be material dependent.
We first consider
such Type II doping, returning below to the case of
a Type I Ginzburg-Landau theory.
 
\subsubsection{Type II Behavior}

The phase diagram of a clean three-dimensional type II superconductor
is well understood.\cite{Tinkham}\ Above the lower critical field,
$H_{c1}$, flux tubes penetrate, and form an Abrikosov flux lattice -
usually triangular.  As the applied field increases the flux tubes
start overlapping, when their separation is closer than the
penetration length.  Upon approaching the upper critical field
$H_{c2}$ their cores start overlapping, the Abrikosov flux lattice
disappears, and the superconductivity is destroyed.  

These results hold equally well
for our dual Ginzburg-Landau superconductor, except
that now the direction parallel
to the applied field is actually imaginary time.
Moreover, the
Ginzburg-Landau order parameter
describes quantum ($hc/e$) vortices, and the penetrating flux
tubes are spinless charge $e$ holons. 
Upon doping the nodal liquid with $\mu > \mu_{c1}$, charge is added
to the 2d system,
which corresponds to the penetration of dual magnetic flux. 
In this dual transcription, the resulting Abrikosov flux-lattice phase
is a Wigner crystal of holons, with one holon
per real space unit cell of the lattice.  
Upon further doping, at $\mu = \mu_{c2}$, the crystal of holons melts, and they condense - this is the d-wave superconductor.

In the holon Wigner crystal phase, translational symmetry is
spontaneously broken.  However, in a real material the Wigner crystal
will have a preferred location, determined by impurities and perhaps
crystal fields, which will tend to pin and immobilize the
crystal.  The resulting phase should be an electrical insulator.

A striking and unusual feature of the holon Wigner crystal 
is that it {\it co-exists} with the nodal liquid.
We thereby arrive at a description of a rather remarkable new phase of matter.
A Wigner crystal of doped holons co-exists with neutral
gapless fermionic excitations -- the nodons.  In this co-existing phase,
low energy spin and thermal
properties will be dominated by the nodons.  The behavior will
be qualitatively similar to that in the undoped nodal liquid phase.
It is possible that 
this phase underlies the physics of
the pseudo-gap
region of the high $T_c$ cuprates.

\subsubsection{Type I Behavior}

In a Type I superconductor, the applied field is expelled
until the critical $H_c$ is exceeded.\cite{Tinkham}\  At this point there
is a {\it first order} phase transition
from the Meissner phase with all the flux expelled,
to a normal metal phase in which (essentially) all
the field penetrates.  
If our dual Ginzburg-Landau theory is type I, then analogous properties are expected.
Specifically, as the chemical potential increases,
the dual field -- which is the holon density -- remains
at zero until a critical chemical potential $\mu_c$ is
reached.
At this point there is a first order phase transition,
between the nodal liquid phase at half-filling, and
a d-wave superconductor at finite doping, $x_c$.
At fixed doping $x < x_c$, phase separation is
impeded by long-ranged Coulomb interactions between the holons.
The system will break apart into co-existing ``micro-phases"
of nodal liquid and d-wave superconductor.
The configuration of the
``micro-phases" will
be determined by a complicated competition between 
the Coulomb energy and the (positive) energy of the
domain walls.  In practice, impurities will also probably
play a very important role.

\subsection{Closing Remarks}

The theoretical framework described 
above gives a skeletal description of the nodal
liquid and, upon doping, the holon Wigner crystal.
There are many important issues
which will need to be addressed in detail
to see if this novel Mott insulating phase
gives a correct description of the
low temperature pseudo-gap regime
in the cuprates.
At very low doping the cuprates
are antiferromagnetic so it will clearly
be necessary to incorporate magnetism
into the theoretical framework.  
Perhaps even more important is assessing
the role of impurities, which are expected
to have rather dramatic effects
both on the holon Wigner crystal and the
gapless nodons.  Impurities will tend
to disorder the Wigner crystal and will
scatter the nodons probably
leading to a finite density of states
and diffusive rather than ballistic motion.  
Since the nodons carry spin but no charge,
a rather exotic ``spin metal" phase is possible
with a finite ``spin conductivity" (but zero 
electrical conductivity) even at zero temperature.
It is also possible that
the impurities will
localize the nodons,
perhaps leading to a random singlet phase
or a spin glass.  
An additional complication is that
some materials might exhibit
phase separation upon doping (Type I rather than Type II
behavior) exhibiting micro-phase co-existence
between the antiferromagnet and the d-wave superconductor,
preempting the nodal liquid phase.
It clearly remains as a future
challenge to fully sort out the mysteries of the pseudo-gap regime.

A more general theme of these notes is that
novel spin liquid phases can sometimes
be more conveniently viewed as descendents
of superconductors  - rather than the more
traditional route via magnetism.
One can imagine quantum disordering 
other exotic superconducting phases besides the $d_{x^2-y^2}$ state,
to obtain new spin liquid phases.
Perhaps some of these phases will
appear in other systems
which exhibit finite angular momentum pairing,
such as $3-He$ and the heavy Fermion materials.

\acknowledgements 

It gives me genuine pleasure to acknowledge my
wonderful collaborators on the research
described above.  The
renormalization group
analysis of the two-leg ladder was
carried out in collaboration with
Hsiu-hau Lin and Leon Balents.  The nodal liquid
phase was introduced and analyzed in a collaboration
with Chetan Nayak and Leon Balents.
This research has been a true collective phenomena,
to which I am deeply appreciative.
I am also extremely 
grateful to Doug Scalapino
for stimulating my interest
in strongly correlated d-wave superconductors and
for numerous discussions about Hubbard ladders.
I would
like to thank Todadri Senthil for
sharing his insights about the effects of impurities
in d-wave superconductors.
This work has
been supported by the National Science Foundation under grants No.
PHY94-07194, DMR94-00142 and DMR95-28578.

\appendix

\section{Lattice Duality}

Duality plays a key role in understanding
how to quantum disorder a superconductor,
both in $1+1$ space-time dimensions (Sec.~\ref{sec:Two-leg})
and in $2+1$ (Sec.~\ref{sec:Duality}).  The key idea involves
exchanging the order parameter phase $\phi$
for vortex degrees of freedom.  In $1+1$ dimensions
these are point-like space-time vortices,\cite{Jose}
whereas in $2+1$ there are 
point like vortices in space which propogate in time.\cite{Duality}
In Sec.~\ref{sec:Duality}
we chose to work directly in the continuum in
implementing
the $2+1$ duality transformation. 
However, the physics of duality is perhaps more
accessible when carried out on the lattice.
In this Appendix we show in some detail how lattice duality
is implemented in both $1+1$ and $2+1$ dimensions.\cite{Jose,Duality}\
For simplicity we 
first Wick rotate
to Euclidian space, and rescale imaginary
time to set the charge velocity to one.
The appropriate lattice model is then 
simply a 2d square lattice or 3d cubic lattice
xy model.  In the latter case, we also want to include
a gauge-field, $A$, which
is a sum of the physical electromagnetic
field and the nodon current, as discussed
in Sec.~\ref{sec:Duality} - see Eq.~\ref{eff-A}.

The degrees of freedom which live on the sites
of the square or cubic lattice (denoted by a
vector of integers $\vec{x}$) are
the phases $\phi_{x} \in  [0,2\pi]$.
As usual, the gauge field lives on the links.
Discrete
lattice derivatives are denoted by
\begin{equation}
\triangle_\mu \phi_x = \phi_{x+ \mu} - \phi_x ,
\end{equation}
where $\mu=x,y$ for the square lattice and $\mu=x,y,z$
for the cubic lattice and $x+\mu$ denotes the nearest
neighbor site to $\vec{x}$ in the $\hat{\mu}$ direction. 
The gauge field is minimally coupled via,
\begin{equation}
\triangle_\mu \phi_x \rightarrow \triangle_\mu \phi_x + A_x^\mu  .
\end{equation}

Consider the partition function,
\begin{equation}
Z = \int_0^{2 \pi} \prod_x d \phi_x exp[\sum_{x, \mu} V_\kappa(\triangle_\mu \phi_x ) ]   .
\end{equation}
Here the periodic ``Villain" potential $V_\kappa$ is given by,
\begin{equation}
exp[V_\kappa (\triangle \phi)] = \sum_{J= - \infty}^{\infty} e^{-\kappa
J^2/2} e^{iJ\triangle \phi }   ,
\end{equation}
with integer $J$.  When $\kappa>>1$ only the
terms with $J=0,\pm1$ contribute appreciably in the sum
and this reduces to the more familiar form:
\begin{equation}
V_\kappa (\triangle \phi) = K \cos(\triangle \phi) ,
\end{equation}
with $K = 2exp(-\kappa/2)$.  

The partition function can thus be expressed
as a sum over both $\phi$ and a
vector of integers, $\vec{J}_x$,
with components $J_x^\mu$ living on the {\it links}
of the lattice:
\begin{equation}
Z = \int \prod_x d\phi \sum_{[\vec{J}]}  e^{-S} \equiv Tr_{\phi, \vec{J}}
\hskip 0.2cm e^{-S}    ,
\end{equation}
with action
\begin{equation}
S = S_0 + \sum_x i (\vec{\triangle} \cdot \vec{J}_x ) \phi_x ,
\end{equation}
\begin{equation}
S_0 = {\kappa \over 2} \sum_x |\vec{J}_x|^2   .
\end{equation}
In this form the integration over $\phi$ can
be explicitly performed giving
\begin{equation}
Z = Tr^\prime_{\vec{J}} \hskip 0.2cm e^{-S_0}  ,
\end{equation}
where the prime on the trace indicates
a divergenceless {\it constraint} at each site of the lattice:
\begin{equation}
\vec{\triangle} \cdot \vec{J}_x = 0  .
\end{equation}

In the presence of a gauge field there is an additional
term in the action of the form,
\begin{equation}
\label{SA}
S_A = i \sum_x \vec{J}_x \cdot \vec{A}_x  .
\end{equation}
It is thus clear that the integer of vectors
$\vec{J}$ can be interpreted
as a conserved electrical current flowing on the links of the lattice.
The divergenceless constraint on this
electrical 3-current can be imposed
automatically by re-expressing $\vec{J}$ as
a curl of an appropriate {\it dual} field.
Consider first the 2d case.

\subsection{Two dimensions}

To guarantee divergenceless we set the current equal
to the (2d) curl of a scalar field, $\theta_x$:
\begin{equation}
2\pi J^\mu_x = \epsilon_{\mu \nu} \triangle_\nu \theta_x  ,
\end{equation}
so that the action becomes
\begin{equation}
S_0(\theta) = {\kappa \over {8\pi^2}} \sum_{x, \mu} 
(\triangle_\mu \theta_x)^2  .
\end{equation}
To insure that $\vec{J}$ is an {\it integer}
field,  $\theta$ must be constrained to be $2\pi$ times an integer.
This additional constraint can be imposed by introduction
of yet another integer field, $n_x$, 
which will be interpreted as the (space-time) vortex
density.
The partition is thereby
re-expressed as (dropping an unimportant multiplicative constant),
\begin{equation}
\tilde{Z} = \int_{- \infty}^{\infty} \prod_x d\theta_x  \sum_{[n_x]} e^{-S}  ,
\end{equation}
with 
\begin{equation}
S = S_0(\theta) + \sum_x [{\tilde{\kappa} \over 2} n_x^2 + in_x \theta_x ] .
\end{equation}
For $\tilde{\kappa} =0$ the summation over $n_x$ gives
a sum of delta functions restricting $\theta_x/2\pi$
to be integer.  But we have softened this constraint,
introducing a vortex ``core" energy 
$\tilde{\kappa} \ne 0$.

At this stage
one could perform the Gaussian integral over $\theta$,
to obtain a logarithmically interacting plasma
of (space-time) vortices.  Alternatively, for $\tilde{\kappa} >>1$
the summation over $n_x$ can be performed
giving,
\begin{equation}
S = S_0(\theta) - u \sum_x cos(\theta_x) ,
\end{equation}
with $u = 2exp(-\tilde{\kappa}/2)$.  Upon taking the continuum limit,
$\theta_x \rightarrow \theta(x)$, one recovers the (Euclidian)
sine-Gordon
theory, $S = \int d^2 x{\cal L}$
with 
\begin{equation}
{\cal L} = {\kappa \over {8\pi^2}} (\vec{\nabla} \theta)^2 - u \cos(\theta)  .
\end{equation}
After Wick rotating back to real time
and restoring the velocity this takes the
identical form to the dual Lagrangian considered
for the 2-leg ladder in Sec.~\ref{sec:Two-leg}.

\subsection{Three dimensions}

In three dimensions the divergenceless integer 3-current
$\vec{J}$ can be written as the curl of a {\it vector} field, $\vec{a}$:
\begin{equation}
\label{Jcurla}
2\pi \vec{J}_x = \vec{\triangle} \times \vec{a}_x .
\end{equation}
As in 2d one imposes the integer constraint (softly)
by introducing an integer vortex field, in this
case a 3-vector $\vec{j}$, to express the partition function as,
\begin{equation}
\tilde{Z} = \int_{- \infty}^{\infty} \prod_x d\vec{a}_x  \sum_{[\vec{j}_x]} e^{-S}  ,
\end{equation}
with
\begin{equation}
S = S_0(\vec{a}) + \sum_x [{\tilde{\kappa} \over 2} |\vec{j}_x|^2 - i\vec{j}_x \cdot \vec{a}_x ] ,
\end{equation}
\begin{equation}
S_0 (\vec{a}) =  {\kappa \over {8\pi^2}} \sum_{x} 
|\vec{\triangle} \times \vec{a}_x|^2  .
\end{equation}
The integer vector field $\vec{j}$ is
the vortex 3-current, ``minimally" coupled
to $\vec{a}$. 
To see that the vortex 3-current is conserved,
it is convenient to decompose
the vector field $\vec{a}$ into
transverse and longitudinal pieces:
$\vec{a} = \vec{a}_t - \vec{\triangle}
\theta$, with $\theta_x$ a scalar field.  The action becomes,
\begin{equation}
S = S_0(\vec{a}) + \sum_x [{\tilde{\kappa} \over 2} |\vec{j}_x|^2 + i\vec{j}_x \cdot (\vec{\triangle} \theta_x - \vec{a}_x) ] ,
\end{equation}
where we have dropped
the subscript ``$t$" on $\vec{a}$.
The partition function follows from integrating over both $\vec{a}$ {\it and} $\theta$ and summing over integer $\vec{j}$.  
Integrating over $\theta$ leads to the expected condition:  
$\vec{\triangle} \cdot \vec{j} =0$.  
Alternatively, for $\tilde{\kappa} >>1$ one can perform
the summation over $\vec{j}$ to arrive
at an action depending on $\theta$ and $\vec{a}$:
\begin{equation}
S = S_0(\vec{a}) - K \sum_{x,\mu} \cos(\triangle_\mu \theta_x -
a^\mu_x )    ,
\end{equation}
with $K = 2 exp(-\tilde{\kappa}/2)$.

In the presence of
a gauge field $A^\mu$ there is an additional
term in the action of the form,
\begin{equation}
S_A = {i \over {2\pi}} \sum_x (\vec{\triangle} \times \vec{a}_x) 
\cdot \vec{A}_x   ,
\end{equation}
which follows directly from Eqn.~\ref{SA} and \ref{Jcurla}.

At this stage one can take the continuum limit,
letting $\vec{a}_x \rightarrow \vec{a}(x)$ and
$\theta_x \rightarrow \theta(x)$.  Upon expanding the cosine
for small argument one obtains
$S = \int d^3x {\cal L}$ with (Euclidian) Lagrangian
\begin{equation}
\label{L-cont}
{\cal L} = {\kappa \over {8\pi^2}} (\vec{\nabla} \times \vec{a})^2
+ {K \over 2} (\vec{\nabla} \theta - \vec{a})^2   .
\end{equation}
In this dual representation, the vortex 3-current
(which follows from $\partial {\cal L} / \partial \vec{a}$)
is given by $\vec{j}^v = K(\vec{\nabla} \theta - \vec{a})$.
Notice that the vortices are minimally coupled 
to the ``vector potential" $\vec{a}$, whose curl
equals the electrical 3-current.  The field $\theta$ can
be interpreted as the phase of a vortex operator.
In fact it is convenient to introduce
such a complex vortex field before taking
the continuum limit:
\begin{equation}
e^{i\theta_x} \rightarrow \Phi (\vec{x})    .
\end{equation}
The continuum limit can then be taking
{\it retaining} the full periodicity of the cosine potential.
The appropriate vortex Lagrangian replacing the second term 
in Eqn.~\ref{L-cont} is,
\begin{equation}
{\cal L}_v = {K \over 2} |(\vec{\nabla} - i \vec{a})\Phi|^2 + V_\Phi(|\Phi|) .
\end{equation}
The vortex current operator becomes,
\begin{equation}
\vec{j}^v = K Im[\Phi^* (\vec{\nabla} - i \vec{a})\Phi]  .
\end{equation}
If the potential is expanded for
small $\Phi$ as $V_\Phi(X) = r_\Phi X^2
+ u_\Phi X^4$, the full
dual theory is equivalent to a Ginzburg-Landau theory
for a classical three-dimensional superconductor. 
Inclusion of the original gauge field $A^\mu$
leads to an additional term in the dual Lagrangian:
\begin{equation}
{\cal L}_A = {i \over {2\pi} } (\vec{\nabla} \times \vec{a}) \cdot \vec{A} .
\end{equation}
After Wick rotating back to real time
and restoring the velocity, ${\cal L} + {\cal L}_A$ becomes
identical to the dual vortex Lagrangian in 
Eqns.~\ref{L-vortex} and \ref{La-dual}.

\end{multicols}

\end{document}